\documentclass[%
 reprint,
 amsmath,amssymb,
 superscriptaddress,
 aps
]{revtex4-2}

\usepackage[whole]{bxcjkjatype}

\usepackage{adjustbox}
\usepackage{enumitem} 
\usepackage{graphicx} 
\usepackage{svg} 
\usepackage{dcolumn} 
\usepackage{bm} 
\usepackage{enumerate}
\usepackage{academicons}
\usepackage{mathtools}
\usepackage{physics}
\usepackage{orcidlink}
\usepackage{soul}
\usepackage{url}
\usepackage{textcomp}

\makeatletter
\let\MYcaption\@makecaption
\makeatother
\usepackage{subcaption}
\makeatletter
\let\@makecaption\MYcaption
\makeatother
\usepackage{hyperref}
\usepackage{multirow}

\usepackage{xcolor}

\begin{document}

\title{Parameter Estimation of Ringdown Quasinormal Modes with Autoencoder}

\author{Momoka Iida\orcidlink{0009-0008-7811-9446}}
\affiliation{Graduate School of Integrative Science and Engineering, Tokyo City University, 1-28-1 Tamazutsumi, Setagaya-ku, Tokyo 158-8557, Japan}%
\affiliation{Research Center for Space Science, Advanced Research Laboratories, Tokyo City University, 1-28-1 Tamazutsumi, Setagaya-ku, Tokyo 158-8557, Japan}

\author{Hayato Motohashi\orcidlink{0000-0002-4330-7024}}
\affiliation{Department of Physics, Graduate School of Science, Tokyo Metropolitan University, 1-1 Minami-Osawa, Hachioji, Tokyo 192-0397, Japan}

\author{Hirotaka Takahashi\orcidlink{0000-0003-0596-4397}}
\affiliation{Research Center for Space Science, Advanced Research Laboratories, Tokyo City University, 1-28-1 Tamazutsumi, Setagaya-ku, Tokyo 158-8557, Japan}
\affiliation{Graduate School of Information and Data Science and Department of Design and Data Science, Tokyo City University, 3-3-1 Ushikubo-Nishi, Tsuzuki-ku, Yokohama, Kanagawa 224-8551, Japan}%
\affiliation{Earthquake Research Institute, The University of Tokyo, 1-1-1 Yayoi, Bunkyo-ku, Tokyo 113-0032, Japan}

\date{\today}

\begin{abstract}
Ringdown gravitational waves from binary black hole mergers can be modeled as superpositions of quasinormal modes (QNMs), whose frequencies and excitation factors encode properties of the remnant Kerr black hole.
Reliable extraction of multiple QNM components is challenging because of mode overlap and noise.
We develop an autoencoder-based framework for multi-component QNM analysis, in which the latent space is trained to represent the physical parameters of individual modes, enabling waveform denoising and parameter estimation within a common framework.
Using controlled model waveforms constructed as finite sums of Kerr QNMs with recently established high-precision frequencies and excitation factors, including their nontrivial spin dependence near resonant excitation, we assess the method across partitioned spin intervals.
The model achieves good in-domain waveform reconstruction and parameter recovery for the two longest-lived components of eight-component input waveforms, while its performance degrades when the validation spins lie far outside the training range.
In a selected spin interval, the framework also recovers the 32 parameters of an eight-component waveform with good overall agreement.
These results demonstrate the feasibility of physics-informed autoencoder-based inference for a prescribed multi-component ringdown waveform family and motivate further tests with progressively more realistic signals.
\end{abstract}


\maketitle

\section{Introduction}\label{sec:intro}
Since the first detection of gravitational waves in 2015, the International Gravitational Wave Observatory Network (IGWN), which includes LIGO, Virgo, and KAGRA, has conducted intermittent observations. More than 400 gravitational wave events from compact binary coalescence have been reported~\cite{GWTC-1,GWTC-2,GWTC-2.1,GWTC-3,GWTC4.0_results,GWTC5.0_results}. 

The ringdown gravitational waves emitted after a binary black hole merger probe the dynamical response of the remnant object in the strong-field regime. 
The importance of ringdown analysis has recently been highlighted by the observation of GW250114, one of the highest signal-to-noise-ratio (SNR) binary black-hole merger events. The post-merger signal was found to be consistent with the Kerr quasinormal mode (QNM) spectrum and enabled precision tests of both the Kerr nature of the remnant black hole and Hawking's area law~\cite{LIGOScientific:2025rid}. Such observations demonstrate the increasing potential of ringdown gravitational waves as probes of strong-field gravity and motivate the development of robust methods for extracting QNM information from observational data.

In general relativity, if the remnant is described by a Kerr black hole, the dominant part of the ringdown signal is modeled within linear perturbation theory as a superposition of exponentially damped oscillations known as QNMs. 
Their complex frequencies are determined solely by the mass and spin of the remnant black hole.
Their complex amplitudes also encode important physical information and can be factorized into initial-data-dependent parts and excitation factors.
The latter, like the QNM frequencies, are determined solely by the black hole mass and spin.
Ringdown gravitational waves therefore provide a valuable arena for black-hole spectroscopy and for tests of gravity in the dynamical, strong-field regime~\cite{Berti:2025hly}.

The QNM frequencies and excitation factors are intrinsic quantities of the Kerr background and are independent of the source or initial data.
Their calculation has a long history, beginning with the pioneering work of Leaver~\cite{Leaver:1985ax,Leaver:1986a,Leaver:1986gd}; see, e.g., Refs.~\cite{Kokkotas:1999bd,Nollert:1999ji,Berti:2009kk,Konoplya:2011qq,Berti:2025hly} for reviews.
While QNM frequencies have long been available to high accuracy, the numerical values reported for the excitation factors showed discrepancies for approximately four decades.
Recently, two independent calculations in Ref.~\cite{Motohashi:2024fwt} yielded mutually consistent high-precision values of the Kerr excitation factors.
The resulting dataset was made publicly available in Ref.~\cite{Motohashi2024Zenodo}.
These values have subsequently been confirmed by two further independent analyses employing distinct methods~\cite{Lo:2025njp,Kubota:2025hjk}.
Together, these studies establish a consistent high-precision dataset of Kerr excitation factors, resolving the long-standing discrepancies among earlier numerical results.

The recent establishment and public availability of high-precision QNM frequencies and excitation factors enable their application to physically motivated ringdown models and, ultimately, black-hole spectroscopy.
Realizing this potential, however, requires reliable extraction of the relevant QNM content and accurate inference of the associated physical parameters from observed or simulated waveforms.
In practice, this extraction is nontrivial. 
Although the precise time window in which linear perturbation theory provides an accurate description is not known \emph{a priori}, the difficulty does not disappear even if one restricts attention to the regime where a QNM superposition is expected to be applicable. 
A central issue is that ringdown signals contain multiple damped components, including the fundamental mode and overtones. 
In particular, the extraction of overtones has attracted considerable attention because identifying multiple QNM components is essential for black-hole spectroscopy: since QNM frequencies depend on the mass and spin of the remnant black hole, resolving more than one mode provides nontrivial consistency tests of the Kerr ringdown spectrum. 

In this context, it was shown that including overtones can significantly improve fits to numerical-relativity waveforms even near the peak amplitude~\cite{Giesler:2019uxc}. At the same time, subsequent studies have emphasized that improved mismatch alone does not guarantee physically reliable mode identification, and have pointed out issues such as sensitivity to the fit start time, model dependence, and possible overfitting~\cite{Baibhav:2023clw,Nee:2023osy}. 
One strategy to address these issues is to use stability as a criterion for determining which modes can be reliably retained in the fit~\cite{Cheung:2023vki}. 
Another strategy is to iteratively subtract the most stable, longest-lived component in order to improve the extraction of shorter-lived overtones~\cite{Takahashi:2023tkb}.
A complementary approach uses rational filters to remove selected QNM contributions and thereby expose subdominant components in ringdown signals~\cite{Ma:2022wpv}, although applying such filters to early-time data can redistribute QNM power in time rather than cleanly remove an isolated QNM component~\cite{Cheung:2026gfd}.
More recent analyses have also focused on the robustness and stability of the extracted mode amplitudes and phases~\cite{Giesler:2024hcr}.
These developments indicate that QNM extraction remains a delicate inference problem, particularly when several damped components overlap strongly.

Against this background, it is worthwhile to explore approaches that are complementary to conventional fitting methods. 
Machine learning techniques offer such possibilities. 
Their potential advantage in the present context is not that they bypass the underlying physics, but that they may provide an efficient data-driven representation of complicated multi-component damped signals and enable fast parameter inference once they are trained. 
This may be useful when several QNM contributions overlap, making the mapping from the waveform morphology to the physical parameters effectively nonlinear.

Among the various machine learning architectures, autoencoders (e.g.,~\cite{ref:Autoencoder_intro,ref:Autoencoder_rev}), originally developed and widely used in areas such as image processing and representation learning, are particularly attractive because they compress the input into a low-dimensional latent representation while retaining the essential features of data.
More recently, autoencoder-based methods have also been applied successfully in gravitational-wave data analysis, for example in glitch rejection for burst searches~\cite{Bini:2023gil}, source-agnostic signal detection~\cite{Moreno:2021fvp}, accelerated parameter inference~\cite{ref:Gabbard_2022,Sun:2025cbo,Bada-Nerin:2024wkn}, and waveform generation~\cite{Liao:2021vec,Sun:2025afb,ref:Garg_2026}.
To our knowledge, however, their application to parameter estimation of multi-component QNMs in ringdown signals has not been investigated in detail.

More broadly, damped oscillations are important phenomena observed in many physical systems.
Their parameter estimation is important in a wide range of applications, including fault detection in machinery~\cite{ref:bearing_fault}, structural health monitoring~\cite{ref:structure_health_monitoring}, nuclear magnetic resonance~\cite{ref:NMR}, free-induction-decay optical magnetometry~\cite{ref:FID1,ref:FID2}, cavity ring-down polarimetry/ellipsometry~\cite{ref:CRDP1,ref:CRDE1}, vibration analysis~\cite{ref:vibration_analysis}, radar~\cite{ref:radar}, sonar~\cite{ref:sonar}, communication channels~\cite{ref:communication}, nuclear collective excitations~\cite{ref:nuclear_dipole_review}, and speech and audio modeling~\cite{Jensen:2004,kafentzis2024}.

In this broader context, Visschers {\it et al.}~\cite{ref:autoencoder} demonstrated that an autoencoder can be used for rapid parameter extraction from single-component damped sinusoidal signals, with performance comparable to or better than conventional methods such as least squares~\cite{ref:Halmer} and fast-Fourier-transform-based approaches~\cite{Visschers:21,ref:Bostrom}.
The same study also highlighted the usefulness of training the latent-space representation to directly encode the physical parameters of interest. 

In Refs.~\cite{ref:Iida2026,Iida:2026jox}, we extended an autoencoder-based method~\cite{ref:autoencoder} to estimate the frequency, phase, decay time, and amplitude of each component in noisy multi-component damped sinusoidal signals. 
These studies suggested that autoencoder-based parameter estimation may provide a useful tool for investigating the physics behind such signals, including possible future applications.

In this study, we adopt an autoencoder-based framework to estimate the QNM parameters from ringdown gravitational waves. 
By training the latent space to represent the physical parameters of interest, we aim to construct a framework in which denoising and parameter estimation are performed in a unified manner while maintaining a clear connection with the underlying signal physics.
As training and validation data, we use waveforms constructed as superpositions of Kerr QNMs, with frequencies and excitation factors taken from the publicly available dataset of Refs.~\cite{Motohashi:2024fwt,Motohashi2024Zenodo}.
This allows us to work with physically motivated multi-mode waveforms whose properties vary nontrivially across the black-hole spin range. 

The rest of this paper is organized as follows. In Sec.~\ref{sec:autoencoder}, we introduce the autoencoder architecture and the latent-space parameter-estimation framework adopted in this study. In Sec.~\ref{sec:dataset-experimental-setup}, we describe the Kerr-QNM-based dataset, the signal model, the preprocessing procedure, and the training and validation setup. In Sec.~\ref{sec:result}, we present the results for both same-range and different-range spin training/validation, as well as for the multi-component setting. Section~\ref{sec:concl} summarizes our findings and discusses future directions.

\begin{figure*}[tb]
    \centering
    \includegraphics[width=0.98\textwidth]{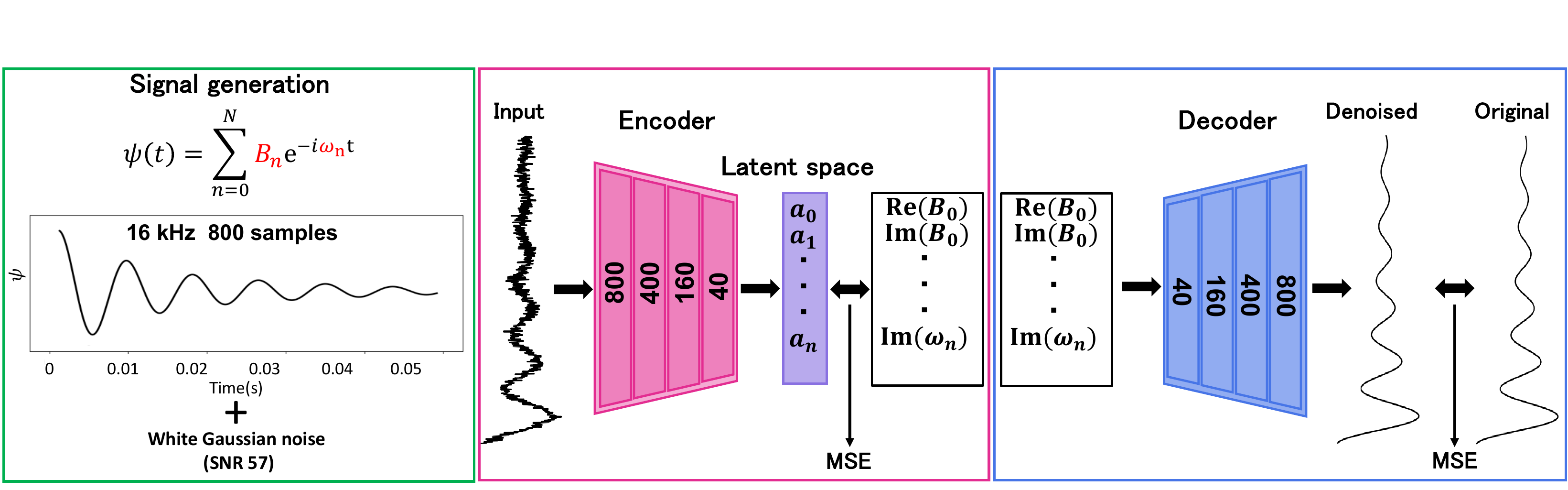}
    \caption{Analysis flow and autoencoder architecture.}
    \label{fig:autoencoder}
\end{figure*}

\section{Latent space parameter estimation framework with autoencoder}\label{sec:autoencoder}
An autoencoder (e.g.,~\cite{ref:Autoencoder_intro}) is a type of neural network with an hourglass-shaped architecture, in which the input is compressed into a lower-dimensional latent space (bottleneck) and then reconstructed back to the original dimensionality. By passing through the latent space, the network can learn the essential features of the data, enabling it to extract important information while suppressing noise. Owing to this property, autoencoders have been widely applied in various fields, including the denoising of image and audio data, anomaly detection, and feature extraction~\cite{ref:Autoencoder_rev}. In this study, we focus on these characteristics and investigate a method to estimate the physical parameters of damped oscillation signals by utilizing the latent space as a low-dimensional representation that characterizes the input signals. This approach aims to efficiently extract physically meaningful parameters from noisy time-series data~\cite{ref:autoencoder,ref:Iida2026,Iida:2026jox}.

Figure~\ref{fig:autoencoder} illustrates the overall analysis flow, including the data generation process (left), the encoder (middle), and the decoder (right). 
The detailed settings for data generation are described in Sec.~\ref{sec:dataset-experimental-setup}. 

The encoder estimates the parameters from the waveform, and the decoder reconstructs the waveform from the estimated parameters, thereby enabling the analysis of damped oscillation signals.
The autoencoder comprises nine layers, including the input and output layers, with the number of neurons in each layer set to 800--400--160--40--$N_{\rm lat}$ (latent space)--40--160--400--800. 
Here, $N_{\rm lat}$ is determined by the number of physical parameters to be estimated. 
The $\tanh$ activation function was applied to each layer except for the latent-space layer.
No activation function was applied to the latent-space layer, so that the latent variables can represent the physical parameters without imposing additional nonlinear constraints.

First, the encoder takes a noisy waveform composed of superposed damped oscillation components as input and maps it to the latent space. 
The input waveforms for the encoder were standardized to stabilize training. 
In general, the latent space of an autoencoder does not have to coincide directly with the physical parameters of interest. 
In the present study, this correspondence was enforced through the training procedure of the encoder. 
Specifically, following the method described in Refs.~\cite{ref:autoencoder,ref:Iida2026,Iida:2026jox}, we set the dimension of the latent space to match the total number of physical parameters $N_{\rm lat}$ shown in the middle of Fig.~\ref{fig:autoencoder}.

The mean squared error (MSE) between the estimated and true parameters was used as the loss function during encoder training, ensuring that the latent variables corresponded to the physical parameters of each damped oscillation component. 
In a previous study~\cite{ref:autoencoder}, stochastic gradient descent (SGD)~\cite{ref:SGD} was used for optimization. 
In this study, we employed Adam~\cite{Kingma:2014vow}, which resulted in faster convergence and reduced computational time.

Subsequently, the decoder was trained to reconstruct the denoised waveform from the parameters of each damped oscillation component represented in the latent space. 
During decoder training, the MSE between the denoised waveform and the original noise-free waveform was used as the loss function, with optimization performed using Adam. 
This approach allowed the encoder and decoder to be trained separately, facilitating parameter estimation within the latent-space framework.

Additionally, to mitigate overfitting, a dropout rate of 0.1 was applied to each layer of both the encoder and decoder in all cases. This technique involves the random deactivation of a fraction of neurons during training, thereby preventing excessive co-adaptation within the network and enhancing the generalization performance.

\section{Dataset and experimental setup}\label{sec:dataset-experimental-setup}

\subsection{Dataset description}\label{subsec:dataset-description}
In this study, we analyze model ringdown waveforms constructed from Kerr QNM data calculated in Ref.~\cite{Motohashi:2024fwt} and publicly available on Zenodo~\cite{Motohashi2024Zenodo}. 
The purpose of this dataset is not to reproduce a generic ringdown waveform with arbitrary initial data or source configuration, but to provide a controlled setting in which the QNM frequencies and excitation factors are known functions of the black hole parameters.

In linear perturbation theory, the prograde QNM part of a ringdown waveform can be schematically written as~\cite{Kokkotas:1999bd,Nollert:1999ji,Berti:2009kk,Konoplya:2011qq,Berti:2025hly}
\begin{equation}\label{eq:IBexp}
\psi_\mathrm{QNM}(t) = \sum_{n=0}^\infty I_n B_n e^{-i\omega_n t}.
\end{equation}
Here, $\omega_n$ is the complex frequency of the QNM, whose real and imaginary parts correspond to the frequency and damping rate, respectively.
$B_n$ denotes the excitation factor and $I_n$ denotes the dependence on the source or initial data, whose product represents the amplitude and phase of each QNM.
For an idealized localized delta-function source, $I_n$ can be set to unity, and the QNM amplitudes are determined by the excitation factors.
A detailed benchmark study of QNM extraction in this setup was performed in Ref.~\cite{Kubota:2025hjk}.
The waveforms used in the present study are a further simplified version of this setup: we truncate the expansion to a finite sum of damped QNM contributions using the public QNM-frequency and excitation-factor data~\cite{Motohashi2024Zenodo}.
Thus, apart from the fixed mass and distance scales specified below, the waveform parameters in this model are determined by the black hole spin and the chosen set of modes.

In Ref.~\cite{Motohashi:2024fwt}, the behavior of QNMs was investigated as a function of continuously varying spin. 
Since the mass dependence can be treated as a simple scaling, the analysis focused on the dependence on the dimensionless spin $a/M$, where $a$ and $M$ denote the black hole spin and mass, respectively. 
The range of the dimensionless spin considered in this study was from $a/M = 0$ to $0.999999$.
It was shown that when two QNM frequencies approach each other as the spin varies, the modes can exhibit an avoided crossing, accompanied by resonant excitation in the corresponding excitation factors~\cite{Motohashi:2024fwt}. 
In the present paper, we focus on the $(l,m)=(2,2)$ modes of gravitational perturbations, which are expected to dominate the ringdown signal for most binary black hole mergers. 
For this sector, such characteristic resonance behavior occurs around $a/M \simeq 0.9$.
Based on these features, we divide the data according to the spin values and perform separate training for each subset.

\subsection{Signal model and data generation}\label{subsec:signal-generation}
For a given black hole spin, we consider a superposed damped sinusoidal waveform expressed as
\begin{equation}\label{eq:Bexp}
\psi(t) = \sum_{n=0}^{N} B_n e^{-i\omega_n t},
\end{equation}
where $N$ is the number of components included in the model waveform.
Here, $\omega_n$ and $B_n$ are the QNM frequency and excitation factor for the chosen spin and overtone index $n$, respectively.
As explained above, Eq.~\eqref{eq:Bexp} corresponds to the idealized localized-source case with $I_n=1$ in Eq.~\eqref{eq:IBexp}. 
We convert the dimensionless quantities $\omega_\mathrm{dimless} \equiv 2M\omega$ and $B_\mathrm{dimless}\equiv (2M)^4B$ with the geometrical units $G = c = 1$ provided in the dataset~\cite{Motohashi2024Zenodo} to physical units by restoring the mass and distance scales as
\begin{align}
\omega_{\mathrm{phys}}
&= \frac{c}{r_s}\frac{M_\odot}{M}\, \omega_{\mathrm{dimless}}, \\
B_{\mathrm{phys}}
&= \frac{r_s}{D} \frac{M}{M_\odot} \, B_{\mathrm{dimless}},
\end{align}
where the Schwarzschild radius of the Sun is defined as
\begin{equation}
r_s \equiv \frac{2GM_\odot}{c^2}.
\end{equation}
In this study, we set a black hole mass $M = 60 \,M_\odot$ and a distance of $D = 400\,\mathrm{Mpc}$.
These values are based on the mass and distance of the black hole estimated for the event GW250114~\cite{LIGOScientific:2025rid}.
The signal duration was set to $0.05\,\mathrm{s}$, and the sampling rate was $16\,\mathrm{kHz}$, resulting in time-series data consisting of 800 points.

For each generated waveform, we retain the real part of the complex waveform as the real-valued time series used for training and validation.
We compute its mean and standard deviation over the $800$ time samples and standardize it as
\begin{equation}
\psi_{\rm ind}(t_j)
= \frac{\mathrm{Re}[\psi(t_j)]-\mu_{\rm ind}}{\sigma_{\rm ind}},
\end{equation}
where
\begin{align}
\mu_{\rm ind}&=\frac{1}{N_t}\sum_{j=1}^{N_t}\mathrm{Re}[\psi(t_j)], \\
\sigma_{\rm ind}^2&=\frac{1}{N_t}\sum_{j=1}^{N_t}
\left|\mathrm{Re}[\psi(t_j)]-\mu_{\rm ind}\right|^2,
\end{align}
with $N_t=800$.

We then add white Gaussian noise with zero mean and a standard deviation of $\sigma_{\rm noise} = 0.5$ to $\psi_{\rm ind}(t)$. 
For the standardized time series, the SNR was evaluated using 
\begin{equation}
\mathrm{SNR}
= \frac{\sqrt{\sum_j \left|\psi_{\rm ind}(t_j)\right|^2}}{\sigma_{\rm noise}}.
\end{equation}
Because $\psi_{\rm ind}$ has unit variance and the number of time samples is $800$, this definition gives $\mathrm{SNR}\simeq \sqrt{800}/0.5 \simeq 57$.
This value should be interpreted as an effective SNR for the standardized synthetic waveforms used in this study and should not be directly compared with the matched-filter post-merger SNR of an observed gravitational-wave event.
After adding the noise, we apply the inverse of the per-waveform standardization to return each waveform to its original scale.

\subsection{Preprocessing}\label{subsec:Preprocessing}
For each spin interval considered below, we generate a set of time-domain waveforms by choosing spin values within that interval, constructing the corresponding QNM superpositions described in Sec.~\ref{subsec:signal-generation}, and adding noise as described above.
Let $\psi_k(t_j)$ denote the $j$th time sample of the $k$th noisy input waveform in this dataset, where $k=1,\ldots,N_{\rm wf}$ and $j=1,\ldots,N_t$.
The input waveforms were standardized at the dataset level as
\begin{equation}
\psi_{{\rm std},k}(t_j)
=
\frac{\psi_k(t_j)-\mu_\psi}{\sigma_\psi},
\label{eq:wave_standardization}
\end{equation}
where
\begin{align}
\mu_\psi &= \frac{1}{N_{\rm wf}N_t}
\sum_{k=1}^{N_{\rm wf}}\sum_{j=1}^{N_t}\psi_k(t_j), \\
\sigma_\psi^2 &= \frac{1}{N_{\rm wf}N_t}
\sum_{k=1}^{N_{\rm wf}}\sum_{j=1}^{N_t}
\left|\psi_k(t_j)-\mu_\psi\right|^2.
\end{align}
Here, $N_{\rm wf}$ denotes the number of waveforms in the dataset for the corresponding spin interval, and $N_t=800$ is the number of time samples in each waveform.
The mean $\mu_\psi$ and standard deviation $\sigma_\psi$ are calculated from the training samples and are also used to standardize the corresponding validation samples.
This dataset-level standardization is intended to stabilize the training, and its purpose differs from that of the per-waveform standardization $\psi_{\rm ind}$ introduced in Sec.~\ref{subsec:signal-generation}, which is used only to set the noise level and evaluate the SNR.

For the physical parameters estimated in the latent space, normalization was performed separately for each parameter to make their numerical scales comparable during training:
\begin{equation}
x_{\rm lat} = \frac{x-\mu_x}{3\sigma_x}.
\label{eq:ls_norm}
\end{equation}
Here, $x$ denotes one of the target parameters, and $\mu_x$ and $\sigma_x$ denote the mean and standard deviation of that parameter over the corresponding dataset.
The mean $\mu_x$ and standard deviation $\sigma_x$ are calculated from the training samples and are also used to normalize the corresponding validation targets.
The factor of $3$ was introduced so that most samples fall within a range of order unity, which was found to be suitable for stable training~\cite{ref:autoencoder}.

\subsection{Data split and evaluation setup}\label{subsec:data-split-evaluation}
With the black hole mass and distance fixed as described in Sec.~\ref{subsec:signal-generation}, the remaining variation in the model waveforms is governed by the black hole spin.
The model must therefore learn this spin dependence accurately over the range considered.
However, the variation of the QNM parameters with spin is not uniform, and training a single model over the entire spin range may lead to insufficient learning in specific spin regions.
In particular, the parameter variation is relatively gradual in the low-spin region, whereas the damping rates and amplitudes of the QNMs vary more rapidly in the high-spin region.
Uniform sampling in spin therefore provides a sparser sampling of the parameter variation in the high-spin region.

As a result, when a single model is trained over the entire spin range, the estimation accuracy is not uniform across all spin regions, with larger variations observed particularly in the high-spin region.
Thus, the learning difficulty varies across spin regions because of differences in the spin dependence of the parameters and waveform properties.
We therefore divide the entire spin range into multiple regions and train a separate model for each region.

To assess this strategy, below we consider three configurations.
First, we train and validate the model within the same spin interval.
Second, we use different spin ranges for training and validation to assess the behavior outside the training domain.
In these cases, we focus on the parameter estimation of the two longest-lived components from an eight-component waveform.
Finally, we evaluate simultaneous parameter estimation for the full eight-component waveform in a selected spin interval.

\subsubsection{Same-spin-range configuration}
Let us begin with the case with the training and validation in the same spin range.
The spin intervals and numbers of training and validation samples are listed in Table~\ref{tab:spin_and_data}.

\begin{table}[tbp]
\centering
\caption{Same-spin-range configuration: ranges of spin parameter $a/M$ and numbers of training and validation samples.}
\begin{tabular}{ccc}
\hline
Spin range ($a/M$) & Training samples & Validation samples \\ 
\hline
$[0.0,\,0.15]$      & 6,000 & 1,500 \\ 
$[0.10,\,0.60]$     & 20,000 & 5,000\\
$[0.55,\,0.80]$     & 10,000 & 2,500\\
$[0.75,\,0.95]$     & 8,000 & 2,000\\
$[0.92,\,0.98]$     & 4,800 & 1,200\\
$[0.975,\,0.998]$   & 8,000 & 2,000\\
$[0.995,\,\mathrm{max}]$& 16,000 & 4,000\\
\hline
\end{tabular}
\label{tab:spin_and_data}
\end{table}

Here, we use overlapping spin intervals.
These intervals are not defined by a theoretical criterion but were determined empirically by examining the training stability and variation in the estimation results.
The maximum spin considered in this study is 0.999999.
For the spin intervals from $[0.0,\,0.15]$ to $[0.75,\,0.95]$, 5,000 data samples were used per spin width of 0.1.
In the high-spin intervals $[0.92,\,0.98]$, $[0.975,\,0.998]$, and $[0.995,\,\mathrm{max}]$, increasing the number of samples improved the estimation results; we therefore used 6,000, 10,000, and 20,000 samples, respectively.
Within each spin interval, the samples are uniformly spaced in the dimensionless spin $a/M$.
For each spin interval, 80\% of the generated waveforms were used for training and the remaining 20\% for validation.

The input is a waveform constructed from the eight components with $n=0$ to $7$.
The model is trained to estimate the parameters of the $n=0$ and $n=1$ components and to reconstruct a waveform containing only these two components.
For each component, the estimated parameters are the real and imaginary parts of both the excitation factor $B_n$ and the complex frequency $\omega_n$, giving a total of eight parameters.
The QNM contribution to a physical ringdown signal generally contains many damped components, from which a few dominant modes must be extracted.
The present setup provides a simplified realization of this task.

\subsubsection{Different-spin-range configuration}
We next investigate how the relation between the training and validation spin ranges affects the estimation results.
The two cases considered are listed in Table~\ref{tab:spin_and_data_2}.
In the first case, the training and validation ranges are $[0.1,\,0.6]$ and $[0.55,\,0.8]$, respectively.
In the second case, they are $[0.75,\,0.95]$ and $[0.0,\,0.15]$, respectively.

In the first setting, the training and validation ranges partially overlap.
It therefore probes generalization to a nearby spin range for which the parameter variation remains continuous across the boundary of the training domain.
In contrast, the second setting uses completely disjoint spin ranges and probes the model behavior in a spin region not seen during training.

\begin{table}[tbp]
\centering
\caption{Different-spin-range configuration: ranges of spin parameter $a/M$ and numbers of training and validation samples.}
\begin{tabular}{cccc}
\hline
\shortstack{Training\\spin range ($a/M$)} & \shortstack{Validation\\spin range ($a/M$)} & \shortstack{Training\\samples} & \shortstack{Validation\\samples}\\
\hline
$[0.1,\,0.6]$ & $[0.55,\,0.8]$ & 25,000 & 2,000 \\ 
$[0.75,\,0.95]$ & $[0.0,\,0.15]$ & 15,000 & 2,000 \\
\hline
\end{tabular}
\label{tab:spin_and_data_2}
\end{table}

These tests characterize the generalization performance and limitations of the model under spin distributions that differ from the training distribution.
This is relevant to applications to real data, for which the spin of an input waveform need not lie within the training range.
In particular, apparently good agreement outside the training range could mask incorrect parameter estimates.

In both cases, the input is a waveform constructed from the eight components with $n=0$ to $7$.
The model is trained to estimate the $n=0$ and $n=1$ components and to reconstruct their superposed waveform.
For each component, the estimated parameters are the real and imaginary parts of both the excitation factor $B$ and the complex frequency $\omega$, giving a total of eight parameters.

\subsubsection{Eight-component configuration}
The above two cases focus on the parameter estimation of the two longest-lived components from an eight-component waveform.
This setup provides a simplified model of the task of extracting a few dominant modes from a multi-component ringdown signal.
Here, we instead estimate the parameters of all eight components.
Even in idealized fitting tests, in which the waveform is constructed as a superposition of damped QNM components with known mode content, reliably recovering a large number of components remains challenging~\cite{Baibhav:2023clw,Nee:2023osy,Takahashi:2023tkb,Cheung:2023vki,Kubota:2025hjk}.

More generally, the simultaneous estimation of frequencies and damping coefficients in multi-component damped signals has also been investigated.
For example, a deep-learning method has been proposed to infer the frequencies and damping coefficients of multiple components from a frequency--damping-coefficient spectrum~\cite{ref:DataDrivenDamped}.

Because the waveform models and experimental settings differ from those in previous studies, a direct performance comparison is not appropriate.
Instead, the present eight-component setting provides a controlled test of the ability of the autoencoder-based framework to infer parameters from highly multi-component damped signals.

Both the input and target waveforms contain the eight components with $n=0$ to $7$.
The model is trained to estimate the parameters of all eight components and to reconstruct their superposed waveform.
For each component, the real and imaginary parts of both $B_n$ and $\omega_n$ are estimated, giving a total of 32 parameters.
This setting is intended to evaluate the parameter-estimation capability of the model for multi-component signals.

The eight-component configuration was trained and validated in the spin interval $[0.75,\,0.95]$, using 8,000 training samples and 2,000 validation samples, as in the corresponding same-spin-range configuration in Table~\ref{tab:spin_and_data}.

\subsection{Evaluation metrics}\label{subsec:metrics}
We evaluate the estimation performance using three complementary diagnostics: waveform agreement, the consistency of the true and estimated parameter distributions, and parameter-wise errors.

First, we evaluate the agreement between the original noise-free waveform and the denoised waveform reconstructed by the decoder using the match score~\cite{ref:Pycbc},
\begin{equation}
\mathrm{Match\ score} = \max_{t_{\mathrm{s}}, \phi_{\mathrm{s}}} 
\frac{
    \left( y_{\mathrm{deno}} \mid y_{\mathrm{org}} \right)
}{
    \left( y_{\mathrm{org}} \mid y_{\mathrm{org}} \right)^{1/2}
    \left( y_{\mathrm{deno}} \mid y_{\mathrm{deno}} \right)^{1/2}
}.
\label{eq:match}
\end{equation}
The frequency-domain inner product is defined as
\begin{equation}
\left( y_{\mathrm{deno}} \mid y_{\mathrm{org}} \right) = 4 \int_{0}^{\infty} 
\frac{
    \tilde{y}_{\mathrm{deno}}(f)\cdot \tilde{y}^{*}_{\mathrm{org}}(f)
}{
    S_n(f)
} \, df.
\end{equation}
Here, $*$ denotes complex conjugation.
$\tilde{y}_{\mathrm{deno}}(f)$ and $\tilde{y}_{\mathrm{org}}(f)$ denote the Fourier transforms of the denoised and original waveforms, respectively.
$S_n(f)$ is the noise power spectral density, which is set to $1$ for the present white-noise tests.
In Eq.~\eqref{eq:match}, $t_{\mathrm{s}}$ and $\phi_{\mathrm{s}}$ are optimized time and phase shifts, respectively.
The match score lies between $0$ and $1$, with values closer to $1$ indicating better waveform agreement.

Second, we compare the distributions of the true parameters in the validation data with those of the estimated parameters derived from the latent-space representation.
This comparison visualizes how well the latent-space representation reproduces the underlying parameter distributions.

Third, we compare the true and estimated parameters in the validation data using relative errors for $\mathrm{Re}(\omega_n)$ and $\mathrm{Im}(\omega_n)$ and absolute errors for $\mathrm{Re}(B_n)$ and $\mathrm{Im}(B_n)$.
The signed relative error and absolute error are defined as
\begin{align}
\delta_{\mathrm{rel}} x &= \frac{x_{\mathrm{est}} - x_{\mathrm{true}}}{x_{\mathrm{true}}}, \\
\delta_{\mathrm{abs}} x &=\left|x_{\mathrm{est}}-x_{\mathrm{true}}\right|,
\end{align}
where $x_{\mathrm{true}}$ and $x_{\mathrm{est}}$ denote the true and estimated values, respectively.
Absolute errors are used for the excitation factors because their real or imaginary parts can approach zero.

\section{Results and Discussions}\label{sec:result}
We evaluate the latent-space parameter-estimation framework using the three configurations introduced in Sec.~\ref{subsec:data-split-evaluation}.
For each configuration, the validation data are assessed using the diagnostics described in Sec.~\ref{subsec:metrics}.
The learning rate was set to 0.0005, and the numbers of epochs were set to 100 for the encoder and 150 for the decoder.
These settings were common to all experimental configurations considered in this section.
Training was performed for each spin range shown in Tables~\ref{tab:spin_and_data} and~\ref{tab:spin_and_data_2}, and learning curves were used to confirm the absence of overfitting.
Training for each spin interval was completed within ten minutes on a workstation equipped with an Intel\textsuperscript{\tiny\textregistered} Xeon\textsuperscript{\tiny\textregistered} CPU E5-2637 v4 @ 3.50GHz, an NVIDIA GeForce RTX 2080 Ti GPU, and 128 GB memory.

For the same-spin-range configuration, Figs.~\ref{fig:0.0-0.15}--\ref{fig:0.995-max} show the following diagnostics:
\begin{itemize}
  \item (a) input waveform (top) and the denoised output and original waveforms (bottom)
  \item (b) parameter estimates and match score
  \item (c) match score as a function of spin
  \item (d) distributions of true and estimated parameters
  \item (e) parameter-estimation errors as a function of the true values
\end{itemize}
The match score shown in (b) represents the agreement between the denoised waveform output by the decoder and the original noise-free waveform.
In contrast, the match score shown in (c) is calculated from the agreement between the original waveform and the waveform reconstructed using the parameters estimated by the encoder.
Presenting both scores distinguishes the fidelity of waveform reconstruction by the decoder from the accuracy with which the encoder infers parameters that retain the waveform information needed for reconstruction.

The following subsections discuss, in turn, the results for training and validation in the same spin range (Sec.~\ref{subsec:result-full}), training and validation in different spin ranges (Sec.~\ref{subsec:result-different}), and simultaneous estimation of all eight components (Sec.~\ref{subsec:result-8components}).

\subsection{Same-spin-range configuration}\label{subsec:result-full}
\begin{figure*}[t]
\centering
\begin{minipage}{0.32\textwidth}
  \centering
  \includegraphics[width=\linewidth]{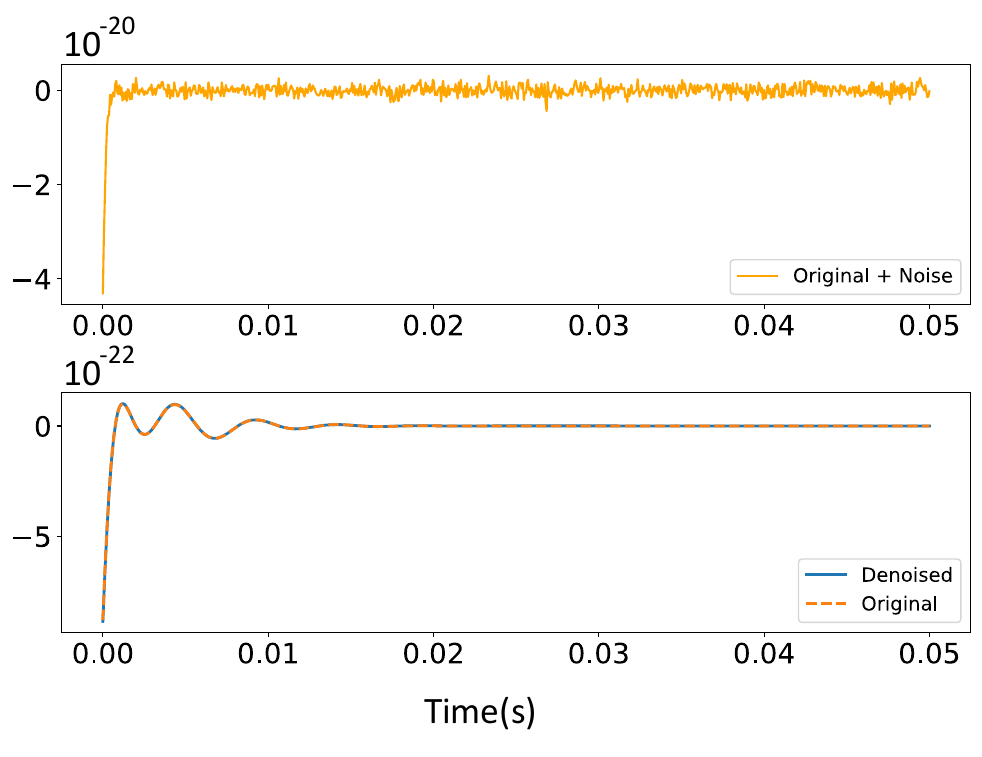}
  (a)
\end{minipage}
\hfill
\begin{minipage}{0.34\textwidth}
  \centering
  \includegraphics[width=\linewidth]{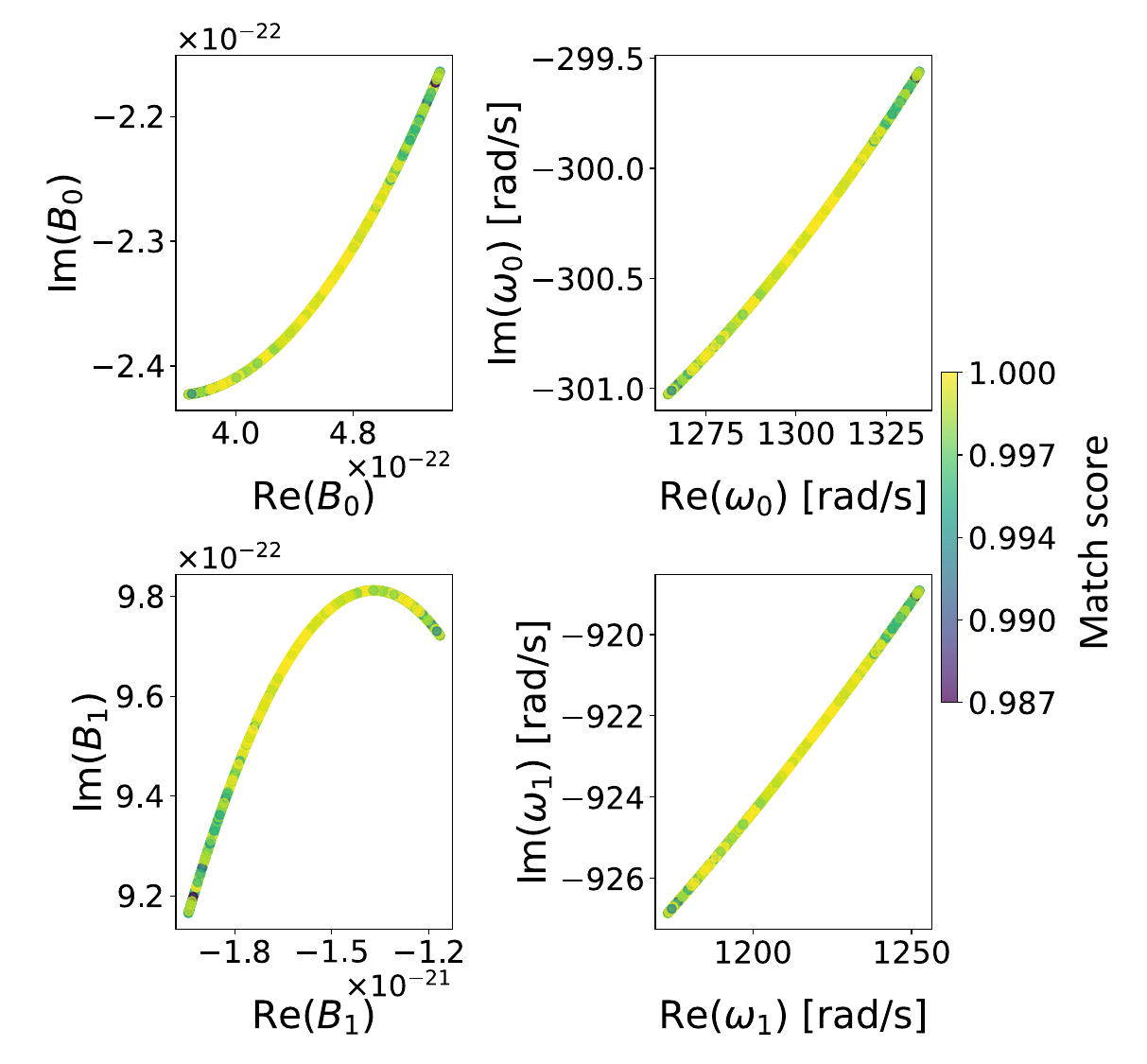}
  (b)
\end{minipage}
\hfill
\begin{minipage}{0.32\textwidth}
  \centering
  \includegraphics[width=\linewidth]{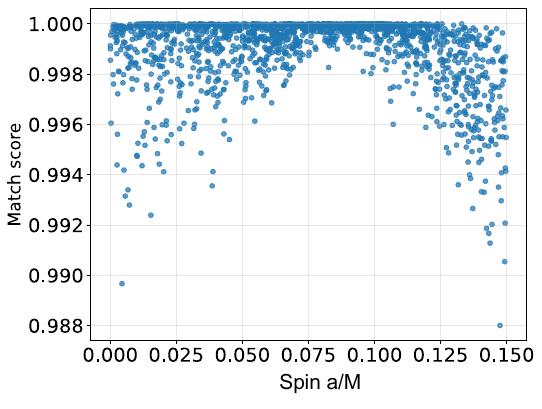}
  (c)
\end{minipage}

\vspace{2mm}

\begin{minipage}{0.9\textwidth}
  \centering
  \includegraphics[width=\linewidth]{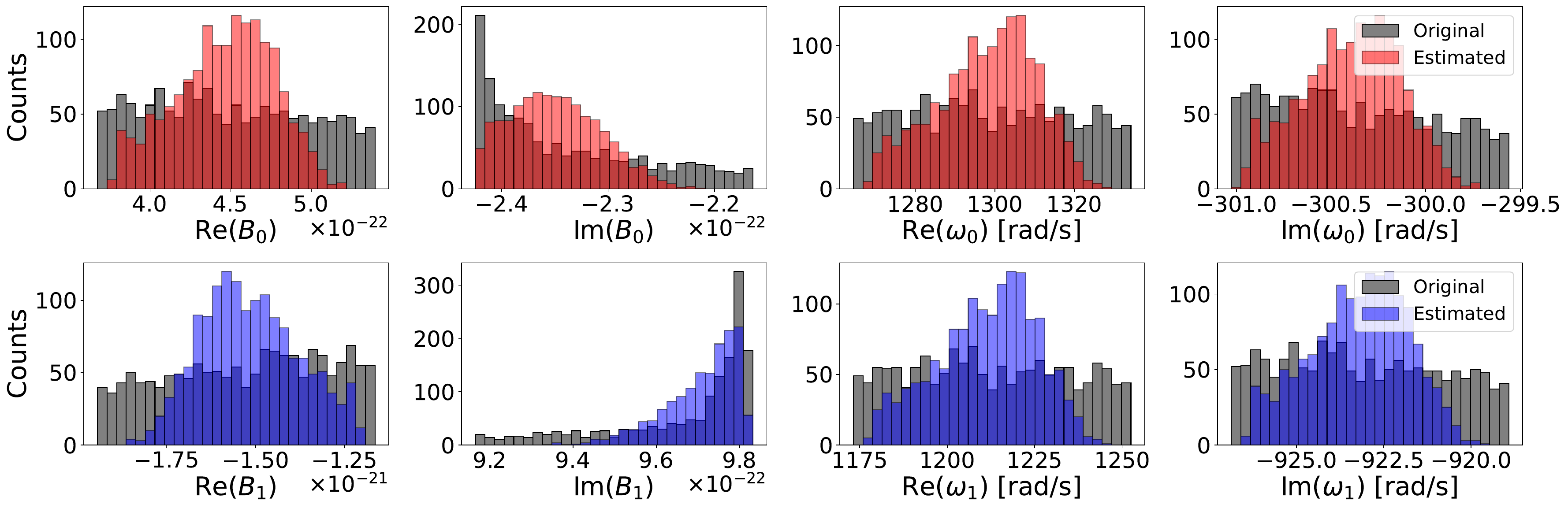}
  (d)
\end{minipage}
\vspace{2mm}

\begin{minipage}{0.9\textwidth}
  \centering
  \includegraphics[width=\linewidth]{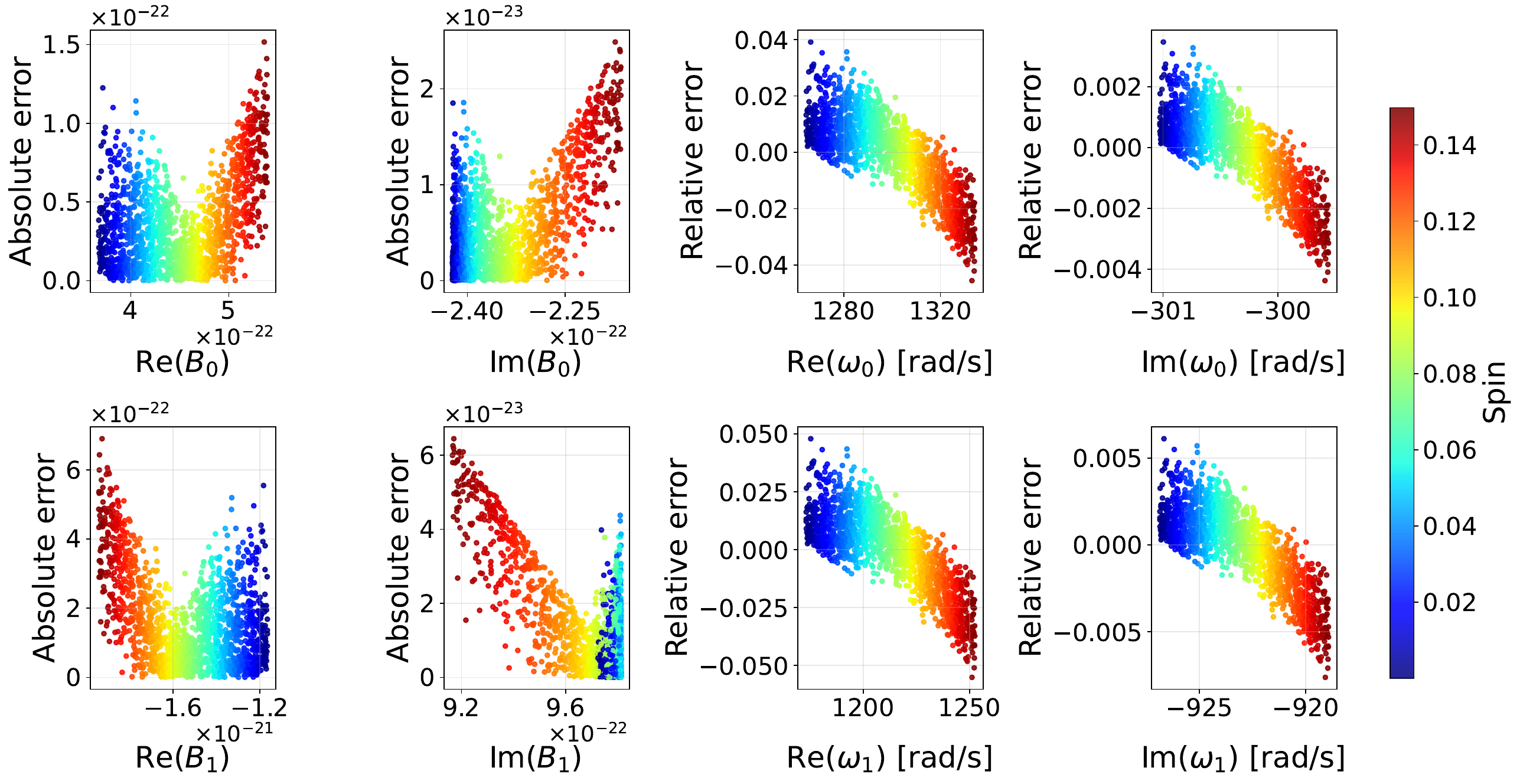}
  (e)
\end{minipage}

\caption{Validation results for spins in the range 0.0–0.15.
(a) Top: input waveform; Bottom: output waveform (Denoised) and original waveform (Original)
(b) Parameters and match score
(c) Relationship between match score and spin parameter
(d) Distributions of true and estimated parameters
(e) Relationship between parameter estimation errors and true values.
}
\label{fig:0.0-0.15}
\end{figure*}

\begin{figure*}[t]
\centering
\begin{minipage}{0.32\textwidth}
  \centering
  \includegraphics[width=\linewidth]{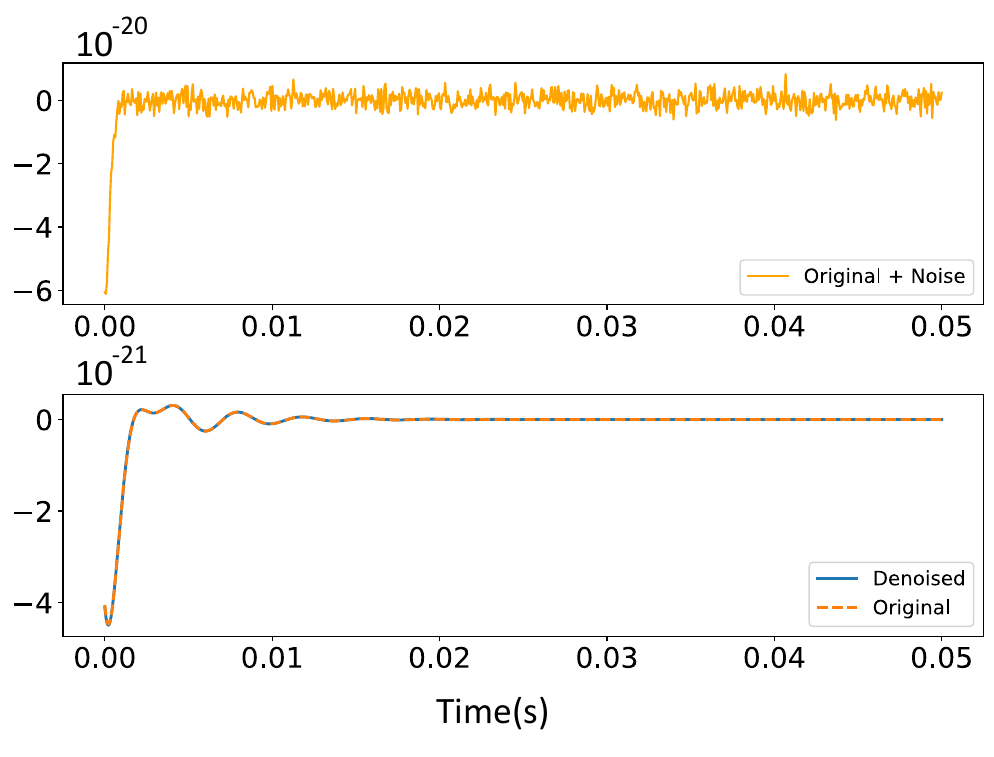}
  (a)
\end{minipage}
\hfill
\begin{minipage}{0.34\textwidth}
  \centering
  \includegraphics[width=\linewidth]{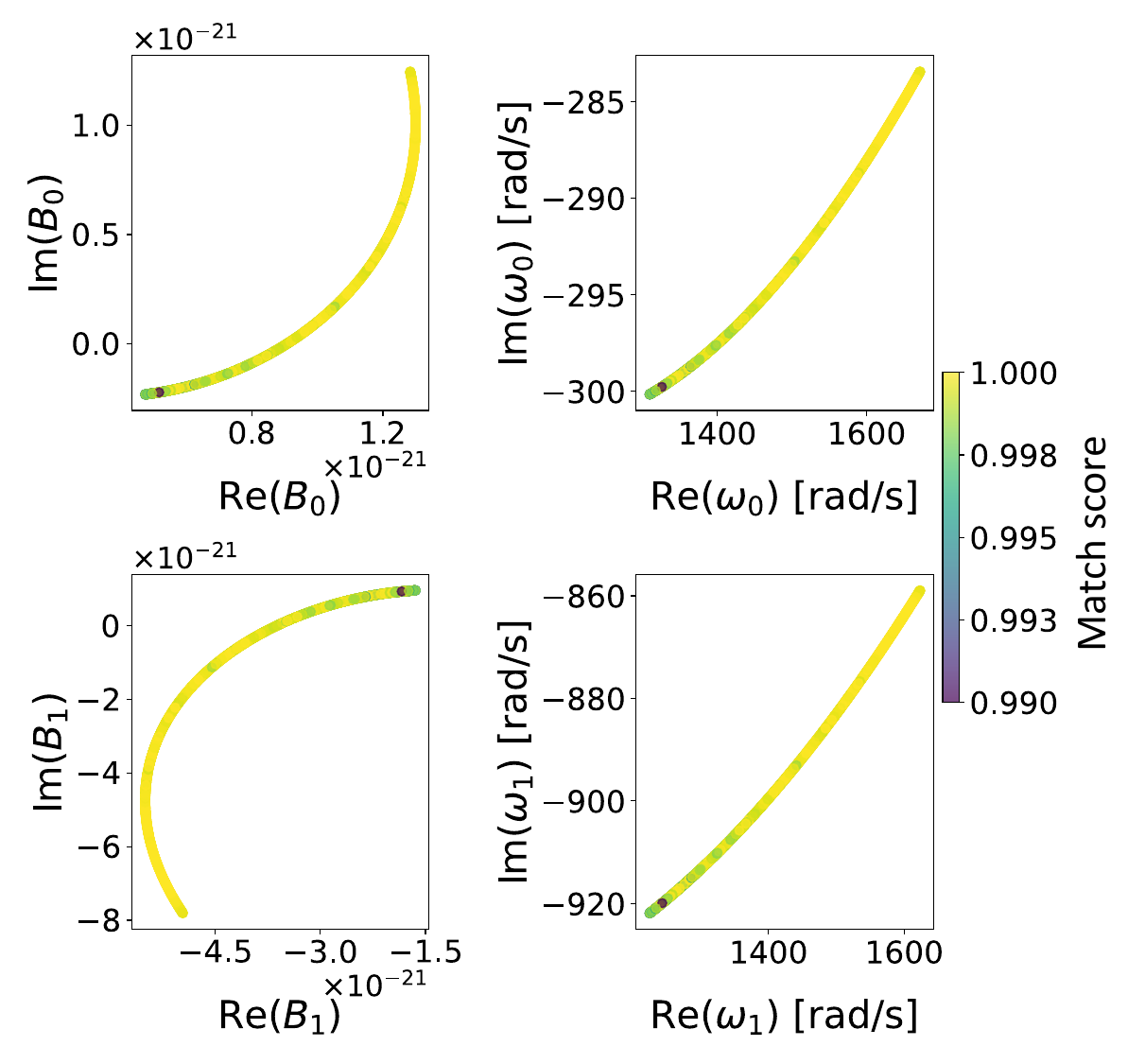}
  (b)
\end{minipage}
\hfill
\begin{minipage}{0.32\textwidth}
  \centering
  \includegraphics[width=\linewidth]{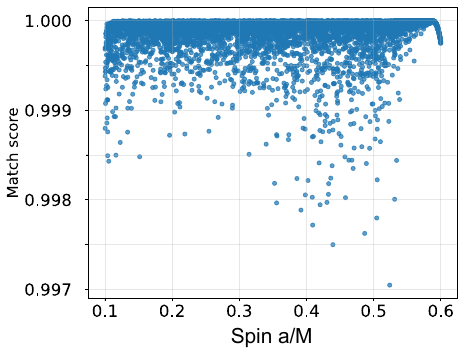}
  (c)
\end{minipage}

\vspace{2mm}

\begin{minipage}{0.9\textwidth}
  \centering
  \includegraphics[width=\linewidth]{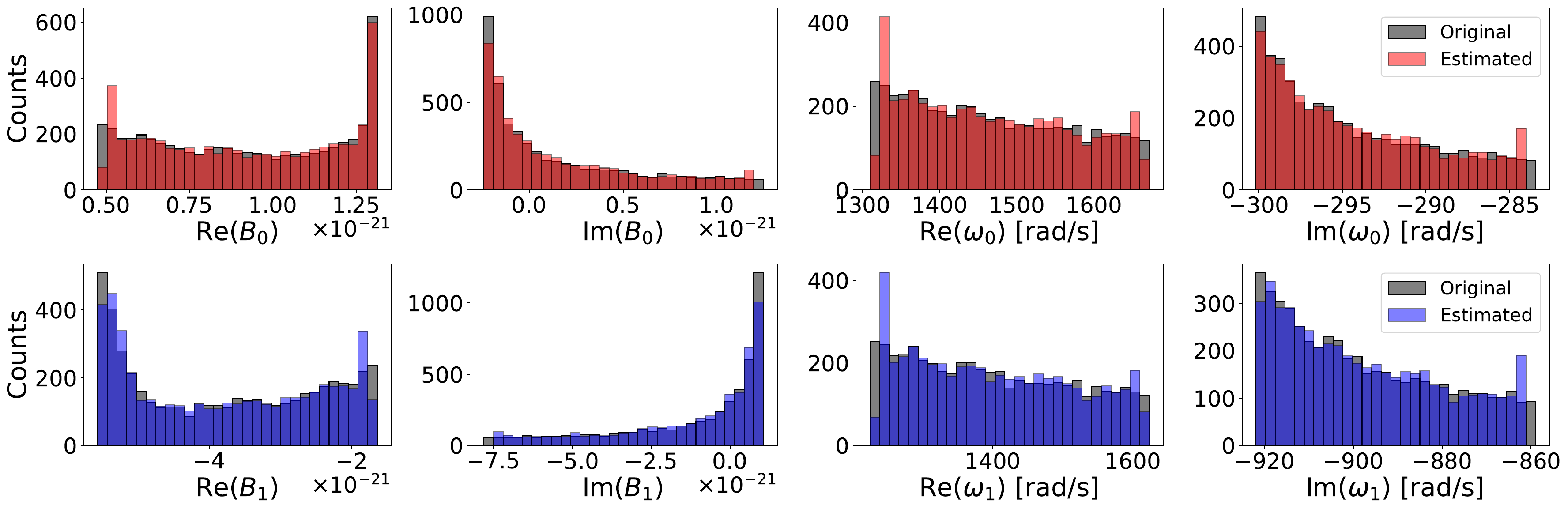}
  (d)
\end{minipage}
\vspace{2mm}

\begin{minipage}{0.9\textwidth}
  \centering
  \includegraphics[width=\linewidth]{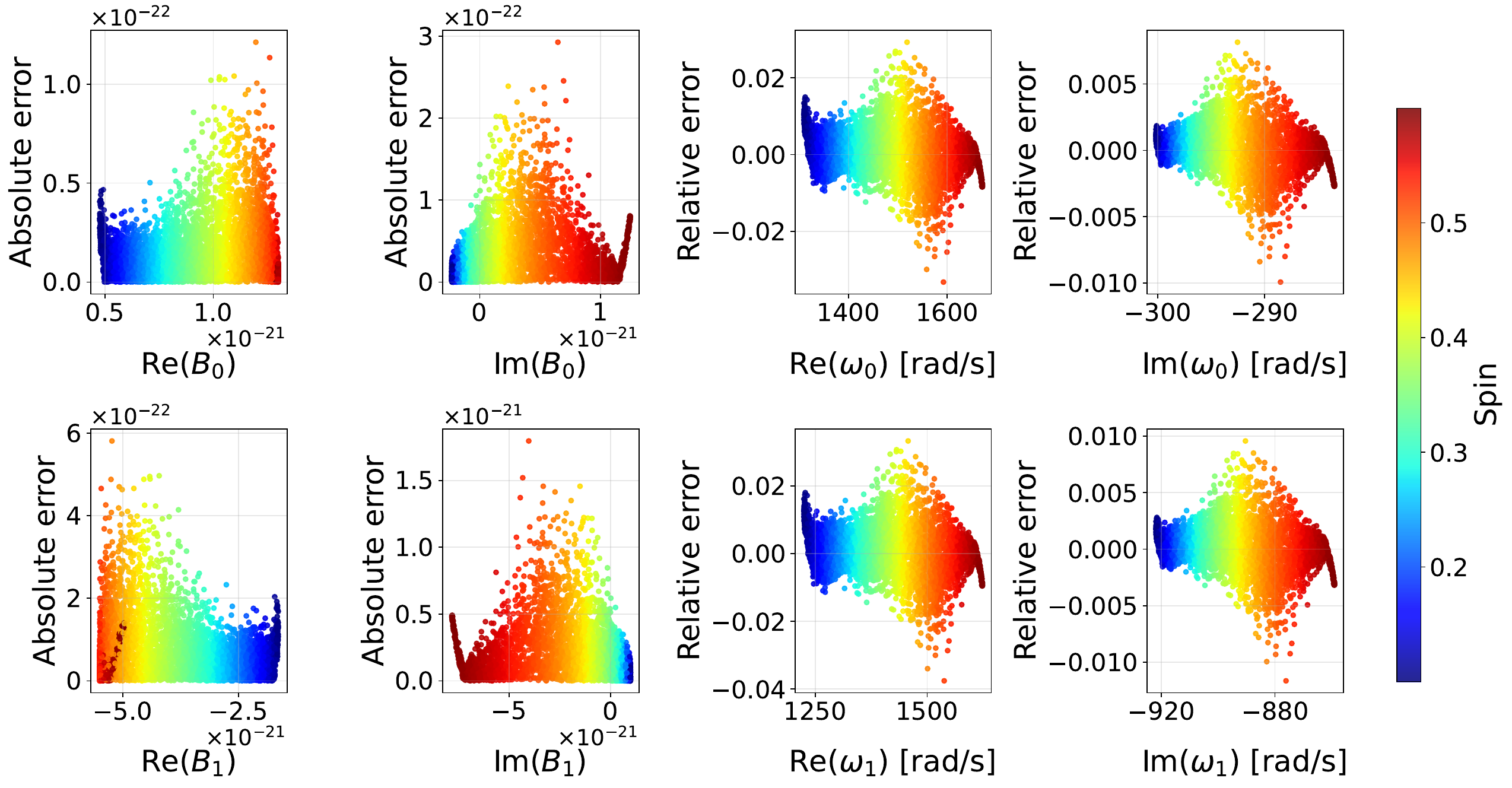}
  (e)
\end{minipage}

\caption{Validation results for spins in the range 0.1--0.6.} 
\label{fig:0.1-0.6}
\end{figure*}

\begin{figure*}[t]
\centering
\begin{minipage}{0.32\textwidth}
  \centering
  \includegraphics[width=\linewidth]{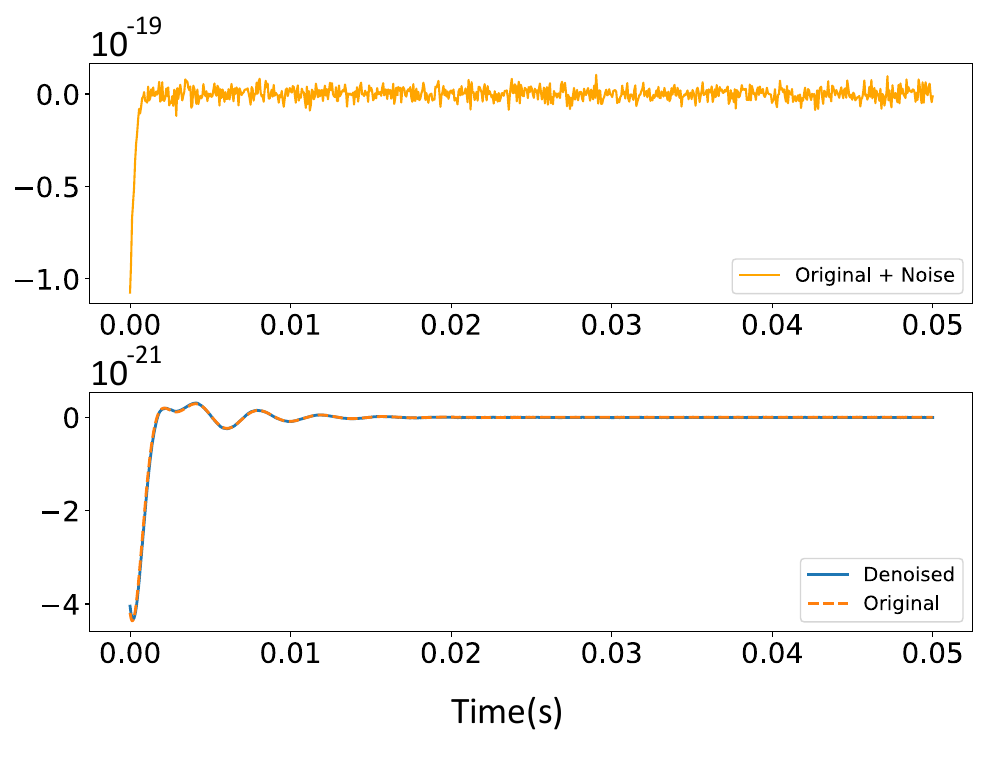}
  (a)
\end{minipage}
\hfill
\begin{minipage}{0.34\textwidth}
  \centering
  \includegraphics[width=\linewidth]{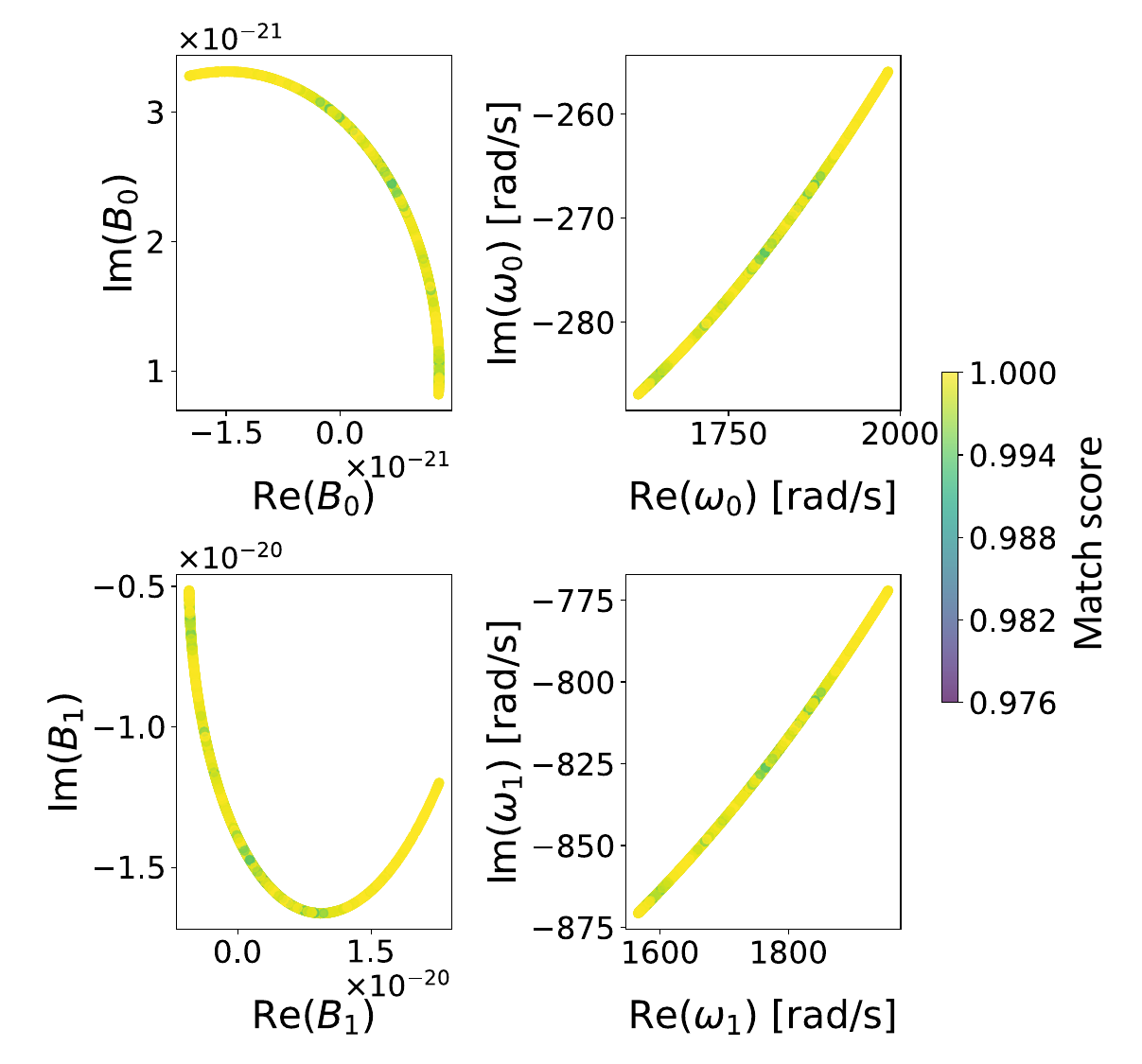}
  (b)
\end{minipage}
\hfill
\begin{minipage}{0.32\textwidth}
  \centering
  \includegraphics[width=\linewidth]{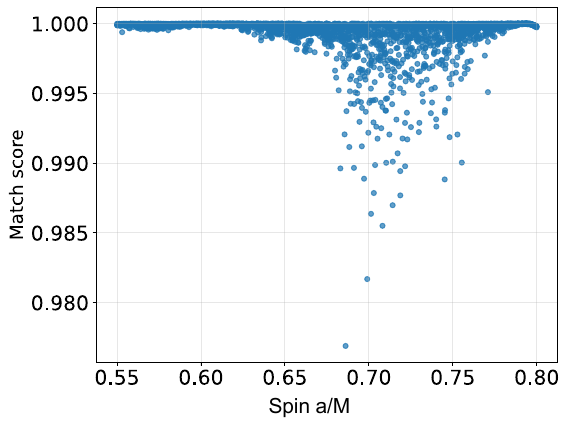}
  (c)
\end{minipage}

\vspace{2mm}

\begin{minipage}{0.9\textwidth}
  \centering
  \includegraphics[width=\linewidth]{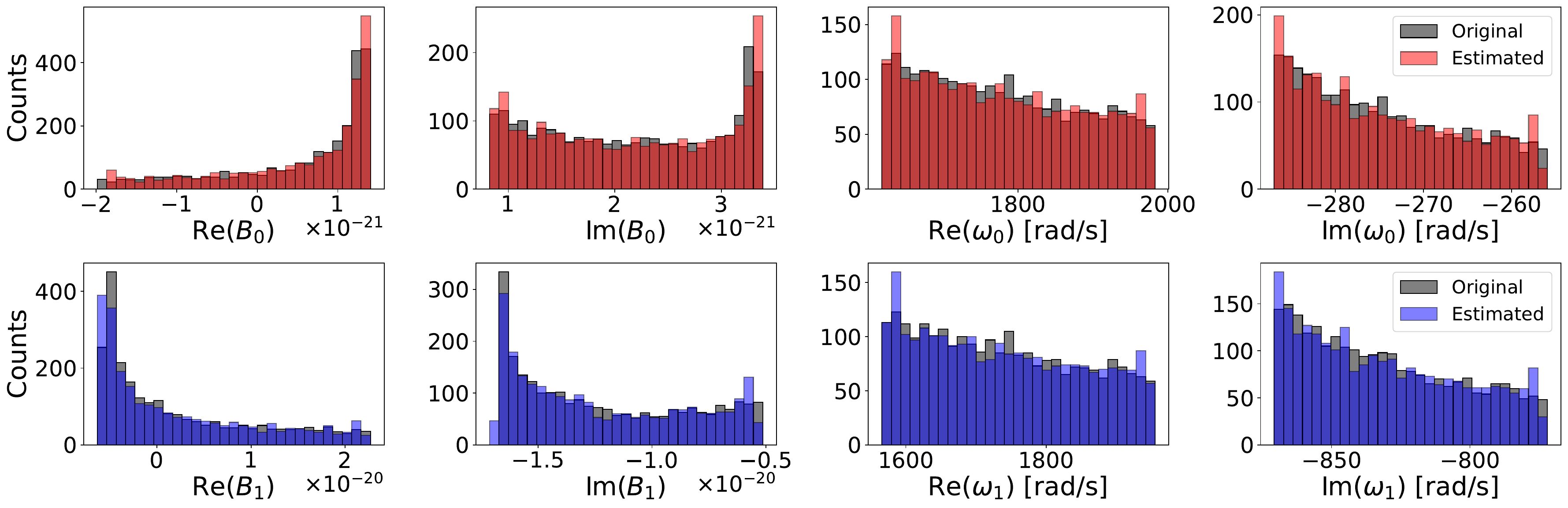}
  (d)
\end{minipage}
\vspace{2mm}

\begin{minipage}{0.9\textwidth}
  \centering
  \includegraphics[width=\linewidth]{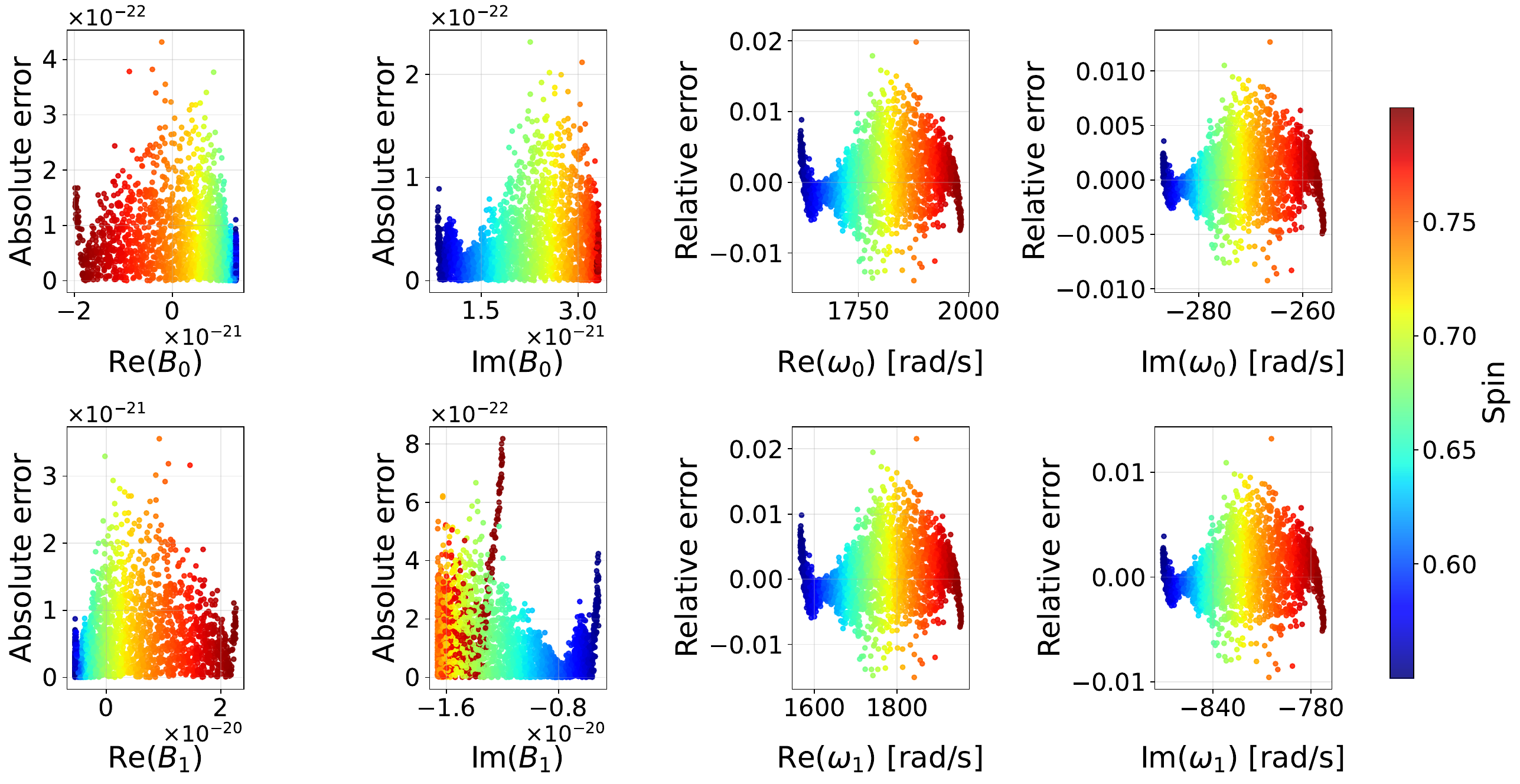}
  (e)
\end{minipage}

\caption{Validation results for spins in the range 0.55–0.8.}
\label{fig:0.55-0.8}
\end{figure*}

\begin{figure*}[t]
\centering
\begin{minipage}{0.3\textwidth}
  \centering
  \includegraphics[width=\linewidth]{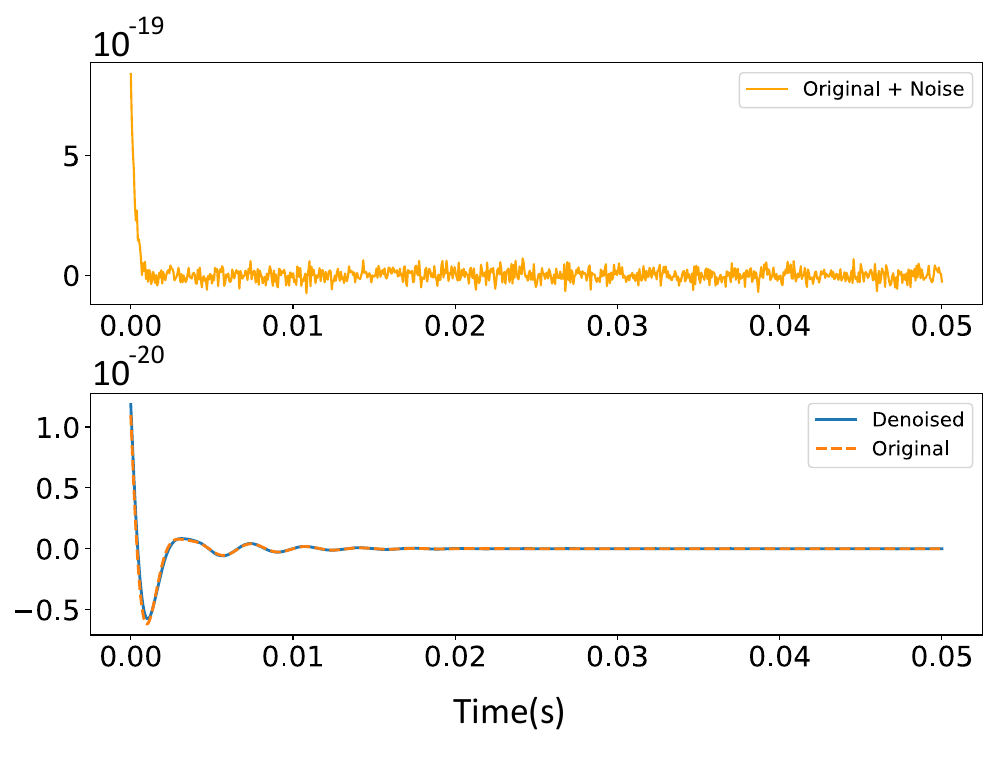}
  (a)
\end{minipage}
\hfill
\begin{minipage}{0.34\textwidth}
  \centering
  \includegraphics[width=\linewidth]{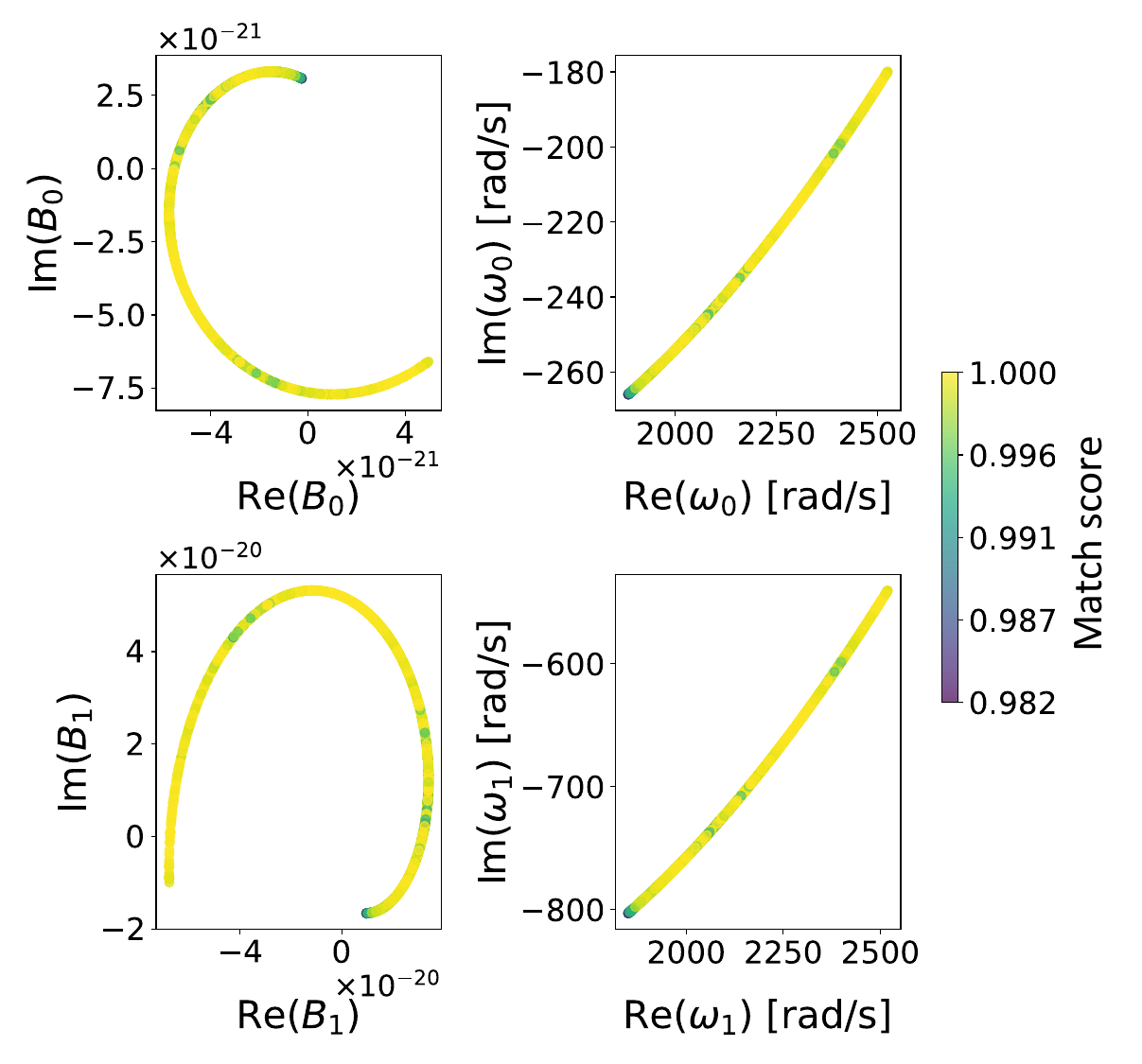}
  (b)
\end{minipage}
\hfill
\begin{minipage}{0.32\textwidth}
  \centering
  \includegraphics[width=\linewidth]{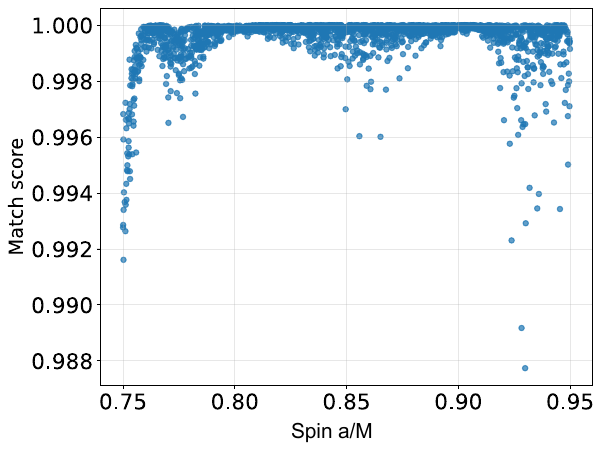}
  (c)
\end{minipage}

\vspace{2mm}

\begin{minipage}{0.9\textwidth}
  \centering
  \includegraphics[width=\linewidth]{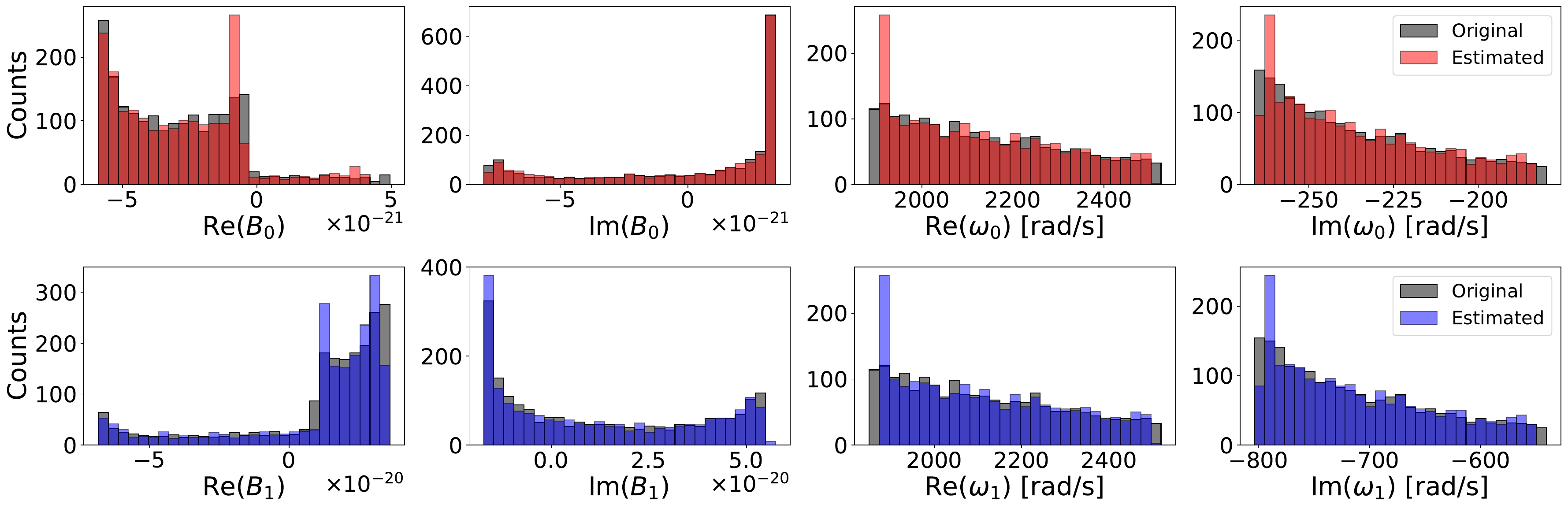}
  (d)
\end{minipage}
\vspace{2mm}

\begin{minipage}{0.9\textwidth}
  \centering
  \includegraphics[width=\linewidth]{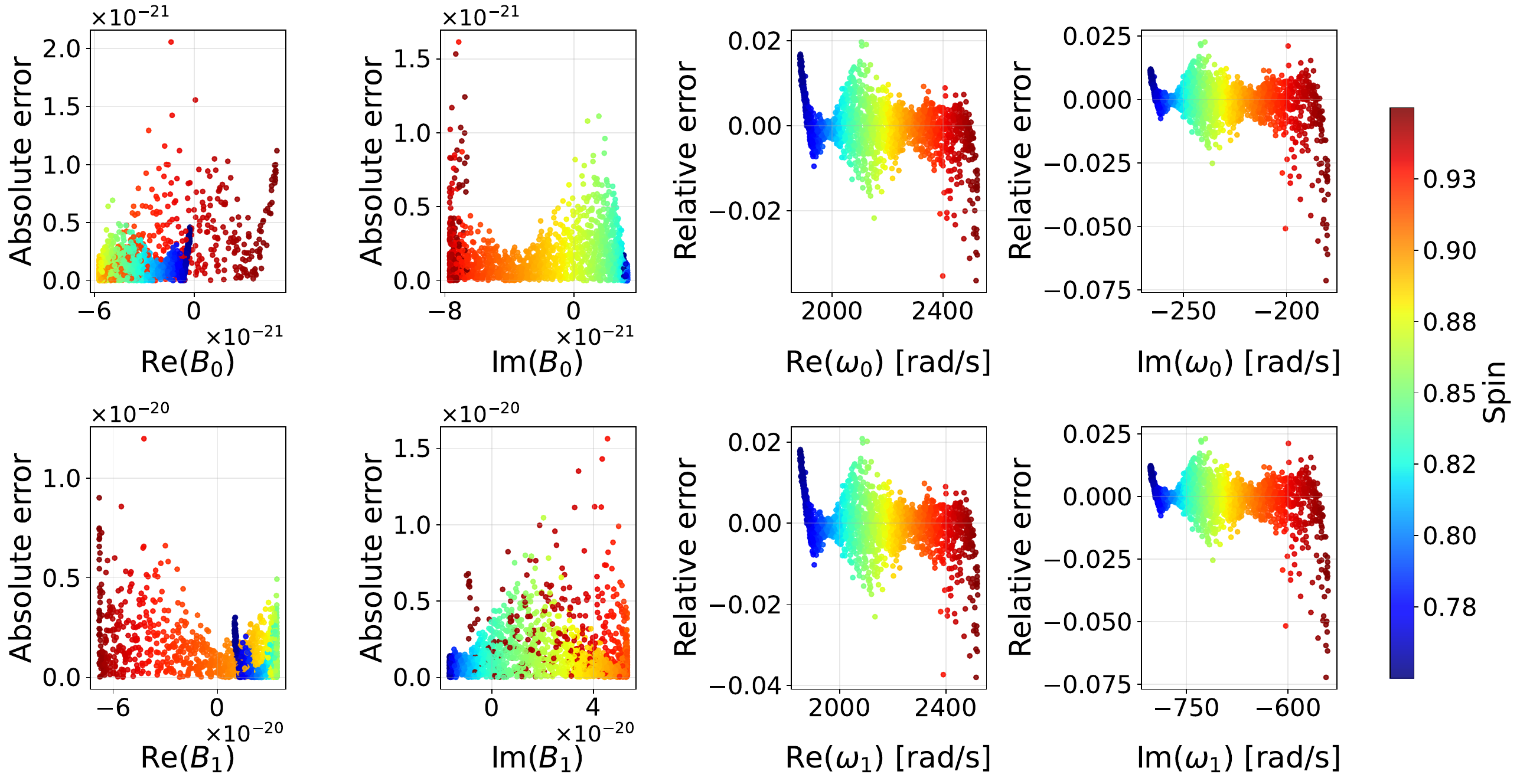}
  (e)
\end{minipage}

\caption{Validation results for spins in the range 0.75–0.95.}
\label{fig:0.75-0.95}
\end{figure*}

\begin{figure*}[t]
\centering
\begin{minipage}{0.32\textwidth}
  \centering
  \includegraphics[width=\linewidth]{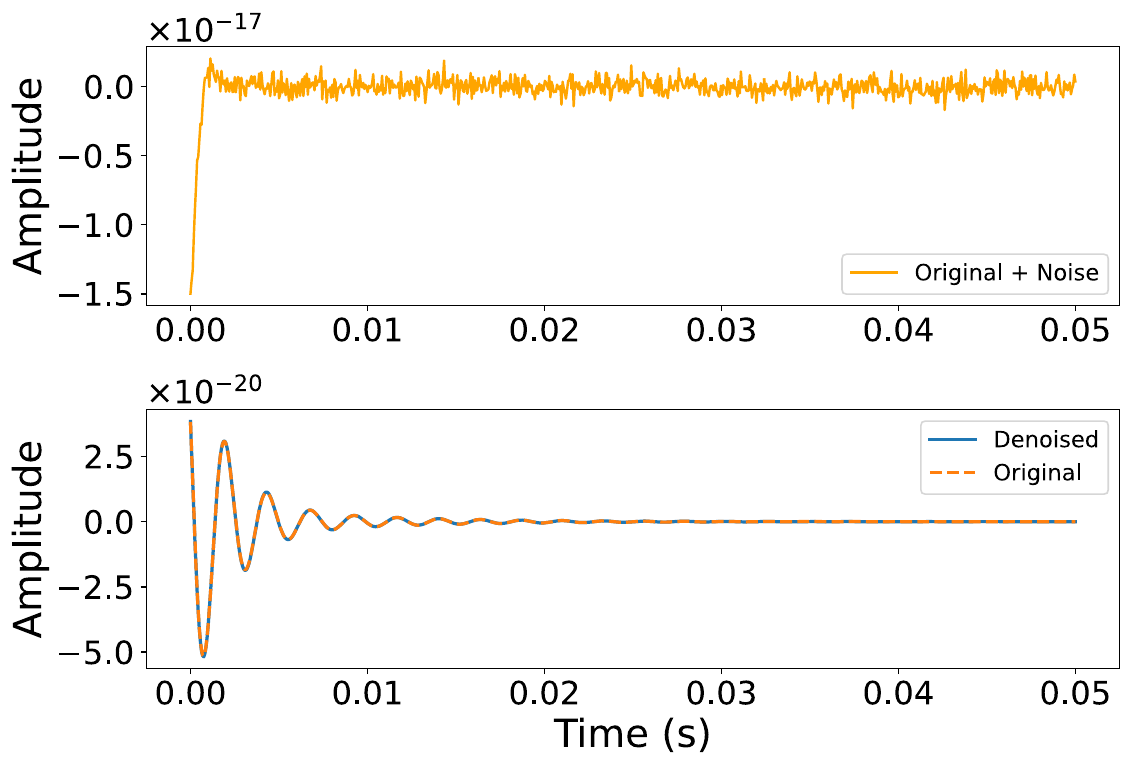}
  (a)
\end{minipage}
\hfill
\begin{minipage}{0.34\textwidth}
  \centering
  \includegraphics[width=\linewidth]{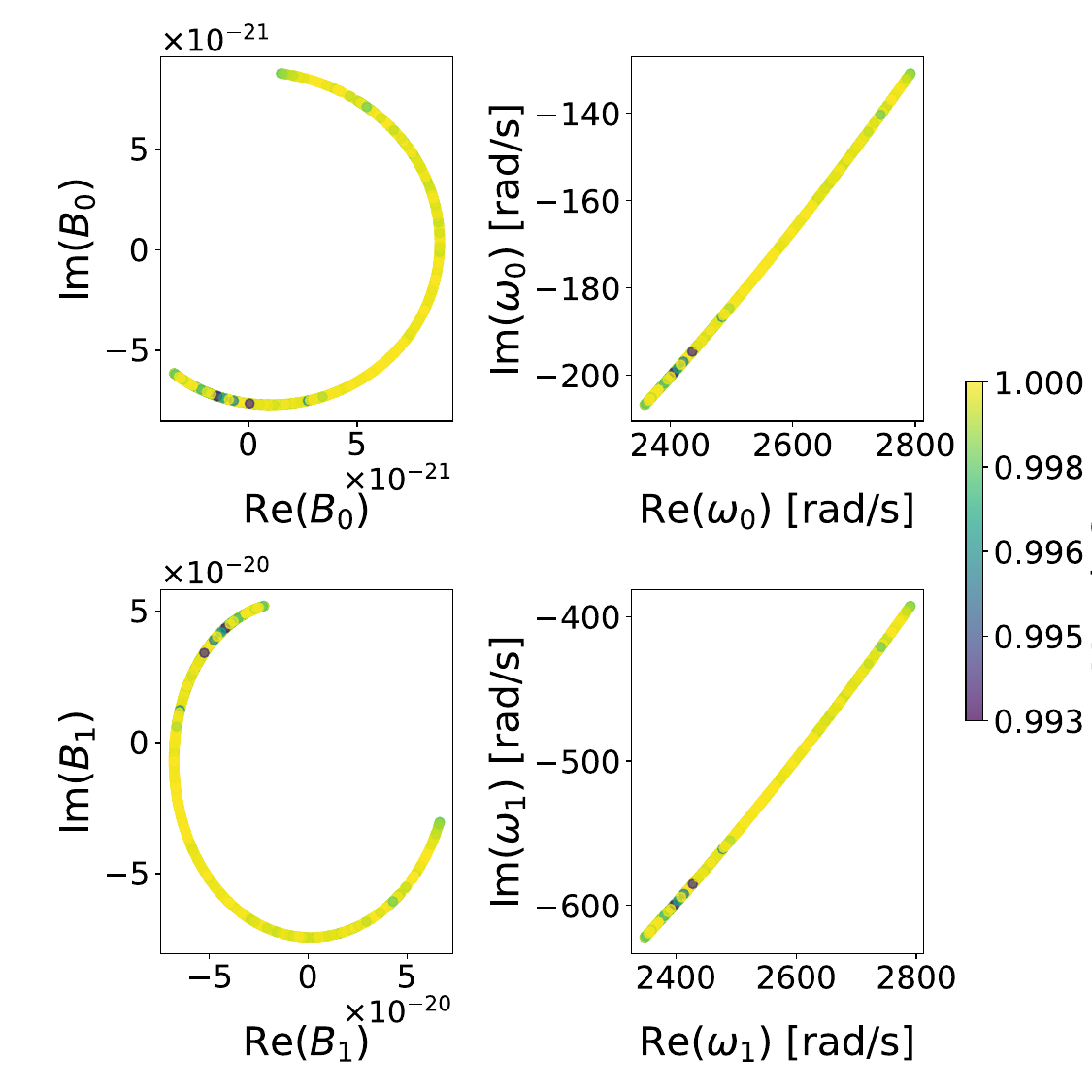}
  (b)
\end{minipage}
\hfill
\begin{minipage}{0.32\textwidth}
  \centering
  \includegraphics[width=\linewidth]{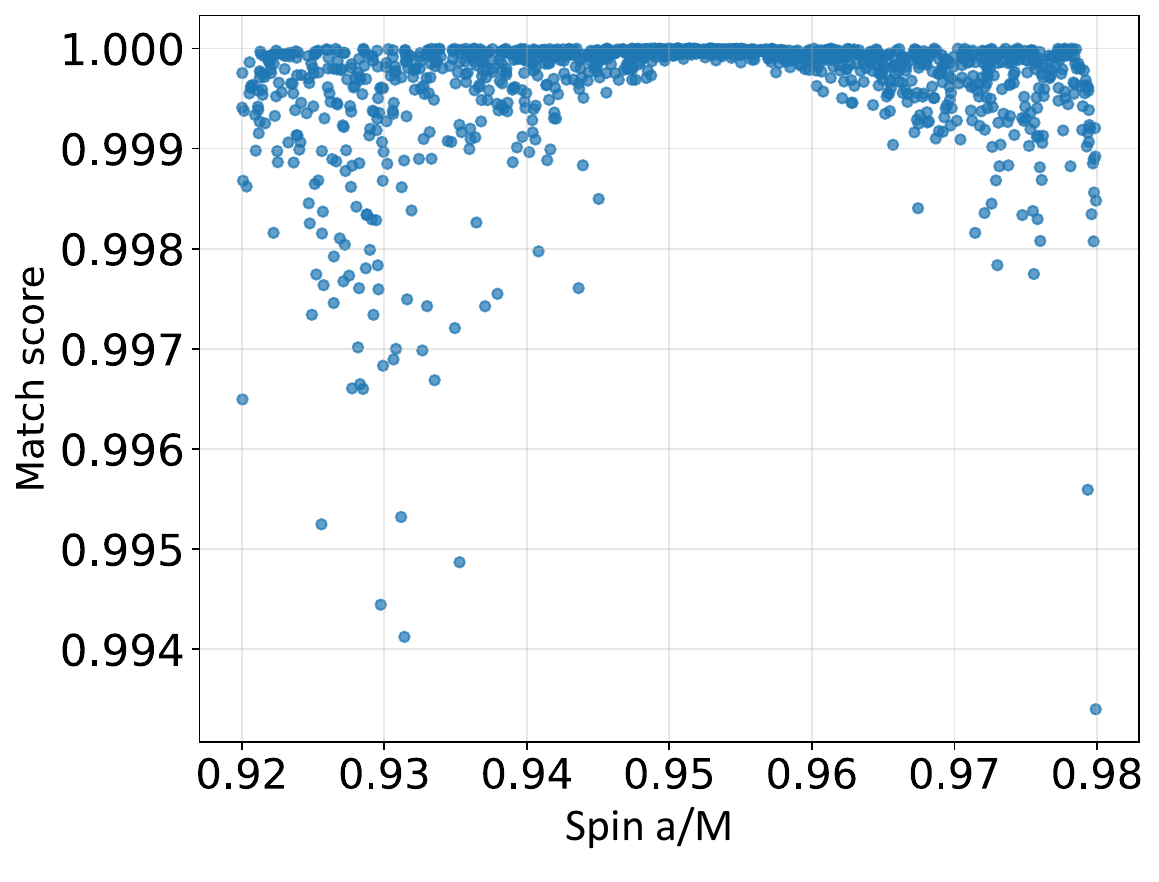}
  (c)
\end{minipage}

\vspace{2mm}

\begin{minipage}{0.9\textwidth}
  \centering
  \includegraphics[width=\linewidth]{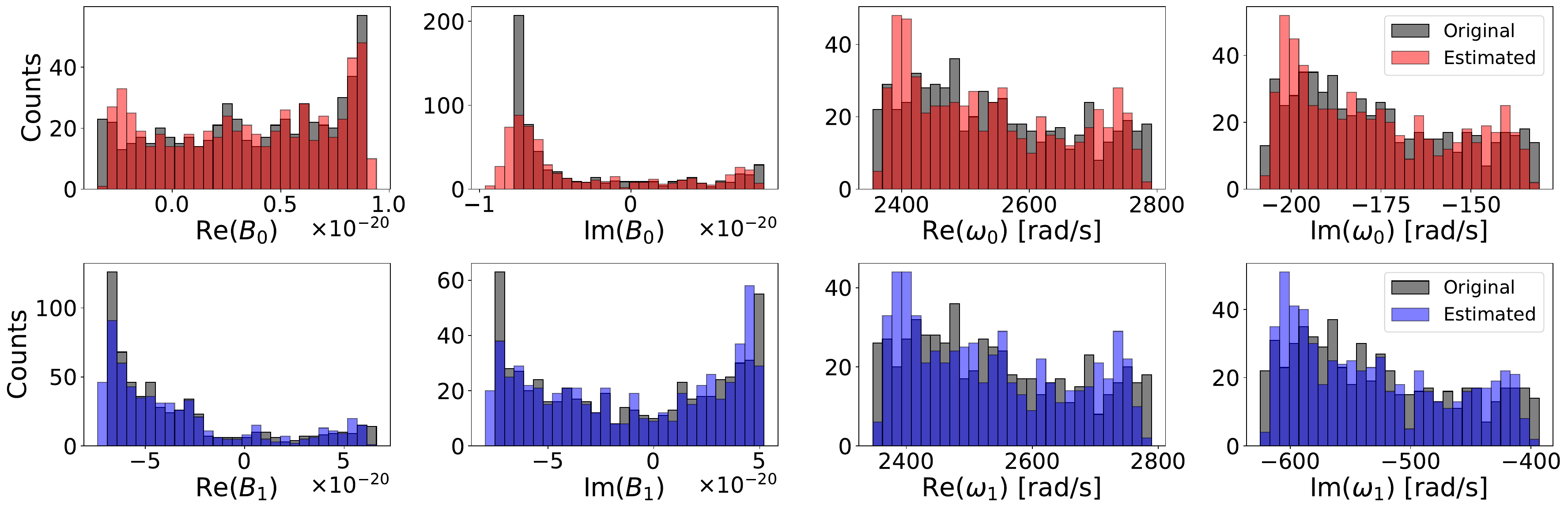}
  (d)
\end{minipage}
\vspace{2mm}

\begin{minipage}{0.9\textwidth}
  \centering
  \includegraphics[width=\linewidth]{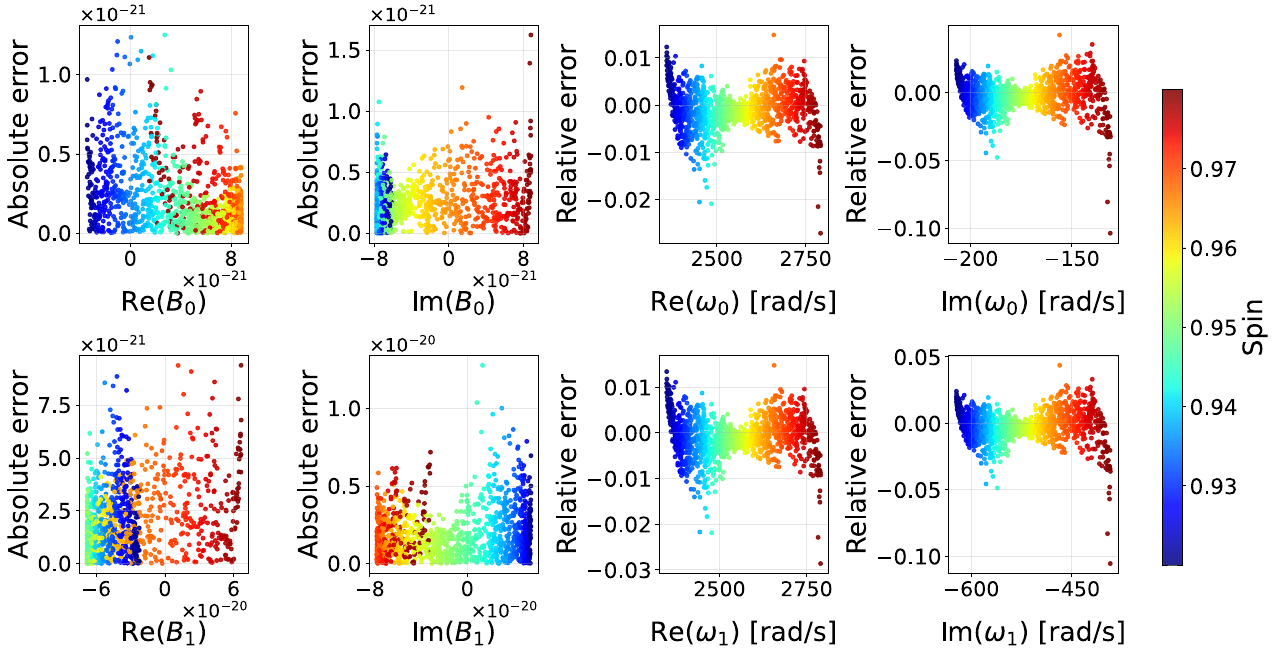}
  (e)
\end{minipage}

\caption{Validation results for spins in the range 0.92–0.98.}
\label{fig:0.92-0.98}
\end{figure*}

\begin{figure*}[t]
\centering
\begin{minipage}{0.33\textwidth}
  \centering
  \includegraphics[width=\linewidth]{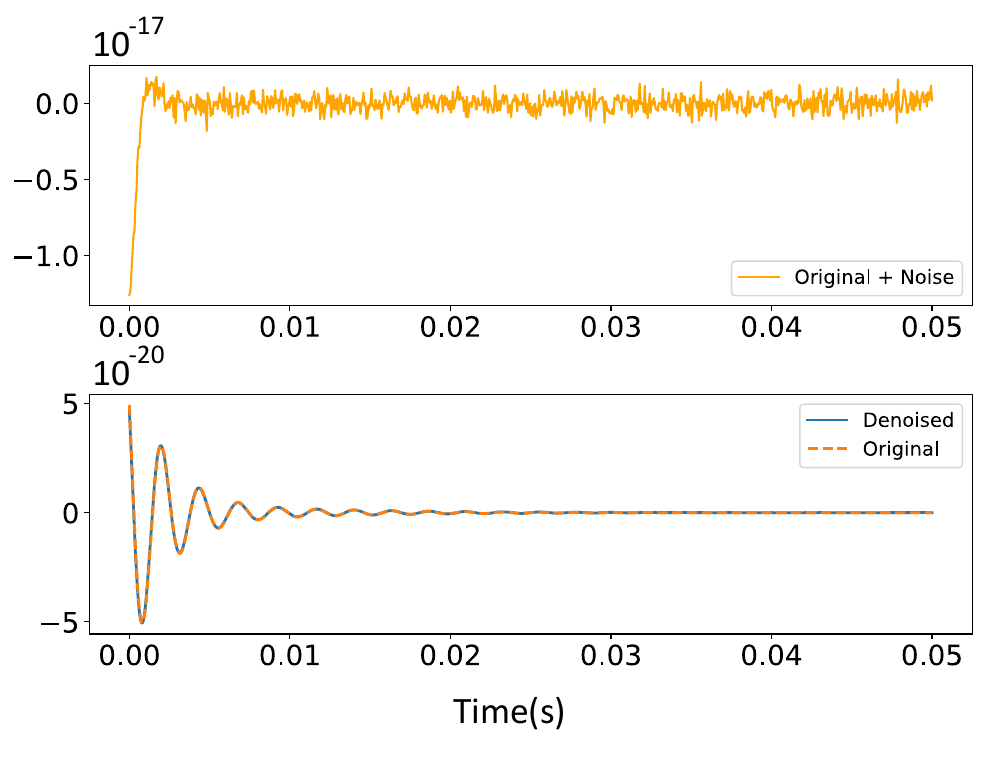}
  (a)
\end{minipage}
\hfill
\begin{minipage}{0.33\textwidth}
  \centering
  \includegraphics[width=\linewidth]{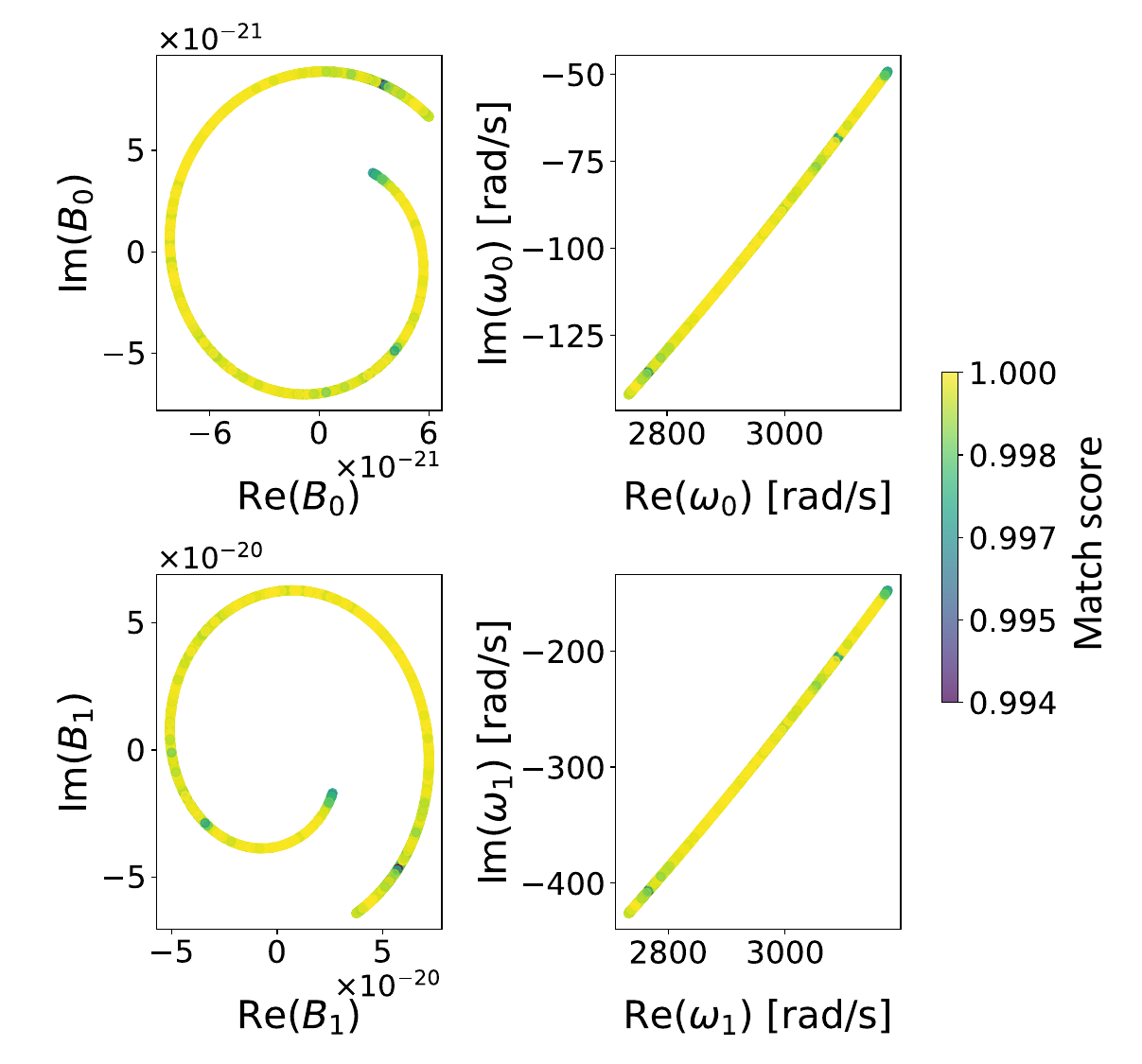}
  (b)
\end{minipage}
\hfill
\begin{minipage}{0.32\textwidth}
  \centering
  \includegraphics[width=\linewidth]{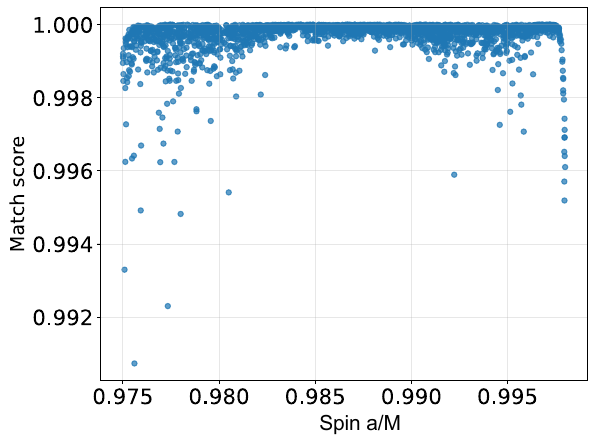}
  (c)
\end{minipage}

\vspace{2mm}

\begin{minipage}{0.9\textwidth}
  \centering
  \includegraphics[width=\linewidth]{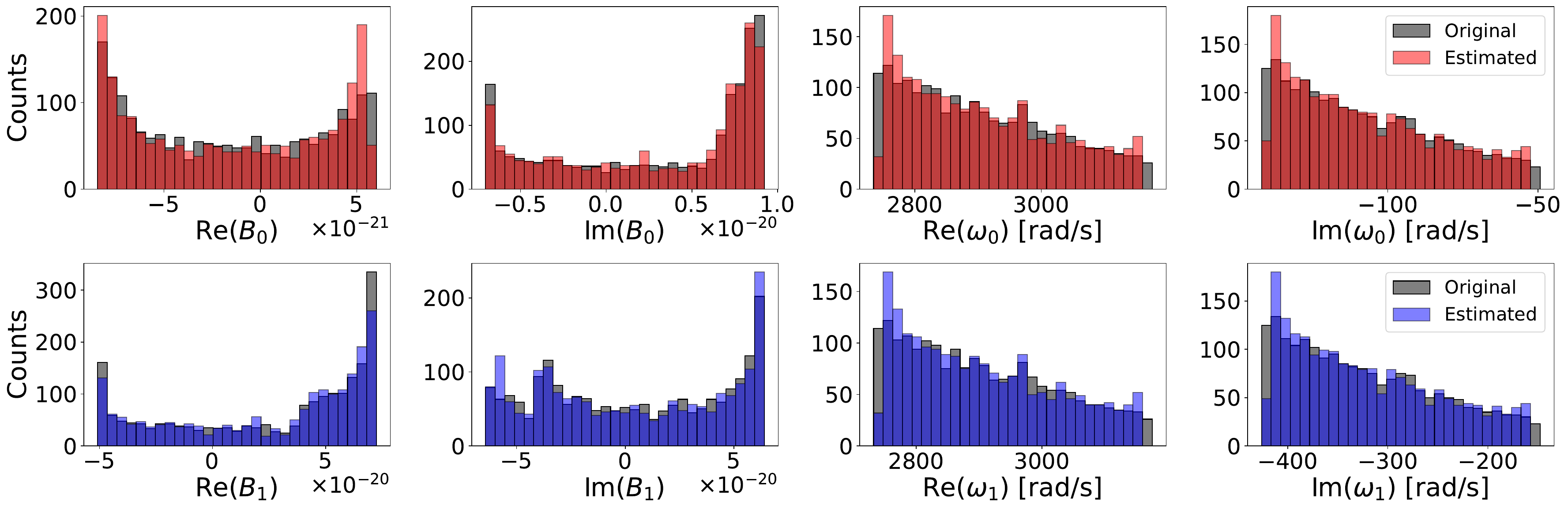}
  (d)
\end{minipage}
\vspace{2mm}

\begin{minipage}{0.9\textwidth}
  \centering
  \includegraphics[width=\linewidth]{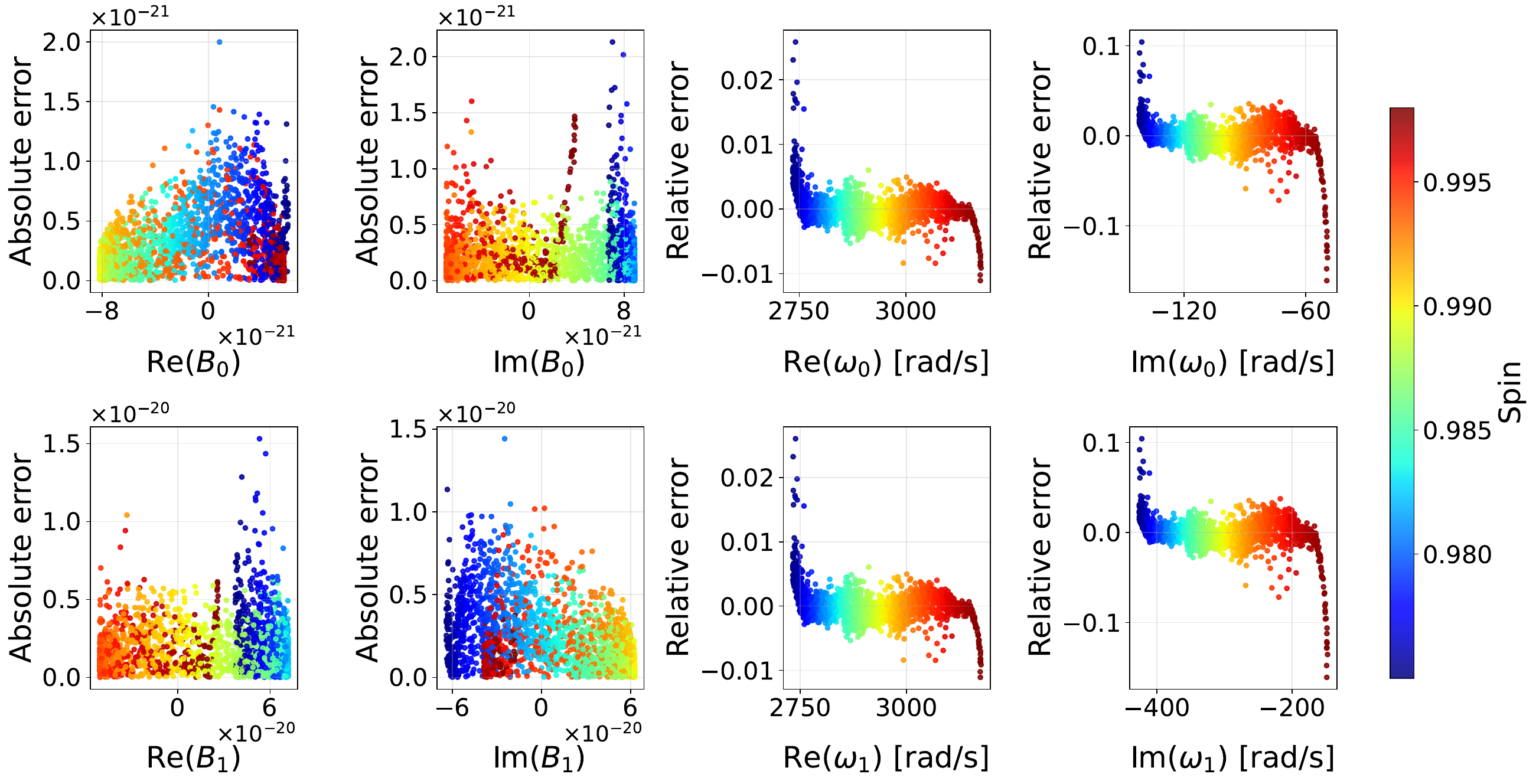}
  (e)
\end{minipage}

\caption{Validation results for spins in the range 0.975–0.998.}
\label{fig:0.975-0.998}
\end{figure*}

\begin{figure*}[t]
\centering
\begin{minipage}{0.32\textwidth}
  \centering
  \includegraphics[width=\linewidth]{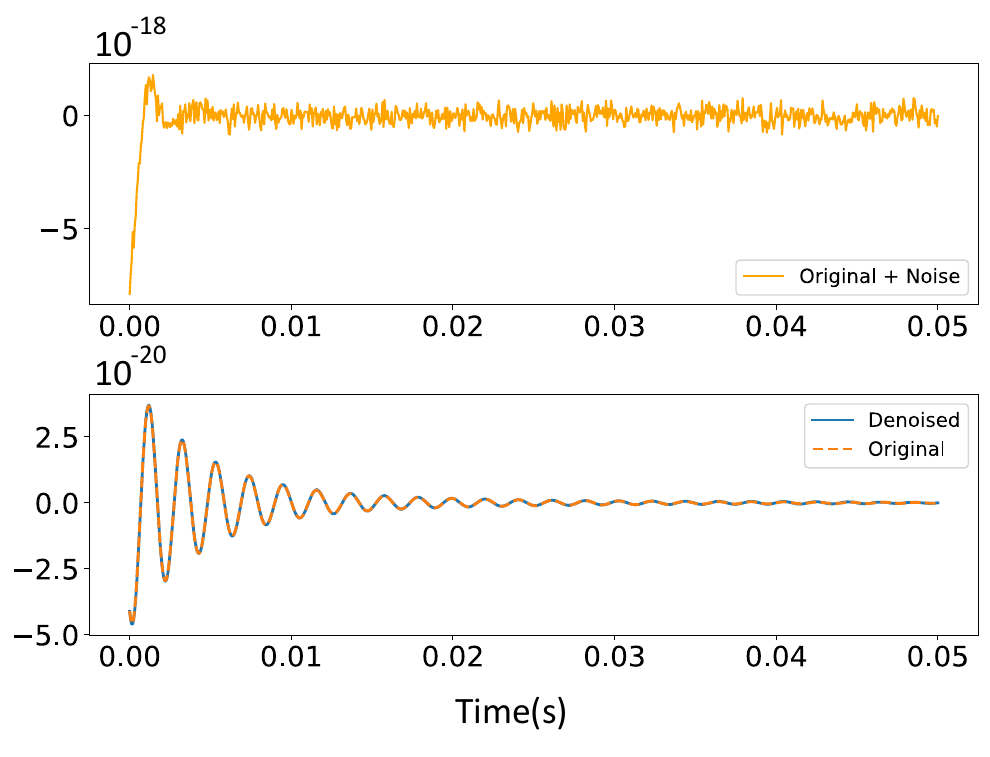}
  (a)
\end{minipage}
\hfill
\begin{minipage}{0.34\textwidth}
  \centering
  \includegraphics[width=\linewidth]{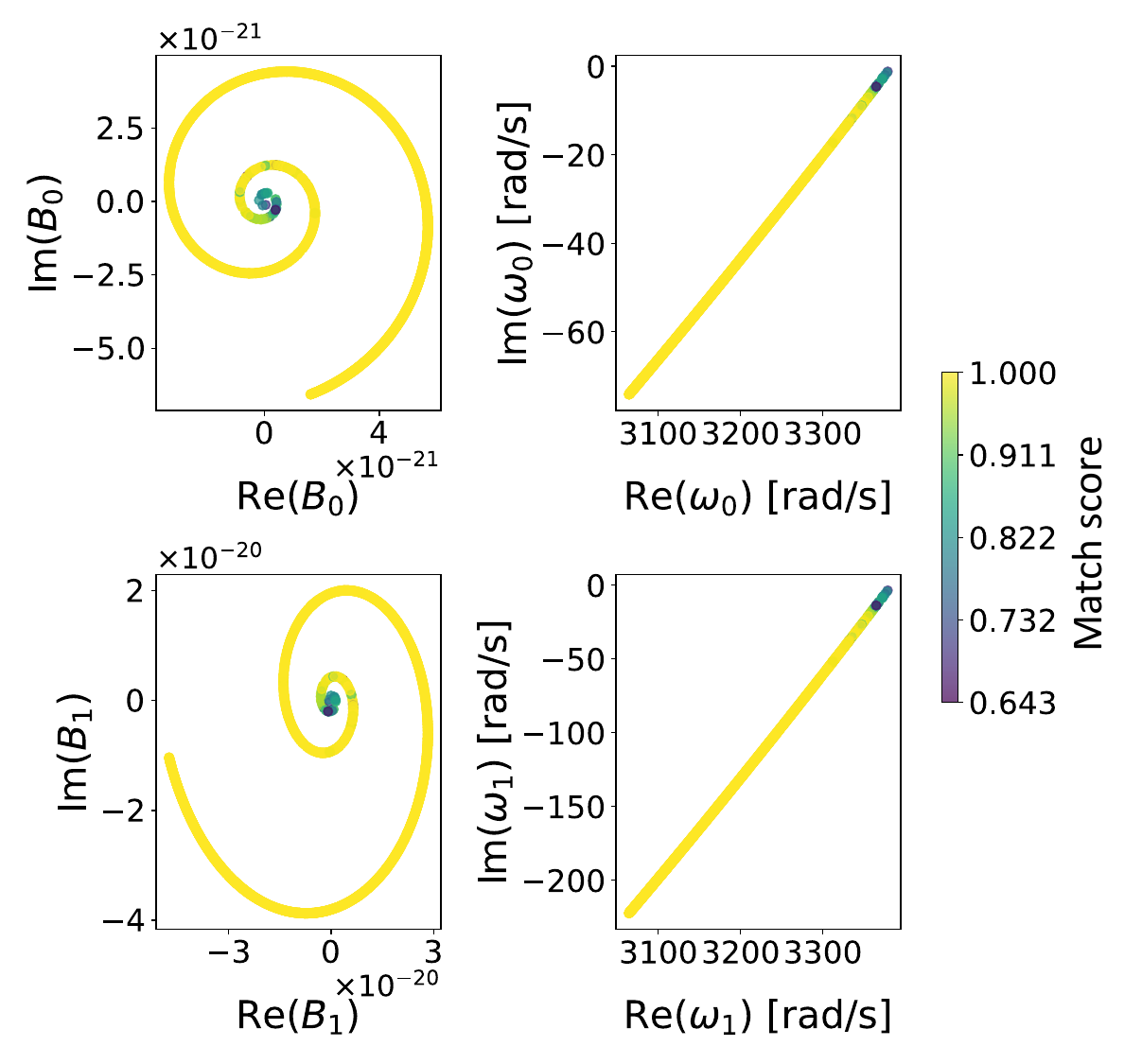}
  (b)
\end{minipage}
\hfill
\begin{minipage}{0.32\textwidth}
  \centering
  \includegraphics[width=\linewidth]{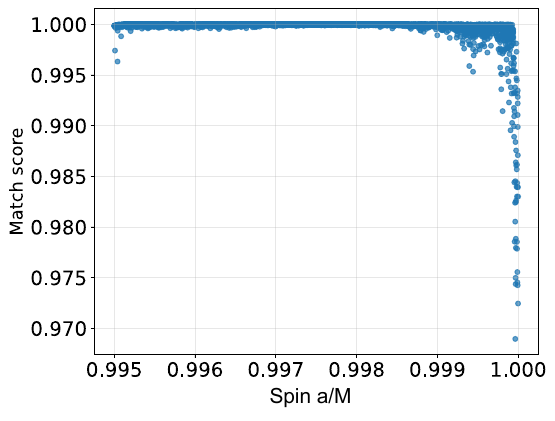}
  (c)
\end{minipage}

\vspace{2mm}

\begin{minipage}{0.9\textwidth}
  \centering
  \includegraphics[width=\linewidth]{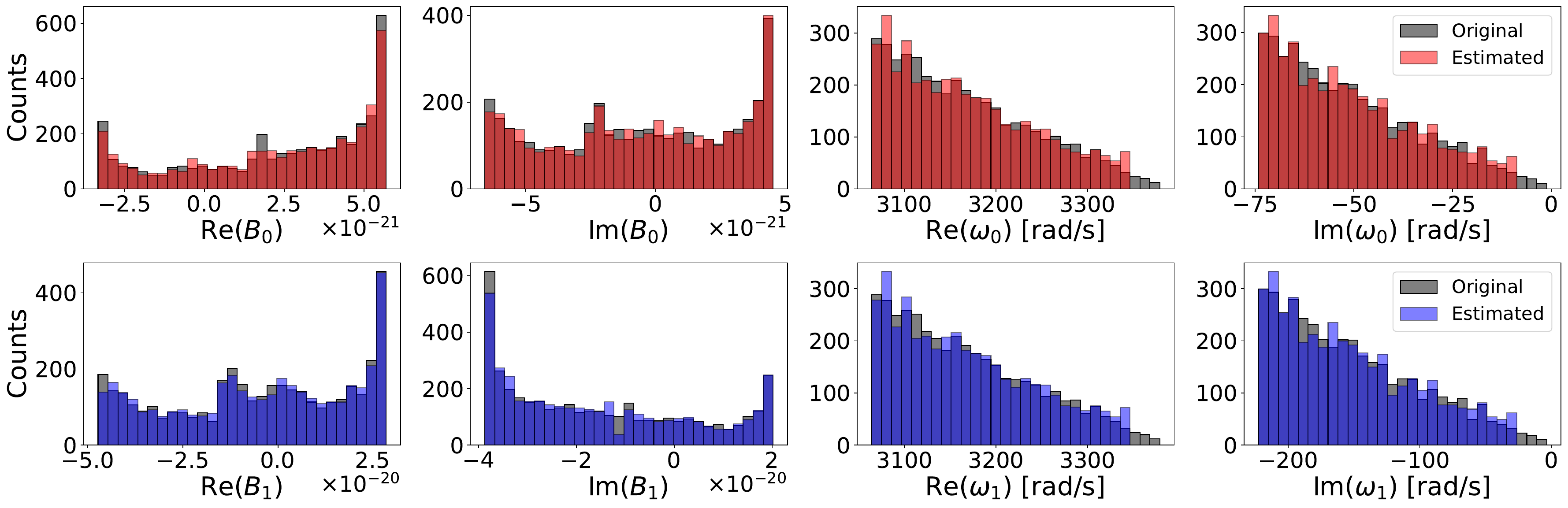}
  (d)
\end{minipage}
\vspace{2mm}

\begin{minipage}{0.9\textwidth}
  \centering
  \includegraphics[width=\linewidth]{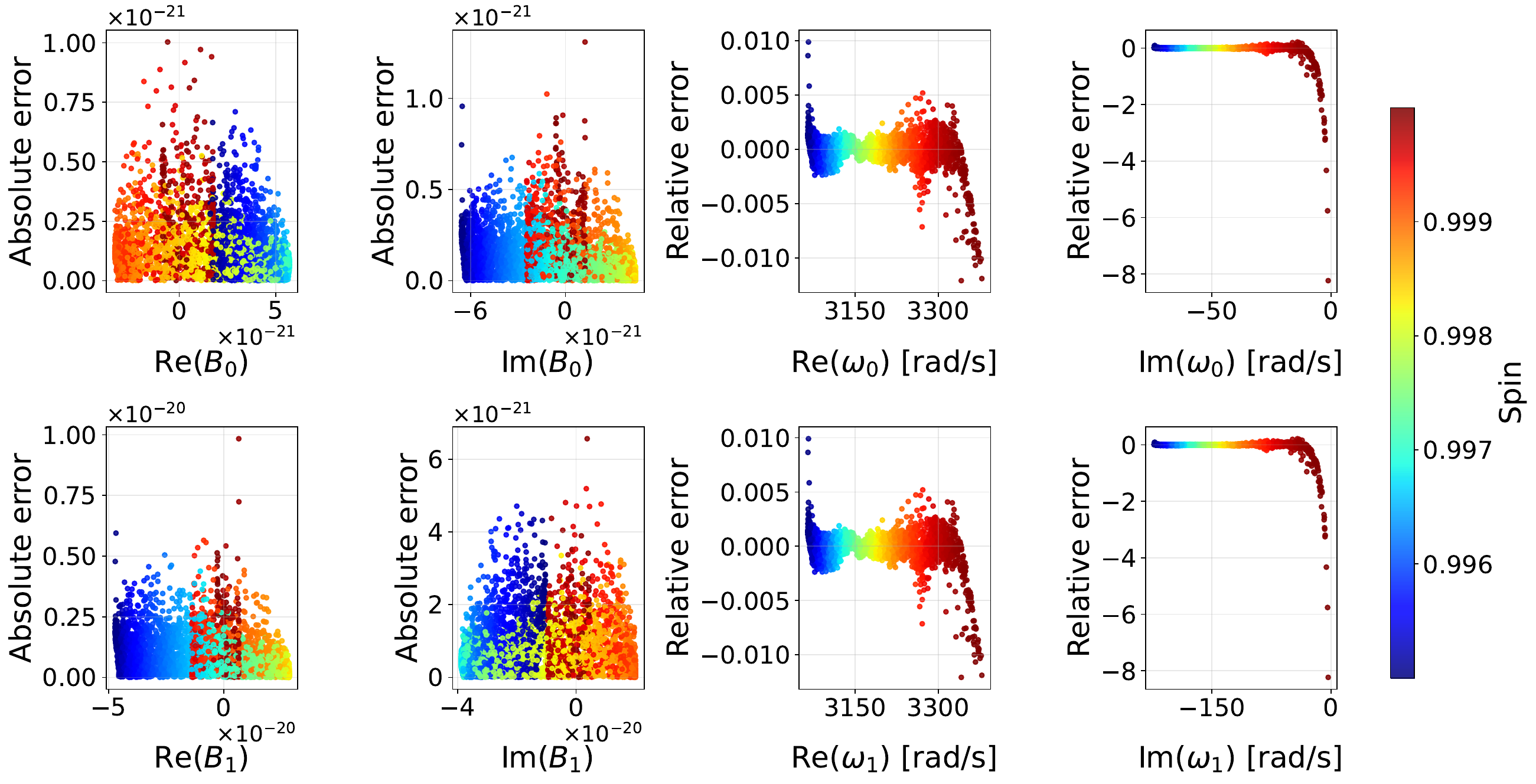}
  (e)
\end{minipage}

\caption{Validation results for spins in the range 0.995–max.}
\label{fig:0.995-max}
\end{figure*}

We first consider the same-spin-range configurations specified in Table~\ref{tab:spin_and_data}.
In this configuration, the input waveform contains eight QNM components ($n=0$--$7$), while the model is trained to estimate the parameters of the dominant two modes ($n=0,1$) and reconstruct the corresponding two-mode waveform.
The validation results for all spin intervals are summarized in Figs.~\ref{fig:0.0-0.15}--\ref{fig:0.995-max} and Table~\ref{tab:summary}.

\begin{figure}[tb]
  \centering
  \includegraphics[width=0.8\linewidth]{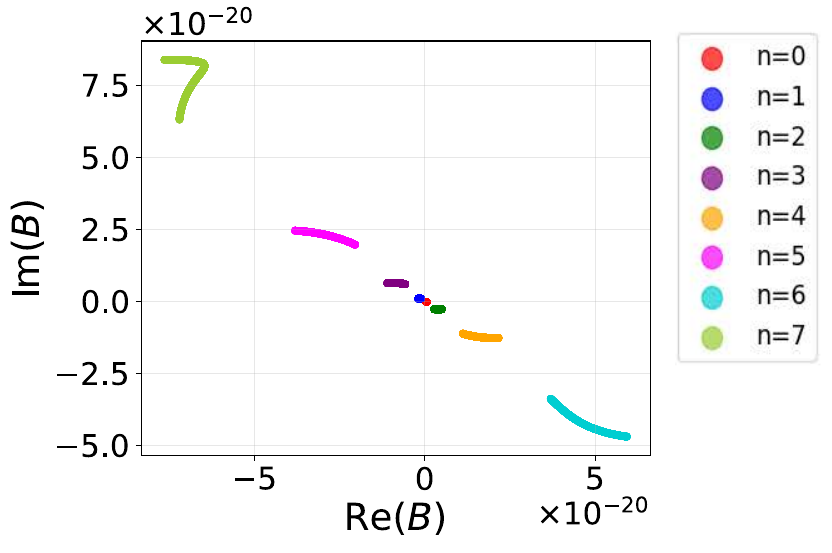}
  \caption{Characteristic behavior of $B$ in the spin interval 0.0--0.15}
  \label{fig:0.0-0.15_n=7}
\end{figure}

Figure~\ref{fig:0.0-0.15} provides a set of diagnostics applied for the same-spin-range configuration with the spin interval ${[0.0,\,0.15]}$.
Panel (a) compares a representative noisy input waveform with the denoised waveform output by the decoder and the original noise-free waveform.
The noise is effectively removed, and the denoised waveform closely matches the original waveform.
Panel (b) displays how the match score between the decoder output and the original noise-free waveform depends on the waveform parameters.
Panel (c) shows the match score for the waveform reconstructed from the parameters estimated by the encoder as a function of spin.
It shows an improvement in the match score around a spin of 0.08.
As shown in Fig.~\ref{fig:0.0-0.15_n=7}, this behavior is considered to originate from the characteristic behavior of $B_7$.

Figure~\ref{fig:0.0-0.15}(d) shows a discrepancy between the distributions of the true and estimated parameters.
However, in this interval, the range of $B_n$ is relatively narrow, and even small estimation errors can appear as large discrepancies in the distributions.
As shown in Table~\ref{tab:summary}, the actual estimation errors are comparable to those in the other intervals, and the overall spread of the distributions is generally consistent.
Additional training was conducted with modified hyperparameters, such as the number of data samples and epochs.
However, the histogram patterns did not change significantly, and similar trends were observed.
Fig.~\ref{fig:0.0-0.15}(e) shows the parameter-estimation errors as functions of the true parameter values.

Together, these panels provide complementary diagnostics of waveform reconstruction by the decoder, parameter inference by the encoder, and the resulting parameter errors.
We apply the same diagnostics below to assess the performance across the remaining spin intervals.

Figure~\ref{fig:0.1-0.6} is the same set of diagnostics for the spin interval ${[0.1,\,0.6]}$. 
The match score exceeds $0.99$, and the distributions of the true and estimated parameters agree well.

For the spin interval ${[0.55,\,0.8]}$, the match score decreases around a spin of $a/M=0.7$, as shown in Fig.~\ref{fig:0.55-0.8}(c).
This decrease coincides with $\mathrm{Re}(B_0)$ and $\mathrm{Re}(B_1)$ simultaneously approaching zero, which may make waveform reconstruction more difficult.

For the spin interval ${[0.75,\,0.95]}$, the match score shows local increases around spins of $a/M=0.8$ and $0.9$, as shown in Fig.~\ref{fig:0.75-0.95}(b) and (c).
Around $a/M=0.8$, the waveform reaches its peak at $t > 0$, and the oscillations before the subsequent decay are clearly visible.
For the spin values examined slightly away from this region, the waveform peaks at $t=0$, making the oscillatory features less apparent and potentially more difficult to estimate.
In addition, the behavior of $\mathrm{Re}(B_0)$ and $\mathrm{Re}(B_1)$ shown in Fig.~\ref{fig:0.75-0.95}(e) suggests that the region around $a/M=0.85$ lies near a turning point at the edge of the data distribution, where larger estimation errors are likely to occur.
The observed behavior around $a/M=0.8$ is therefore likely affected by multiple factors, including the characteristic waveform features and the nearby turning region in the parameter distribution.

\begin{figure}[tb]
  \centering
  \includegraphics[width=0.99\linewidth]{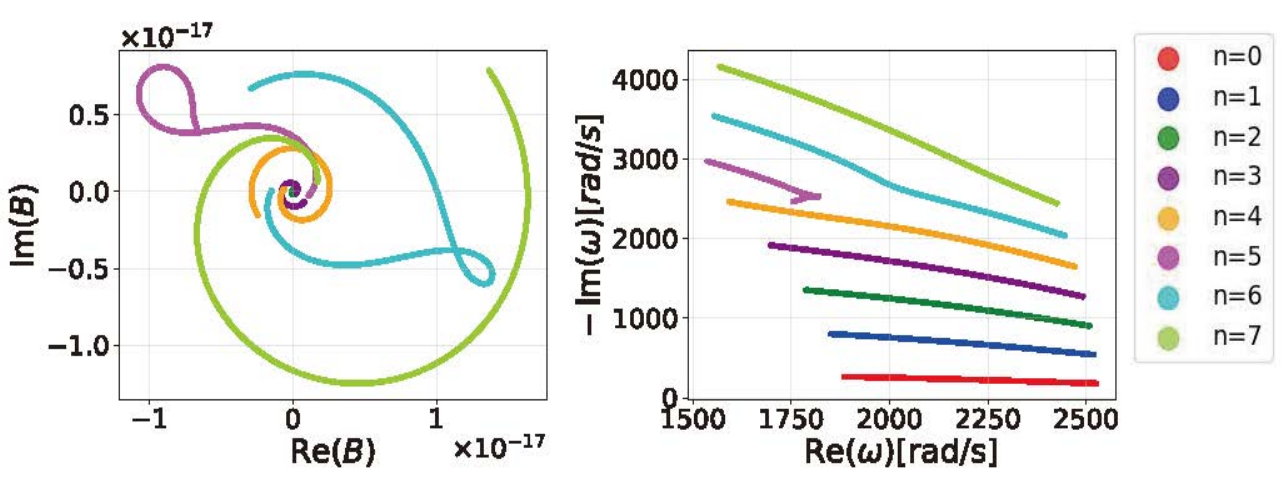}
  \caption{Resonance around a spin of 0.9.}
  \label{fig:0.75-0.95_resonance}
\end{figure}

On the other hand, the improved match score around a spin of $a/M=0.9$ is considered to be affected by a resonance between the $n=5$ and $6$ modes, as shown in Fig.~\ref{fig:0.75-0.95_resonance}.
Because the samples are uniformly spaced in $a/M$, the turning of the QNM-frequency trajectory near the avoided crossing can produce a locally higher density of samples in the frequency plane.
This effect may contribute to the improved match score.
Figure~\ref{fig:0.75-0.95}(e) shows that, although the estimation errors tend to increase near the boundaries of the spin interval, the overall estimation errors remain small.

For the spin intervals ${[0.92,\,0.98]}$ (Fig.~\ref{fig:0.92-0.98}), ${[0.975,\,0.998]}$ (Fig.~\ref{fig:0.975-0.998}), and ${[0.995,\,\mathrm{max}]}$ (Fig.~\ref{fig:0.995-max}), the match score remains high overall, and the waveform reconstruction performance is stable.
On the other hand, panels (e) of Figs.~\ref{fig:0.92-0.98}, \ref{fig:0.975-0.998}, and \ref{fig:0.995-max} show a tendency for the estimation errors to increase near the boundaries of the spin intervals.
This behavior is considered to result from two factors.
In the high-spin region, rapid parameter variations lead to a relatively low data density.
In addition, near the boundaries of the distribution, no data exist outside the interval, making it difficult for the model to sufficiently learn the parameter variations.

Overall, when the training and validation data are drawn from the same spin interval, the model achieves generally high waveform-reconstruction and parameter-estimation performance across the partitioned spin range.
The main limitations are localized increases in the errors near rapidly varying QNM structures, where uniform sampling in spin gives a sparser sampling in parameter space, and at the boundaries of the spin intervals, rather than systematic failures within the training domain.
This in-domain performance provides a baseline for the different-spin-range configuration below, which probes behavior outside the training domain, and for the more demanding simultaneous estimation of all eight components.

\begin{table*}[tbp]
\centering
\caption{Same-spin-range configuration: mean values and standard deviations of the match score, relative errors ($\mathrm{Re}(\omega_n), \mathrm{Im}(\omega_n)$), and absolute errors ($\mathrm{Re}(B_n), \mathrm{Im}(B_n)$) in the validation data.}
\scalebox{0.83}{
\begin{tabular}{c|c|c|cccc}
\hline
Spin range ($a/M$) & Match score & $n$ &
Re$(B_n)$ (abs.)& Im$(B_n)$ (abs.)&
Re$(\omega_n)$ [rad/s] (rel.)& Im$(\omega_n)$ [rad/s] (rel.)\\
\hline

\multirow{2}{*}{$[0.0,\,0.15]$}
& \multirow{2}{*}{$0.999 \pm 1.6\times10^{-3}$}
& 0
& $3.6\times10^{-23} \pm 2.8\times10^{-23}$
& $5.8\times10^{-24} \pm 5.1\times10^{-24}$
& $-5.6\times10^{-4} \pm 1.4\times10^{-2}$
& $9.7\times10^{-5} \pm 1.3\times10^{-3}$ \\
& & 1
& $1.6\times10^{-22} \pm 1.3\times10^{-22}$ 
& $1.4\times10^{-23} \pm 1.4\times10^{-23}$
& $-6.3\times10^{-4} \pm 1.7\times10^{-2}$ 
& $1.4\times10^{-4} \pm 2.3\times10^{-3}$ \\
\hline

\multirow{2}{*}{$[0.10,\,0.60]$}
& \multirow{2}{*}{$0.999 \pm 1.1\times10^{-3}$}
& 0 
& $1.5\times10^{-23} \pm 1.5\times10^{-23}$ 
& $3.0\times10^{-23} \pm 3.3\times10^{-23}$ 
& $1.1\times10^{-3} \pm 5.9\times10^{-3}$ 
& $-2.4\times10^{-4} \pm 1.5\times10^{-3}$ \\
& 
& 1 
& $7.2\times10^{-23} \pm 7.2\times10^{-23}$ 
& $2.0\times10^{-22} \pm 2.2\times10^{-22}$ 
& $1.2\times10^{-3} \pm 6.8\times10^{-3}$ 
& $-3.0\times10^{-4} \pm 1.8\times10^{-3}$ \\
\hline

\multirow{2}{*}{$[0.55,\,0.80]$}
& \multirow{2}{*}{$0.999 \pm 1.9\times10^{-3}$}
& 0 
& $6.1\times10^{-23} \pm 5.9\times10^{-23}$ 
& $3.5\times10^{-23} \pm 3.2\times10^{-23}$ 
& $3.4\times10^{-4} \pm 4.0\times10^{-3}$ 
& $-2.1\times10^{-4} \pm 2.4\times10^{-3}$ \\
& 
& 1 
& $5.2\times10^{-22} \pm 5.0\times10^{-22}$ 
& $1.3\times10^{-22} \pm 1.2\times10^{-22}$ 
& $3.7\times10^{-4} \pm 4.3\times10^{-3}$ 
& $-2.2\times10^{-4} \pm 2.5\times10^{-3}$ \\
\hline

\multirow{2}{*}{$[0.75,\,0.95]$}
& \multirow{2}{*}{$0.999 \pm 2.0\times10^{-3}$}
& 0
& $1.6\times10^{-22} \pm 1.8\times10^{-22}$
& $1.4\times10^{-22} \pm 1.7\times10^{-22}$
& $-2.8\times10^{-4} \pm 5.6\times10^{-3}$
& $5.6\times10^{-4} \pm 7.5\times10^{-3}$ \\
& 
& 1
& $1.1\times10^{-21} \pm 1.1\times10^{-21}$
& $1.4\times10^{-21} \pm 1.6\times10^{-21}$
& $-3.0\times10^{-4} \pm 5.9\times10^{-3}$
& $5.7\times10^{-4} \pm 7.6\times10^{-3}$ \\
\hline

\multirow{2}{*}{$[0.92,\,0.98]$}
& \multirow{2}{*}{$0.999 \pm 1.1\times10^{-3}$}
& 0
& $3.3\times10^{-22} \pm 3.0\times10^{-22}$
& $4.0\times10^{-22} \pm 3.3\times10^{-22}$
& $-1.7\times10^{-3} \pm 8.6\times10^{-3}$
& $5.4\times10^{-3} \pm 2.2\times10^{-2}$ \\
& 
& 1
& $2.4\times10^{-21} \pm 1.9\times10^{-21}$
& $2.9\times10^{-21} \pm 2.6\times10^{-21}$
& $-1.7\times10^{-3} \pm 8.8\times10^{-3}$
& $5.0\times10^{-3} \pm 2.2\times10^{-2}$ \\
\hline

\multirow{2}{*}{$[0.975,\,0.998]$}
& \multirow{2}{*}{$0.9997 \pm 5.1\times10^{-4}$}
& 0
& $3.3\times10^{-22} \pm 2.7\times10^{-22}$
& $2.4\times10^{-22} \pm 2.4\times10^{-22}$
& $2.6\times10^{-5} \pm 2.4\times10^{-3}$
& $9.2\times10^{-4} \pm 1.6\times10^{-2}$ \\
& 
& 1
& $2.0\times10^{-21} \pm 1.7\times10^{-21}$
& $2.5\times10^{-21} \pm 2.0\times10^{-21}$
& $3.0\times10^{-5} \pm 2.4\times10^{-3}$
& $9.0\times10^{-4} \pm 1.6\times10^{-2}$ \\
\hline

\multirow{2}{*}{$[0.995,\,\mathrm{max}]$}
& \multirow{2}{*}{$0.998 \pm 1.7\times10^{-2}$}
& 0
& $1.3\times10^{-22} \pm 1.2\times10^{-22}$
& $1.3\times10^{-22} \pm 1.3\times10^{-22}$
& $-1.3\times10^{-4} \pm 1.2\times10^{-3}$
& $2.0\times10^{-2} \pm 2.4\times10^{-1}$ \\
& 
& 1
& $8.9\times10^{-22} \pm 8.6\times10^{-22}$
& $8.0\times10^{-22} \pm 7.6\times10^{-22}$
& $-1.3\times10^{-4} \pm 1.2\times10^{-3}$
& $2.0\times10^{-2} \pm 2.4\times10^{-1}$ \\
\hline

\end{tabular}
}
\label{tab:summary}
\end{table*}

\subsection{Different-spin-range configuration}\label{subsec:result-different}
\begin{figure*}[tbp]
  \centering
  \begin{minipage}{0.4\linewidth}
    \centering
    \includegraphics[width=\linewidth]{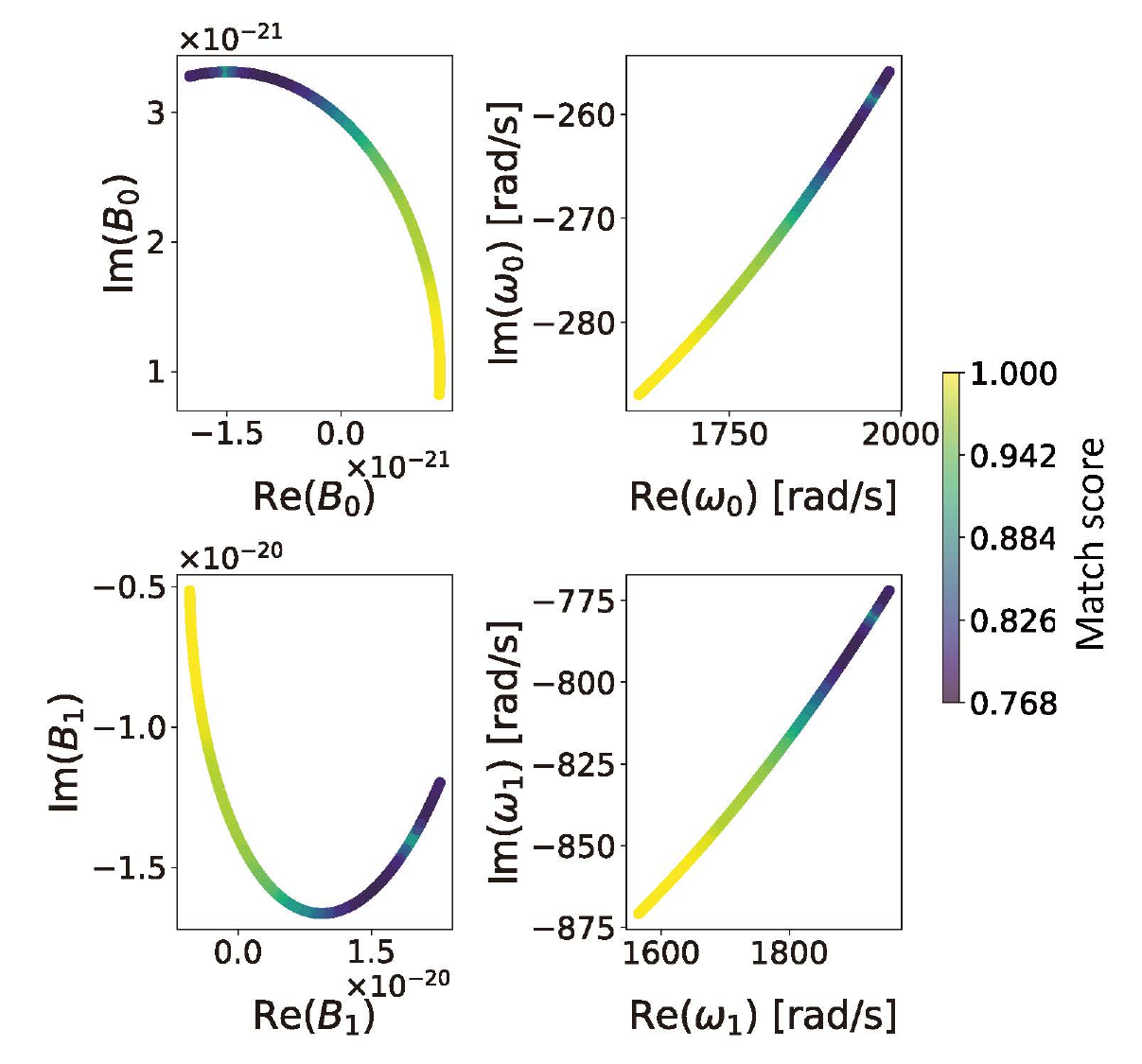}
    \\(a)
  \end{minipage}
  \hfill
  \begin{minipage}{0.5\linewidth}
    \centering
    \includegraphics[width=\linewidth]{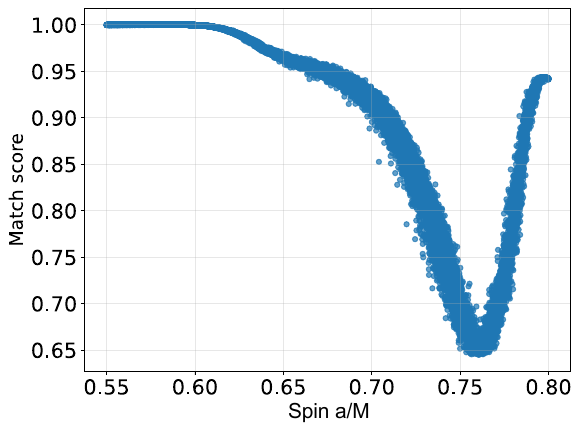}
    \\(b)
  \end{minipage}
  \vspace{1mm}

  \begin{minipage}{0.9\linewidth}
    \centering
    \includegraphics[width=\linewidth]{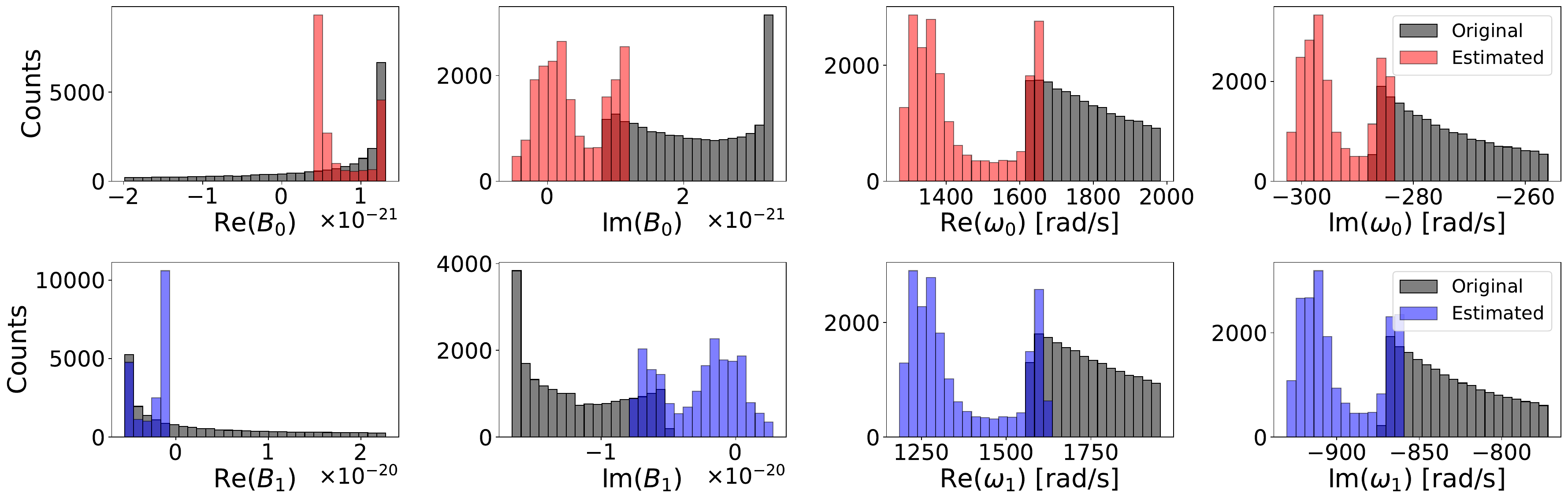}
    \\(c)
  \end{minipage}
  
  \vspace{1mm}
  \begin{minipage}{0.9\linewidth}
    \centering
    \includegraphics[width=\linewidth]{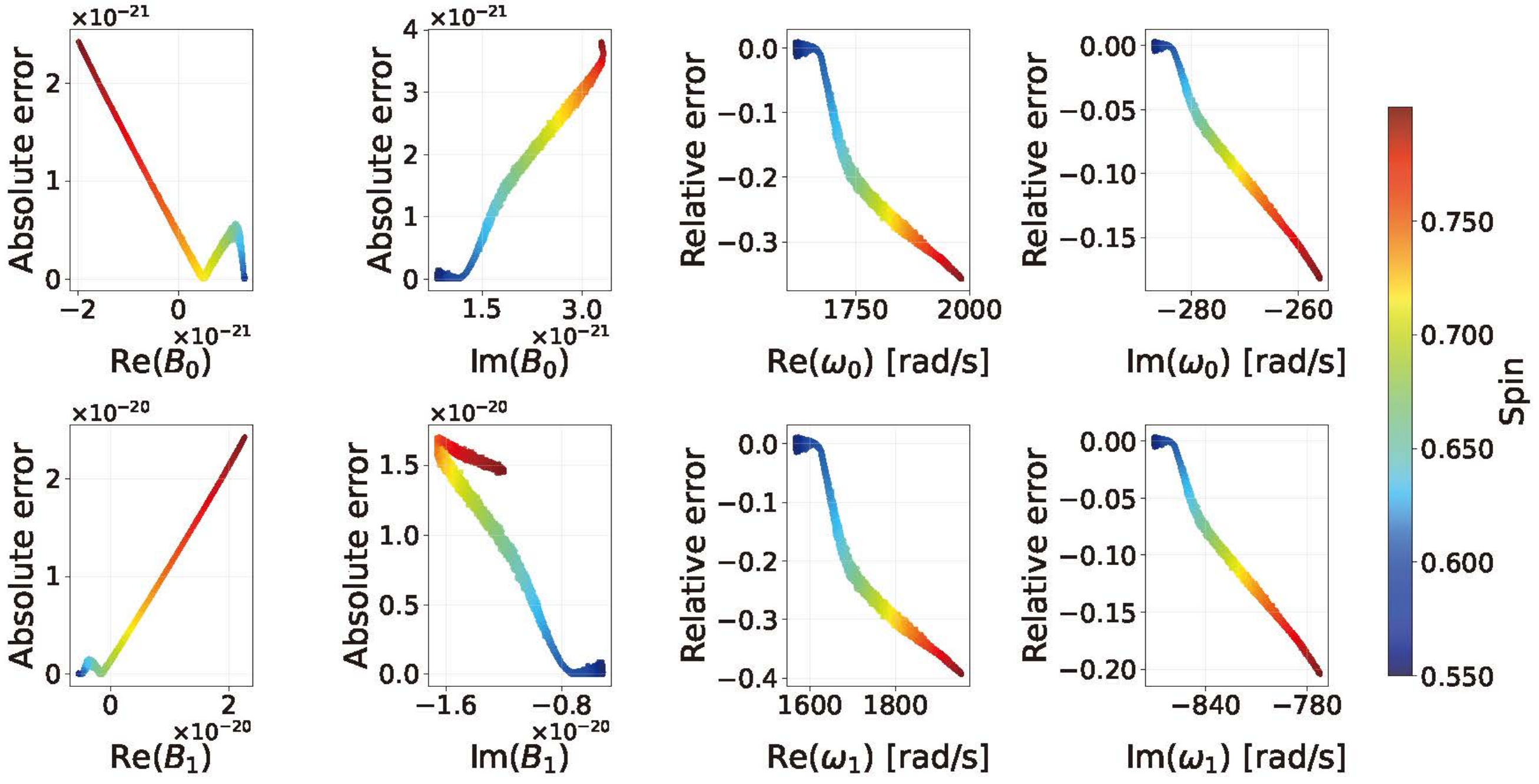}
    \\(d)
  \end{minipage}
  \caption{Different-spin-range configuration with training on spins 0.1--0.6 and validation on spins 0.55--0.8.
  }
  \label{fig:t0.1-0.6_v0.55-0.8}
\end{figure*}

\begin{figure*}[tbp]
  \centering
  \begin{minipage}{0.45\linewidth}
    \centering
    \includegraphics[width=\linewidth]{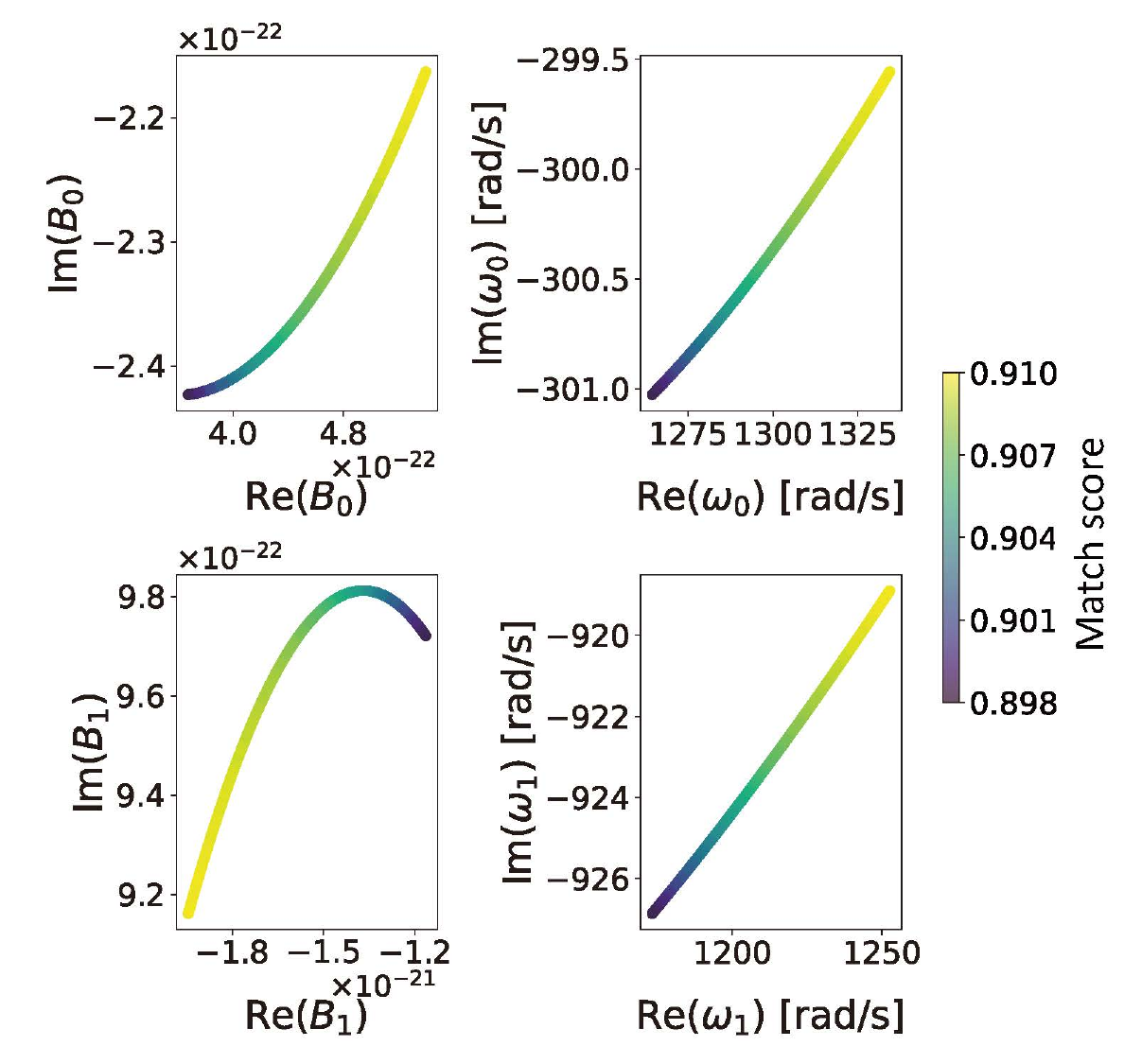}
    \\(a)
  \end{minipage}
  \hfill
  \begin{minipage}{0.45\linewidth}
    \centering
    \includegraphics[width=\linewidth]{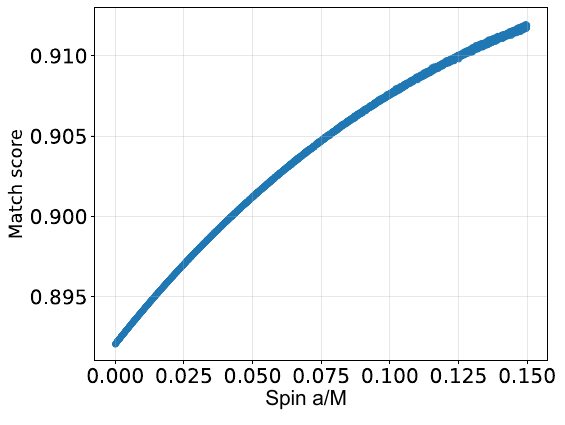}
    \\(b)
  \end{minipage}
  \vspace{1mm}

  \begin{minipage}{0.9\linewidth}
    \centering
    \includegraphics[width=\linewidth]{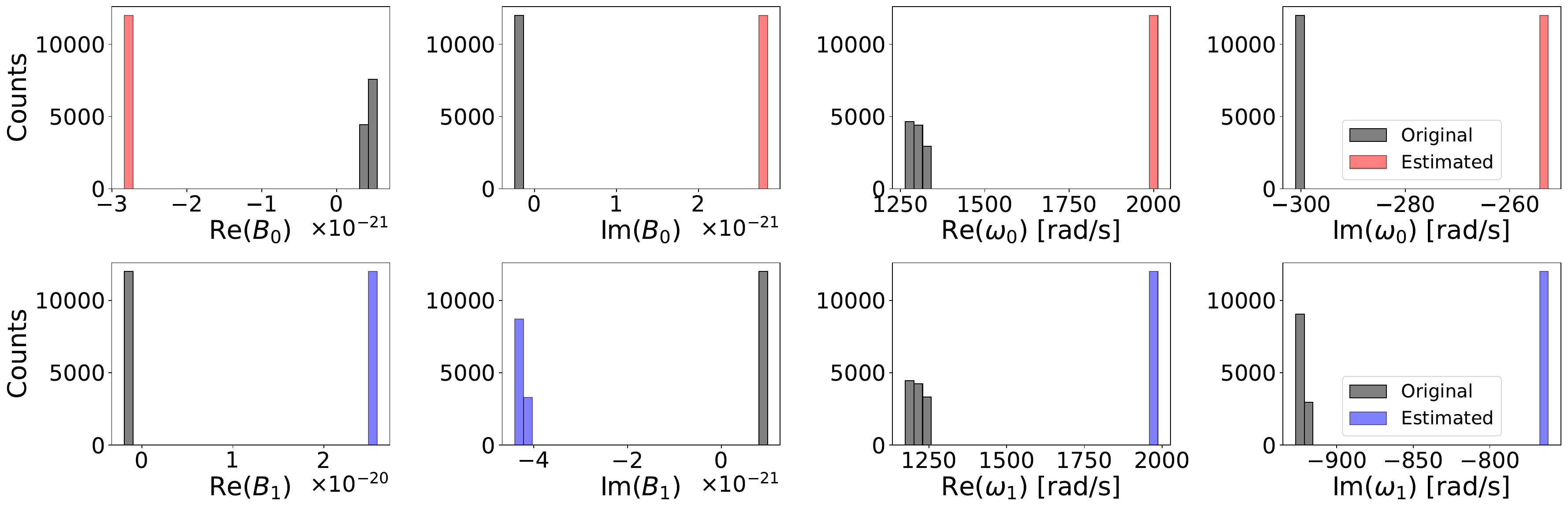}
    \\(c)
  \end{minipage}
  
  \vspace{1mm}
  \begin{minipage}{0.9\linewidth}
    \centering
    \includegraphics[width=\linewidth]{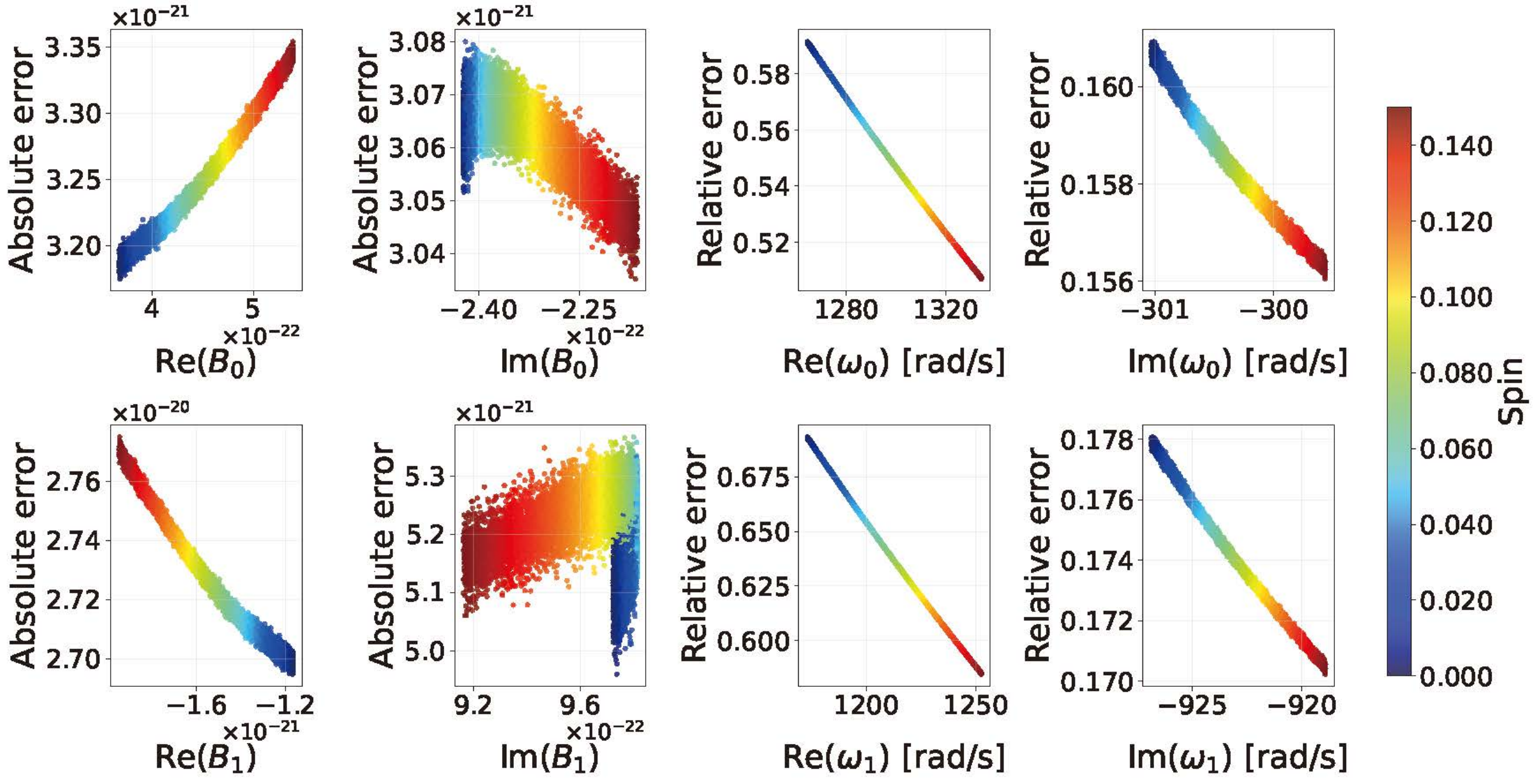}
    \\(d)
  \end{minipage}
  \caption{Different-spin-range configuration with training on spins 0.75--0.95 and validation on spins 0.0--0.15.
  }
  \label{fig:t0.75-0.95_v0.0-0.15}
\end{figure*}

We next consider the different-spin-range configurations specified in Table~\ref{tab:spin_and_data_2}.
As in the same-spin-range configuration, the input waveform contains eight QNM components with $n=0,\ldots,7$, while the network estimates the parameters of the $n=0$ and $n=1$ components and reconstructs their two-component waveform.
The results for the partially overlapping and completely disjoint training and validation ranges are shown in Figs.~\ref{fig:t0.1-0.6_v0.55-0.8} and~\ref{fig:t0.75-0.95_v0.0-0.15}, respectively.

Figure~\ref{fig:t0.1-0.6_v0.55-0.8} shows the estimation results for the case with the training data in the spin interval ${[0.1,\,0.6]}$ and the validation data in ${[0.55,\,0.8]}$.
Figure~\ref{fig:t0.1-0.6_v0.55-0.8}(a) shows the match score for the decoder output waveform.
Around a spin of 0.55, the match score remains close to 1, whereas it decreases as the spin increases and reaches approximately 0.65 around a spin of 0.8.

Figure~\ref{fig:t0.1-0.6_v0.55-0.8}(b) shows the relationship between the match score of the waveform reconstructed from the estimated parameters and the spin. 
The match score begins to decrease when the spin exceeds approximately 0.6.
However, it increases again around a spin of 0.75.
This recovery coincides with a bend in the $\mathrm{Im}(B_1)$ trajectory around $a/M=0.75$, as shown in Fig.~\ref{fig:t0.1-0.6_v0.55-0.8}(d).
As $\mathrm{Im}(B_1)$ turns toward values represented near the lower-spin edge of the training interval, the reconstructed waveform acquires features closer to those learned from the training data.
The error in $\mathrm{Im}(B_1)$ shows a corresponding bend, suggesting that this parameter variation may contribute to the recovery of the match score even though the parameter error remains non-negligible.

Figure~\ref{fig:t0.1-0.6_v0.55-0.8}(c) compares the distributions of the true and estimated parameters.
The true and estimated distributions partially overlap, corresponding to the region around ${[0.55,\,0.6]}$, where the training and validation spin intervals overlap.
This result indicates that the model reproduces the parameter distributions more accurately in spin regions close to the training data.

Figure~\ref{fig:t0.1-0.6_v0.55-0.8}(d) shows the relative and absolute errors.
The errors remain small in the region where the training and validation spin intervals overlap, whereas they increase as the spin becomes larger.

Thus, in this partially overlapping case, the model recovers the validation waveforms and parameters most accurately in the overlap region.
The accuracy degrades as the validation spins move beyond the training range.

Next, Fig.~\ref{fig:t0.75-0.95_v0.0-0.15} shows the results for training on the spin interval ${[0.75,\,0.95]}$ and validation on ${[0.0,\,0.15]}$, where the training and validation spin intervals are completely separated.
Figure~\ref{fig:t0.75-0.95_v0.0-0.15}(a) shows the match score for the decoder output waveform. 
The match score remains around 0.9 overall, which is lower than the values obtained in the cases where the training and validation spin intervals overlap.

Figure~\ref{fig:t0.75-0.95_v0.0-0.15}(b) shows the relationship between the match score of the waveform reconstructed from the estimated parameters and the spin. 
The match score increases slightly as the spin approaches 0.15.
This trend is consistent with the validation spins becoming closer to the training spin interval, where the waveform features are more similar to those of the training data.

Figure~\ref{fig:t0.75-0.95_v0.0-0.15}(c) compares the distributions of the true and estimated parameters.
The two distributions do not overlap, indicating that the model does not reproduce the parameter distributions in regions far from the training spin interval.

Finally, Fig.~\ref{fig:t0.75-0.95_v0.0-0.15}(d) shows the relative and absolute errors.
In this case, the errors remain large across the entire spin range, and no region shows sufficiently small errors.

Overall, the different-spin-range results contrast with the consistently high in-domain performance obtained in the same-spin-range configuration.
When the validation range partially overlaps the training range, the estimation remains accurate near the overlap but degrades outside it.
When the two ranges are completely disjoint, both the waveform reconstruction and parameter estimation deteriorate across the validation range.
These results indicate that the present model is effective primarily within, or close to, the spin range represented in the training data and should not be interpreted as providing reliable extrapolation to distant spin regions.
The absence of uniformly high match scores outside the training range also makes this limitation visible rather than masking it through apparently accurate out-of-domain predictions.

 \subsection{Eight-component configuration}\label{subsec:result-8components}
\begin{figure*}[tbp]
  \centering

  \begin{minipage}{0.45\linewidth}
    \centering
    \includegraphics[width=\linewidth]{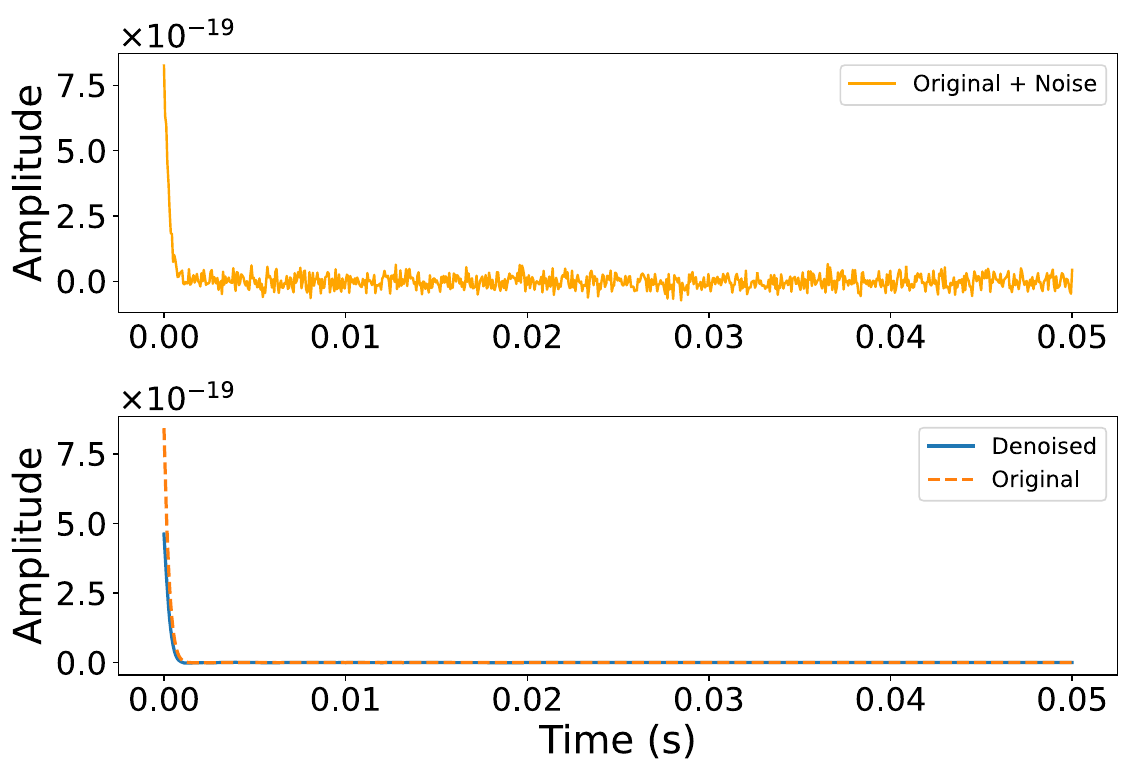}
    \caption{Representative denoising result for an eight-component superposed signal.
    The upper panel shows the noisy input waveform, and the lower panel compares the denoised waveform with the noise-free target waveform.}
    \label{fig:8waves_denoised}
  \end{minipage}
  \hfill
  \begin{minipage}{0.45\linewidth}
    \centering
    \includegraphics[width=\linewidth]{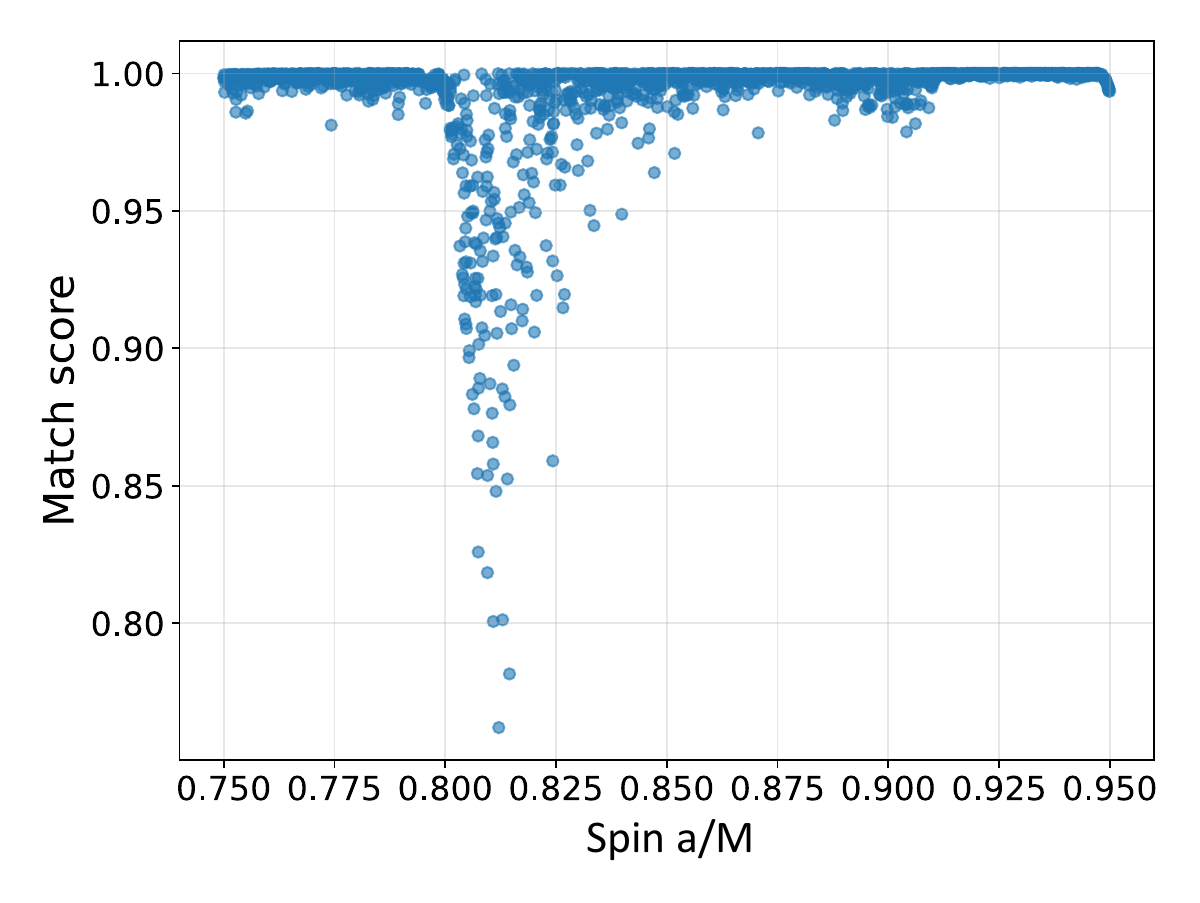}
    \caption{Match score between the noise-free target waveform and the waveform reconstructed from the parameters estimated by the encoder, as a function of spin.}
    \label{fig:8waves_match_spin}
  \end{minipage}

  \vspace{0.8em}

  \begin{minipage}{0.78\linewidth}
    \centering
    \includegraphics[width=\linewidth]{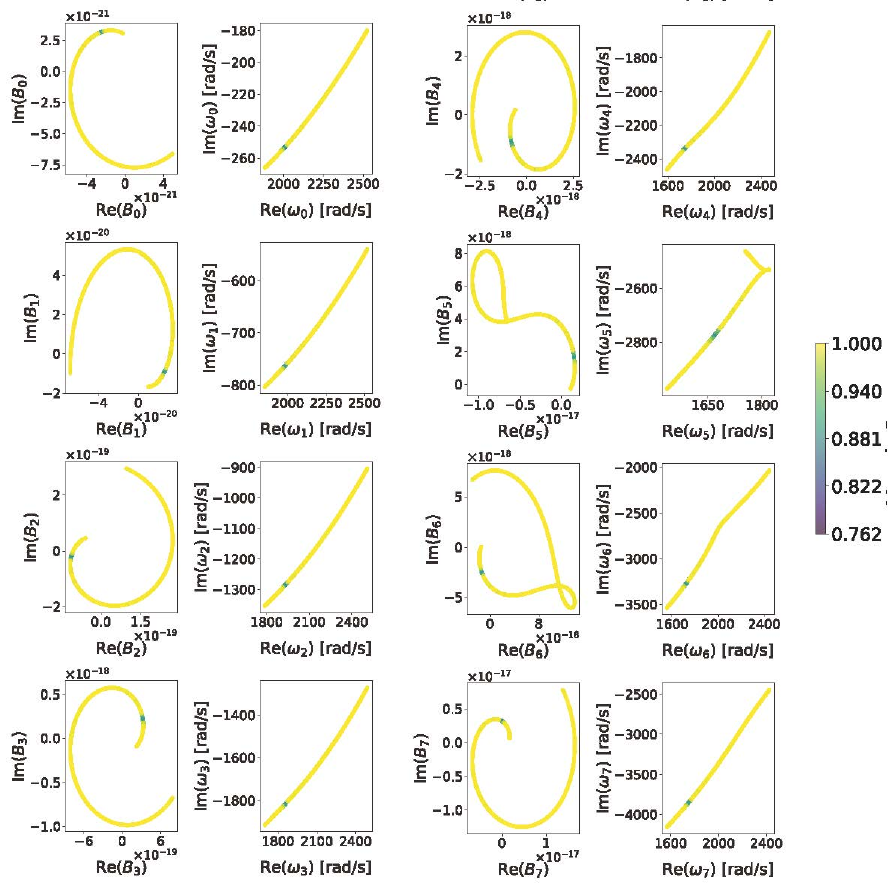}
    \caption{Match score between the decoder output waveform and the noise-free target waveform as a function of the physical parameters.}
    \label{fig:8waves_match}
  \end{minipage}

\end{figure*}

\begin{figure*}[tbp]
  \centering
  \includegraphics[width=0.85\linewidth]{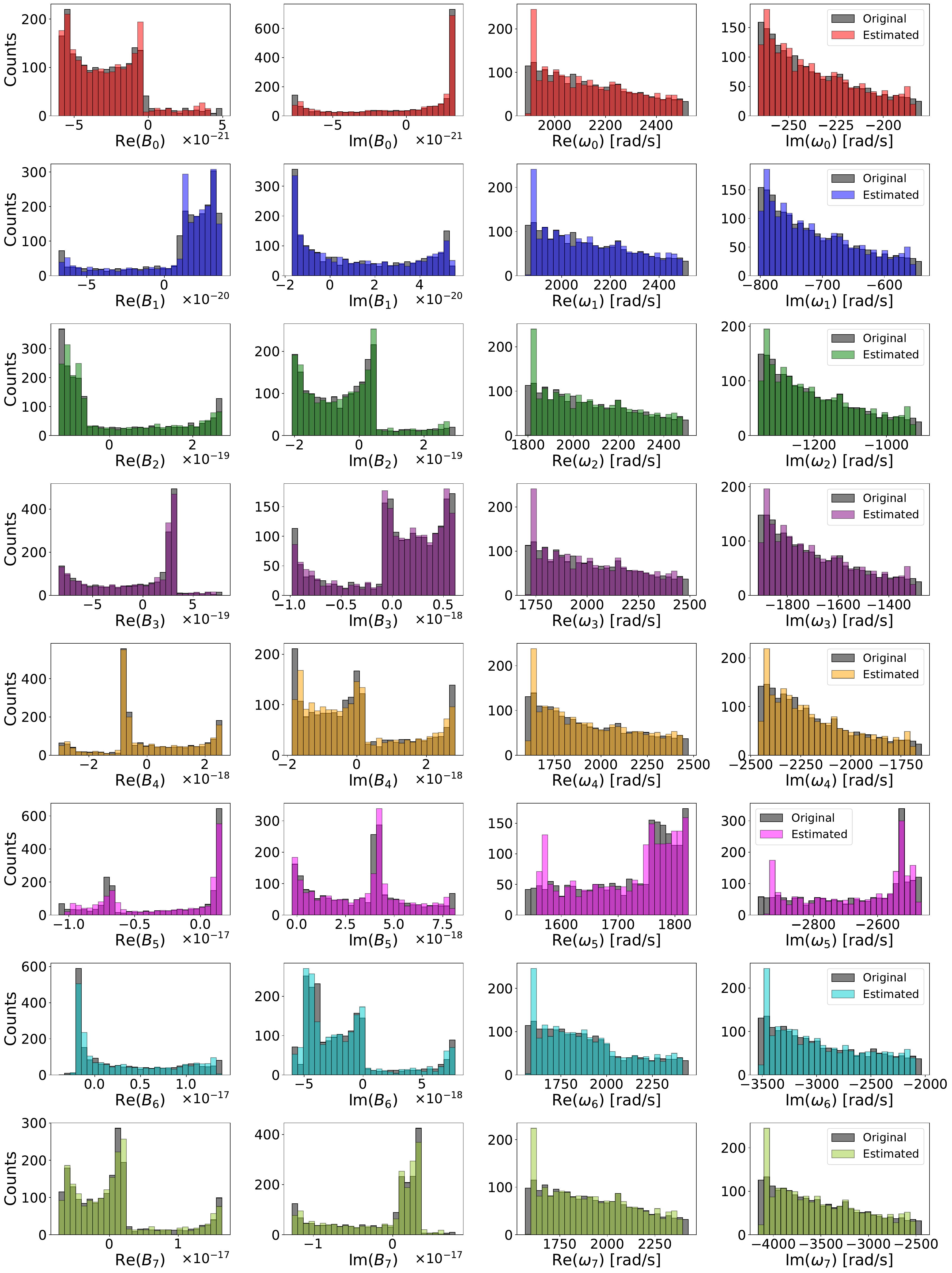}
  \caption{Comparison between the distributions of the true parameters (gray) and the estimated parameters for all 32 parameters.}
  \label{fig:8waves_hist}
\end{figure*}

\begin{figure*}[tbp]
  \centering
  \includegraphics[width=0.65\linewidth]{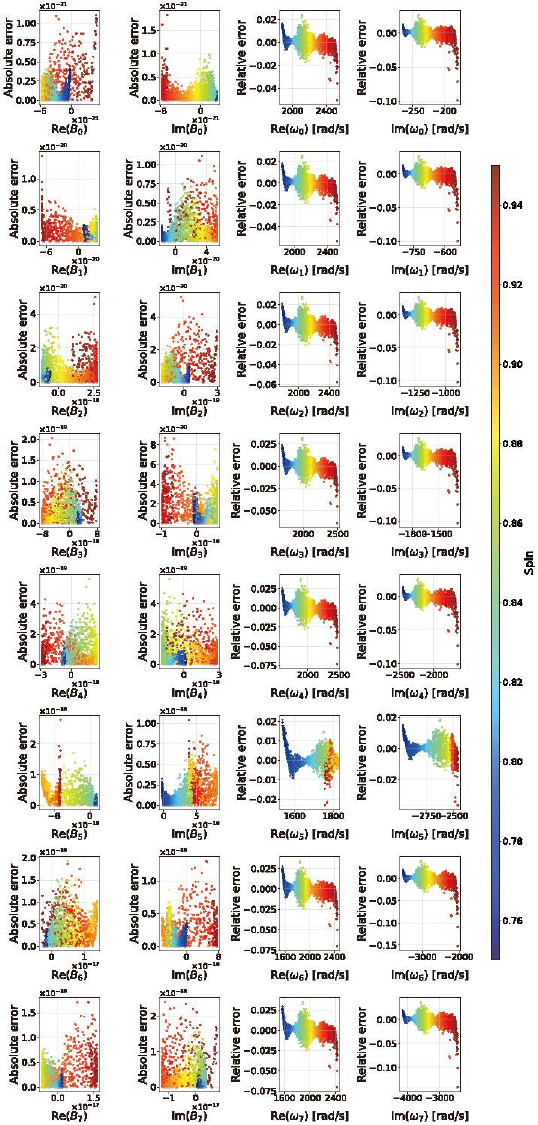}
  \caption{Parameter-estimation errors for the 32 inferred parameters.
  The absolute errors are shown for the excitation factors, and the relative errors are shown for the QNM frequencies.}
  \label{fig:8waves_err}
\end{figure*}

We finally consider simultaneous estimation of all eight QNM components.
Both the input and target waveforms contain the components with $n=0,\ldots,7$.
The model estimates the real and imaginary parts of $B_n$ and $\omega_n$ for all eight components, giving a total of 32 parameters.
Training and validation are performed over the spin interval $[0.75,\,0.95]$, using 8,000 training samples and 2,000 validation samples.
The denoising example, the encoder-based match score, the decoder-based match score, and the parameter distributions and errors are shown in Figs.~\ref{fig:8waves_denoised}, \ref{fig:8waves_match_spin}, \ref{fig:8waves_match}, \ref{fig:8waves_hist}, and~\ref{fig:8waves_err}, respectively.

Figure~\ref{fig:8waves_denoised} presents a representative noisy input waveform and the corresponding denoised waveform.
The denoised waveform closely reproduces the noise-free target waveform, showing that the decoder can reconstruct the eight-component signal in this example.

\begin{figure}[tb]
  \centering
  \includegraphics[width=0.75\linewidth]{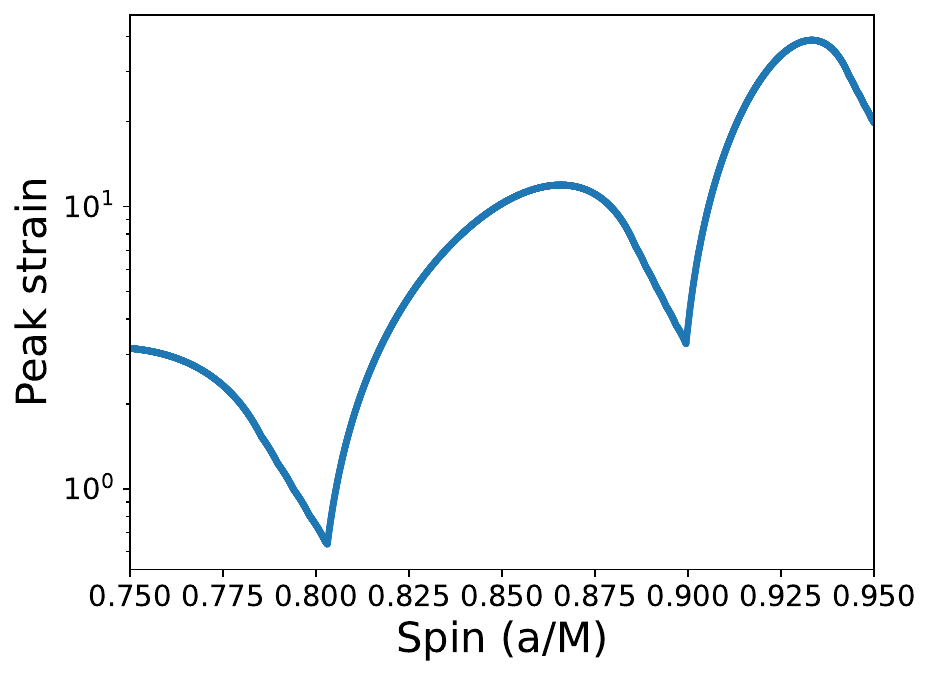}
  \caption{Relationship between spin and the peak amplitude of the standardized input data used for training.}
  \label{fig:8waves_peak}
\end{figure}

As shown in Fig.~\ref{fig:8waves_match_spin}, the match score between the waveform generated from the estimated parameters and the true waveform exceeds 0.95 over most of the spin range.
However, the match score decreases around a spin of 0.8.
Figure~\ref{fig:8waves_peak} shows the relationship between the spin and the peak amplitude of the standardized waveform.
The peak amplitude has a local minimum around a spin of 0.8, coinciding with the decrease in the match score.
This coincidence suggests that the smaller waveform amplitude may make it more difficult for the model to capture the signal features sufficiently.
The peak amplitude in Fig.~\ref{fig:8waves_peak} is calculated from the standardized input data used for training, and relative amplitude variations therefore remain after dataset-level standardization.

The match score between the decoder-generated waveform and the target waveform, shown in Fig.~\ref{fig:8waves_match}, also decreases around a spin value of 0.8.
This decrease occurs in the same spin region as that in Fig.~\ref{fig:8waves_match_spin}.
Furthermore, the histogram in Fig.~\ref{fig:8waves_hist}, which overlays the distributions of the true and estimated parameters, shows good overall agreement between the two distributions.
In addition, the error plots in Fig.~\ref{fig:8waves_err} indicate a tendency for the estimation error to increase slightly in the high-spin region.
Because the spin samples are uniformly spaced in $a/M$, the more rapid parameter variation in the high-spin region gives a lower sampling density in parameter space, which likely contributes to the increase in the errors.
These results demonstrate that the proposed model maintains generally good estimation performance even for multi-component signals containing eight components.

Overall, the model reconstructs the waveforms and recovers the 32 parameters with good agreement over most of the selected spin interval.
The localized decrease in the match score around $a/M\simeq0.8$ and the larger errors toward high spin identify regions in which the performance is less uniform.
Within this controlled, physics-informed setting, these results demonstrate the capability of the autoencoder-based framework to analyze highly multi-component damped signals.


\section{Conclusions}\label{sec:concl}
In this study, we applied an autoencoder-based latent-space framework~\cite{ref:autoencoder,ref:Iida2026,Iida:2026jox} to parameter estimation for Kerr QNM model waveforms.
The latent variables were trained to represent the physical parameters of individual QNM components, allowing denoising and parameter estimation to be performed within a common framework.
We used the recently established high-precision dataset of Kerr QNM frequencies and excitation factors from Ref.~\cite{Motohashi:2024fwt}, which is publicly available via Ref.~\cite{Motohashi2024Zenodo}.
The model waveforms are finite sums of these QNM contributions in the idealized localized-source case.

For waveforms containing eight QNM components, we first considered the estimation of the two longest-lived components across partitioned spin intervals.
When the training and validation data were drawn from the same spin interval, the model achieved generally high waveform-reconstruction and parameter-estimation performance across the spin range considered.
In contrast, when the validation interval differed from the training interval, the performance was highest near the overlap between the two intervals and degraded as the validation spins moved farther outside the training domain.
For completely disjoint intervals, both the reconstruction and parameter estimates deteriorated substantially.
These results show that the present framework primarily learns the prescribed waveform family within, or close to, its training domain, rather than providing reliable extrapolation to distant spins.
In the present controlled validation setting, this degradation makes the limitation of extrapolation explicit rather than yielding uniformly high agreement that could conceal incorrect parameter estimates.

We also investigated simultaneous inference of all 32 parameters of an eight-component waveform in the spin interval $0.75\leq a/M\leq0.95$.
Despite the noisy input and the overlapping damped components, the model obtained good waveform agreement and parameter-distribution agreement over most of this interval, while exhibiting localized reductions in performance around $a/M\simeq0.8$ and toward high spin.
The training data include the rapidly varying QNM frequencies and excitation factors associated with avoided crossings and resonant excitation in the $(l,m)=(2,2)$ sector near $a/M\simeq0.9$~\cite{Motohashi:2024fwt}.
The present results therefore show that the framework can accommodate such spin-dependent structures when they are represented in the training data.
They do not, however, constitute an independent identification of resonant excitation from an otherwise unknown ringdown waveform.
Instead, this controlled setting provides a useful testbed for determining how learned parameter-estimation methods behave near these physically distinctive QNM structures.

The good performance in the eight-component setting should not be interpreted as a direct resolution of the longstanding problem of overtone extraction from generic ringdown data.
Conventional ringdown fitting methods generally infer unknown mode amplitudes and, in agnostic analyses, may also infer the mode frequencies directly from the data.
By contrast, the present network is trained on a restricted, physics-informed family of finite Kerr-QNM sums, with spin-dependent frequencies and excitation factors supplied as labels.
In addition, the waveforms assume the localized-source limit and contain neither generic source dependence nor the non-QNM contributions expected in more realistic signals.
A one-to-one comparison with conventional extraction methods is therefore not appropriate.
Nevertheless, the recovery of 32 parameters from noisy, highly multi-component signals demonstrates that the autoencoder can retain detailed information about a structured damped-signal family in its learned representation.
Together with the earlier demonstration of this capability for multi-component damped signals~\cite{Iida:2026jox}, the present work motivates further tests of physics-informed machine-learning methods for QNM inference.

It should be noted that all analyses presented here were performed at a single noise level. While the results demonstrate the feasibility of the proposed approach under these conditions, a systematic investigation of its performance as a function of SNR remains an important topic for future study.
The next step is to evaluate the framework under more realistic conditions.
This includes colored and non-Gaussian detector noise, numerical-relativity waveforms with generic source dependence and non-QNM contributions, and methods to quantify parameter uncertainties and identify inputs outside the training distribution.
Such studies will determine whether the controlled performance demonstrated here can be extended to progressively less idealized ringdown signals.
Our longer-term goal is to develop a reliable framework for applying these methods to observational ringdown data.

\begin{acknowledgments}
This research was supported in part by the Japan Society for the Promotion of Science (JSPS) Grant-in-Aid for Scientific Research [Nos.\ 22K03639 and 26K21813] (H.\ Motohashi) and [Nos. 23H01176, 23K25872 and 23K22499] (H.\ Takahashi). This research was supported by the Joint Research Program of the Institute for Cosmic Ray Research, University of Tokyo, and Tokyo City University Prioritized Studies and Research Equipment Program.
\end{acknowledgments}

\bibliography{main}

@article{GWTC-1,
  title = {{GWTC-1: A Gravitational-Wave Transient Catalog of Compact Binary Mergers Observed by LIGO and Virgo during the First and Second Observing Runs}},
  author = {Abbott, B. P. and others},
  collaboration = {LIGO Scientific Collaboration and Virgo Collaboration},
  journal = {Phys. Rev. X},
  volume = {9},
  issue = {3},
  pages = {031040},
  numpages = {49},
  year = {2019},
  month = {Sep},
  publisher = {American Physical Society},
  doi = {10.1103/PhysRevX.9.031040},
  url = {https://link.aps.org/doi/10.1103/PhysRevX.9.031040}
}

@article{GWTC-2,
  title = {{GWTC-2: Compact Binary Coalescences Observed by LIGO and Virgo during the First Half of the Third Observing Run}},
  author = {Abbott, R. and others},
  collaboration = {LIGO Scientific Collaboration and Virgo Collaboration},
  journal = {Phys. Rev. X},
  volume = {11},
  issue = {2},
  pages = {021053},
  numpages = {52},
  year = {2021},
  month = {Jun},
  publisher = {American Physical Society},
  doi = {10.1103/PhysRevX.11.021053},
  url = {https://link.aps.org/doi/10.1103/PhysRevX.11.021053}
}

@article{GWTC-2.1,
  title = {{GWTC-2.1: Deep extended catalog of compact binary coalescences observed by LIGO and Virgo during the first half of the third observing run}},
  author={R. Abbott and others},
  collaboration = {The LIGO Scientific Collaboration and the Virgo Collaboration},
  journal = {Phys. Rev. D},
  volume = {109},
  issue = {2},
  pages = {022001},
  numpages = {45},
  year = {2024},
  month = {Jan},
  publisher = {American Physical Society},
  doi = {10.1103/PhysRevD.109.022001},
  url = {https://link.aps.org/doi/10.1103/PhysRevD.109.022001}
}

@article{GWTC-3,
  title = {{GWTC-3: Compact Binary Coalescences Observed by LIGO and Virgo during the Second Part of the Third Observing Run}},
  author = {Abbott, R. and others},
  collaboration = {LIGO Scientific Collaboration, Virgo Collaboration, and KAGRA Collaboration},
  journal = {Phys. Rev. X},
  volume = {13},
  issue = {4},
  pages = {041039},
  numpages = {89},
  year = {2023},
  month = {Dec},
  publisher = {American Physical Society},
  doi = {10.1103/PhysRevX.13.041039},
  url = {https://link.aps.org/doi/10.1103/PhysRevX.13.041039}
}

@article{GWTC4.0_results,
doi = {10.3847/2041-8213/ae2c74},
url = {https://doi.org/10.3847/2041-8213/ae2c74},
year = {2026},
month = {jun},
publisher = {The American Astronomical Society},
volume = {1004},
number = {2},
pages = {L22},
author={R. Abbott and others},
collaboration={The LIGO Scientific Collaboration and the Virgo Collaboration and the KAGRA Collaboration},
title = {{GWTC-4.0: Updating the Gravitational-wave Transient Catalog with Observations from the First Part of the Fourth LIGO–Virgo–KAGRA Observing Run}},
journal = {The Astrophysical Journal Letters}
}

@misc{GWTC5.0_results,
 title={{GWTC-5.0: Observations from the Second Part of the Fourth LIGO-Virgo-KAGRA Observing Run and Updates to the Gravitational-Wave Transient Catalog}}, 
 author={R. Abbott and others},
 collaboration={The LIGO Scientific Collaboration and the Virgo Collaboration and the KAGRA Collaboration},
 year={2026},
 eprint={2605.27225},
 rchivePrefix={arXiv},
 primaryClass={gr-qc},
 url={https://arxiv.org/abs/2605.27225}
}

@misc{ref:Autoencoder_intro,
      title={{An Introduction to Autoencoders}}, 
      author={Umberto Michelucci},
      year={2022},
      eprint={2201.03898},
      archivePrefix={arXiv},
      primaryClass={cs.LG},
      url={https://arxiv.org/abs/2201.03898}, 
}

@article{ref:Autoencoder_rev,
  author    = {K. Berahmand and F. Daneshfar and E. S. Salehi and Y. Li and Y. Xu},
  title     = {Autoencoders and their applications in machine learning: a survey},
  journal   = {Artificial Intelligence Review},
  volume    = {57},
  number    = {2},
  pages     = {28},
  year      = {2024},
  DOI       = {10.1007/s10462-023-10662-6},
  url       = {https://link.springer.com/article/10.1007/s10462-023-10662-6}
}

@article{Visschers:21,
author = {Jim C. Visschers and Emma Wilson and Thomas Conneely and Andrey Mudrov and Lykourgos Bougas},
journal = {Opt. Express},
number = {5},
pages = {6863--6878},
publisher = {Optica Publishing Group},
title = {Rapid parameter determination of discrete damped sinusoidal oscillations},
volume = {29},
month = {Mar},
year = {2021},
url = {https://opg.optica.org/oe/abstract.cfm?URI=oe-29-5-6863},
doi = {10.1364/OE.411972},
}

@article{ref:Halmer,
    author = {Halmer, Daniel and von Basum, Golo and Hering, Peter and Mürtz, Manfred},
    title = {Fast exponential fitting algorithm for real-time instrumental use},
    journal = {Review of Scientific Instruments},
    volume = {75},
    number = {6},
    pages = {2187-2191},
    year = {2004},
    month = {06},
    issn = {0034-6748},
    doi = {10.1063/1.1711189},
    url = {https://doi.org/10.1063/1.1711189}
}

@article{ref:Bostrom,
    author = {Bostrom, G. and Atkinson, D. and Rice, A.},
    title = {The discrete Fourier transform algorithm for determining decay constants—Implementation using a field programmable gate array},
    journal = {Review of Scientific Instruments},
    volume = {86},
    number = {4},
    pages = {043106},
    year = {2015},
    month = {04},
    issn = {0034-6748},
    doi = {10.1063/1.4916709},
    url = {https://doi.org/10.1063/1.4916709},
}

@article{ref:Iida2026,
  title={{Parameter Estimation of Two-Component Superposed Decaying Oscillation Signal Using Autoencoders}},
  author={Momoka Iida and Hayato Motohashi and Hirotaka Takahashi},
  journal={Journal of Japan Society for Fuzzy Theory and Intelligent Informatics},
  volume={38},
  number={1},
  pages={528-531},
  year={2026},
  doi={10.3156/jsoft.38.1_528}
}

@article{Iida:2026jox,
    author = "Iida, Momoka and Motohashi, Hayato and Takahashi, Hirotaka",
    title = "{Autoencoder-Based Parameter Estimation for Superposed Multi-Component Damped Sinusoidal Signals}",
    journal={IEEE Access, in press},
    eprint = "2604.03985",
    archivePrefix = "arXiv",
    primaryClass = "cs.LG",
    month = "4",
    year = "2026"
}

@inproceedings{Kingma:2014vow,
    author = "Kingma, Diederik P. and Ba, Jimmy",
    title = "{Adam: A Method for Stochastic Optimization}",
    booktitle = "{International Conference on Learning Representations}",
    eprint = "1412.6980",
    archivePrefix = "arXiv",
    primaryClass = "cs.LG",
    month = "12",
    year = "2014"
}

@article{ref:SGD,
title = {Backpropagation and stochastic gradient descent method},
journal = {Neurocomputing},
volume = {5},
number = {4},
pages = {185-196},
year = {1993},
issn = {0925-2312},
doi = {https://doi.org/10.1016/0925-2312(93)90006-O},
url = {https://www.sciencedirect.com/science/article/pii/092523129390006O},
author = {{Shun-ichi} Amari},
}

@misc{ref:Pycbc,
  doi = {10.5281/ZENODO.8340277},
  url = {https://zenodo.org/record/8340277},
  author = {Nitz,  Alex and others},
  title = {gwastro/pycbc: v2.2.2 release of PyCBC},
  publisher = {Zenodo},
  year = {2023},
  copyright = {Open Access}
}

@article{ref:bearing_fault,
  author    = {M. Kanemaru and M. Tsukima and T. Miyauchi and K. Hayashi},
  title     = {Bearing fault detection in induction machine based on stator current spectrum monitoring},
  journal   = {IEEJ Journal of Industry Applications},
  volume    = {7},
  number    = {3},
  pages     = {282--288},
  year      = {2018},
  url       = {https://www.jstage.jst.go.jp/article/ieejjia/7/3/7_282/_article/-char/ja}
}

@article{ref:structure_health_monitoring,
  author    = {J. Ndambi and B. Peeters and J. Maeck and J. De Visscher and M. Wahab and J. Vantomme and G. De Roeck and W. P. De Wilde},
  title     = {Comparison of techniques for modal analysis of concrete structures},
  journal   = {Engineering Structures},
  volume    = {22},
  number    = {9},
  pages     = {1159--1166},
  year      = {2000},
  url       = {https://www.sciencedirect.com/science/article/pii/S0141029699000541}
}

@book{ref:NMR,
  author    = {Harald G{\"u}nther},
  title     = {{NMR Spectroscopy: Basic Principles, Concepts and Applications in Chemistry}},
  edition   = {3},
  publisher = {Wiley-VCH},
  address   = {Weinheim},
  year      = {2013},
  isbn      = {9783527330003}
}

@article{ref:FID1,
  author  = {I. M. Savukov and M. V. Romalis},
  title   = {{NMR Detection with an Atomic Magnetometer}},
  journal = {Physical Review Letters},
  volume  = {94},
  number  = {12},
  pages   = {123001},
  year    = {2005},
  doi     = {10.1103/PhysRevLett.94.123001},
  url = {https://link.aps.org/doi/10.1103/PhysRevLett.94.123001}
}

@article{ref:FID2,
  author  = {D. Hunter and S. Piccolomo and J. D. Pritchard and N. L. Brockie and T. E. Dyer and E. Riis},
  title   = {{Free-Induction-Decay Magnetometer Based on a Microfabricated Cs Vapor Cell}},
  journal = {Physical Review Applied},
  volume  = {10},
  number  = {1},
  pages   = {014002},
  year    = {2018},
  doi     = {10.1103/PhysRevApplied.10.014002},
  url = {https://link.aps.org/doi/10.1103/PhysRevApplied.10.014002}
}

@article{ref:CRDP1,
  author  = {Thomas M{\"u}ller and Kenneth B. Wiberg and Patrick H. Vaccaro},
  title   = {{Cavity Ring-Down Polarimetry ({CRDP}): A New Scheme for Investigating Circular Birefringence and Circular Dichroism in the Gas Phase}},
  journal = {The Journal of Physical Chemistry A},
  volume  = {104},
  number  = {25},
  pages   = {5959--5968},
  year    = {2000},
  doi     = {10.1021/jp000705n},
  url     = {https://pubs.acs.org/doi/10.1021/jp000705n}
}

@inproceedings{ref:CRDE1,
  author    = {V. Papadakis and M. A. Everest and K. Stamataki and S. Tzortzakis and B. Loppinet and T. P. Rakitzis},
  title     = {{Development of Cavity Ring-Down Ellipsometry with Spectral and Submicrosecond Time Resolution}},
  booktitle = {{Instrumentation, Metrology, and Standards for Nanomanufacturing, Optics, and Semiconductors V}},
  editor    = {Michael T. Postek},
  volume    = {8105},
  pages     = {81050L},
  year      = {2011},
  publisher = {SPIE},
  doi       = {10.1117/12.903053},
  url ={https://www.spiedigitallibrary.org/conference-proceedings-of-spie/8105/1/Development-of-cavity-ring-down-ellipsometry-with-spectral-and-submicrosecond/10.1117/12.903053.short}
}

@article{ref:vibration_analysis,
  author  = {Irina Trendafilova},
  title   = {{Singular Spectrum Analysis for the Investigation of Structural Vibrations}},
  journal = {Engineering Structures},
  volume  = {242},
  pages   = {112531},
  year    = {2021},
  doi     = {10.1016/j.engstruct.2021.112531},
  url = {https://www.sciencedirect.com/science/article/abs/pii/S0141029621006817}
}

@article{ref:radar,
  author  = {J. Andrew Zhang and Fan Liu and Christos Masouros and Robert W. Heath Jr. and Zhiyong Feng and Le Zheng and Athina Petropulu},
  title   = {{An Overview of Signal Processing Techniques for Joint Communication and Radar Sensing}},
  journal = {IEEE Journal of Selected Topics in Signal Processing},
  volume  = {15},
  number  = {6},
  pages   = {1295--1315},
  year    = {2021},
  doi     = {10.1109/JSTSP.2021.3113120},
  url = {https://ieeexplore.ieee.org/document/9540344?signout=success}
}

@article{ref:sonar,
  author  = {Peixuan Yang},
  title   = {{An Imaging Algorithm for High-Resolution Imaging Sonar System}},
  journal = {Multimedia Tools and Applications},
  volume  = {83},
  number  = {11},
  pages   = {31957--31973},
  year    = {2024},
  doi     = {10.1007/s11042-023-16757-0},
  url={https://link.springer.com/article/10.1007/s11042-023-16757-0}
}

@article{ref:communication,
  author  = {Anas Chaaban and Zouheir Rezki and Mohamed{-}Slim Alouini},
  title   = {{On the Capacity of Intensity-Modulation Direct-Detection Gaussian Optical Wireless Communication Channels: A Tutorial}},
  journal = {IEEE Communications Surveys \& Tutorials},
  volume  = {24},
  number  = {1},
  pages   = {455--491},
  year    = {2022},
  doi     = {10.1109/COMST.2021.3120087},
  url = {https://ieeexplore.ieee.org/document/9570774}
}

@article{ref:nuclear_dipole_review,
  author  = {A. Bracco and E. G. Lanza and A. Tamii},
  title   = {Isoscalar and isovector dipole excitations: Nuclear properties from low-lying states and from the isovector giant dipole resonance},
  journal = {Progress in Particle and Nuclear Physics},
  volume  = {106},
  pages   = {360--433},
  year    = {2019},
  doi     = {10.1016/j.ppnp.2019.02.001},
  url ={https://www.sciencedirect.com/science/article/pii/S0146641019300031}  
}

@article{ref:autoencoder,
  author    = {J. C. Visschers and D. Budker and L. Bougas},
  title     = {Rapid parameter estimation of discrete decaying signals using autoencoder networks},
  journal   = {Machine Learning: Science and Technology},
  volume    = {2},
  number    = {4},
  pages     = {045024},
  year      = {2021},
  url       = {https://iopscience.iop.org/article/10.1088/2632-2153/ac1eea}
}

@inproceedings{ref:DataDrivenDamped,
  author    = {Y. Xie and M. B. Wakin and G. Tang},
  title     = {{Data-driven Parameter Estimation Of Contaminated Damped Exponentials}},
  booktitle = {55th Asilomar Conference on Signals, Systems, and Computers},
  pages     = {800--804},
  year      = {2021},
  url       = {https://ieeexplore.ieee.org/document/9723189}
}

@article{Giesler:2019uxc,
    author = "Giesler, Matthew and Isi, Maximiliano and Scheel, Mark A. and Teukolsky, Saul",
    title = "{Black Hole Ringdown: The Importance of Overtones}",
    eprint = "1903.08284",
    archivePrefix = "arXiv",
    primaryClass = "gr-qc",
    doi = "10.1103/PhysRevX.9.041060",
    journal = "Phys. Rev. X",
    volume = "9",
    number = "4",
    pages = "041060",
    year = "2019"
}

@article{Baibhav:2023clw,
    author = "Baibhav, Vishal and Cheung, Mark Ho-Yeuk and Berti, Emanuele and Cardoso, Vitor and Carullo, Gregorio and Cotesta, Roberto and Del Pozzo, Walter and Duque, Francisco",
    title = "{Agnostic black hole spectroscopy: Quasinormal mode content of numerical relativity waveforms and limits of validity of linear perturbation theory}",
    eprint = "2302.03050",
    archivePrefix = "arXiv",
    primaryClass = "gr-qc",
    doi = "10.1103/PhysRevD.108.104020",
    journal = "Phys. Rev. D",
    volume = "108",
    number = "10",
    pages = "104020",
    year = "2023"
}

@article{Nee:2023osy,
    author = {Nee, Peter James and V{\"o}lkel, Sebastian H. and Pfeiffer, Harald P.},
    title = "{Role of black hole quasinormal mode overtones for ringdown analysis}",
    eprint = "2302.06634",
    archivePrefix = "arXiv",
    primaryClass = "gr-qc",
    doi = "10.1103/PhysRevD.108.044032",
    journal = "Phys. Rev. D",
    volume = "108",
    number = "4",
    pages = "044032",
    year = "2023"
}

@article{Cheung:2023vki,
    author = "Cheung, Mark Ho-Yeuk and Berti, Emanuele and Baibhav, Vishal and Cotesta, Roberto",
    title = "{Extracting linear and nonlinear quasinormal modes from black hole merger simulations}",
    eprint = "2310.04489",
    archivePrefix = "arXiv",
    primaryClass = "gr-qc",
    doi = "10.1103/PhysRevD.109.044069",
    journal = "Phys. Rev. D",
    volume = "109",
    number = "4",
    pages = "044069",
    year = "2024",
    note = "[Erratum: Phys.Rev.D 110, 049902 (2024), Erratum: Phys.Rev.D 112, 049901 (2025)]"
}

@article{Takahashi:2023tkb,
    author = "Takahashi, Kazuto and Motohashi, Hayato",
    title = "{Iterative extraction of overtones from black hole ringdown}",
    eprint = "2311.12762",
    archivePrefix = "arXiv",
    primaryClass = "gr-qc",
    doi = "10.1088/1361-6382/ad72c9",
    journal = "Class. Quant. Grav.",
    volume = "41",
    number = "19",
    pages = "195023",
    year = "2024"
}

@article{Motohashi:2024fwt,
    author = "Motohashi, Hayato",
    title = "{Resonant Excitation of Quasinormal Modes of Black Holes}",
    eprint = "2407.15191",
    archivePrefix = "arXiv",
    primaryClass = "gr-qc",
    doi = "10.1103/PhysRevLett.134.141401",
    journal = "Phys. Rev. Lett.",
    volume = "134",
    number = "14",
    pages = "141401",
    year = "2025"
}

@dataset{Motohashi2024Zenodo,
  author       = {Motohashi, H.},
  title        = {Kerr quasinormal mode frequencies and excitation factors},
  year         = {2024},
  publisher    = {Zenodo},
  doi          = {10.5281/zenodo.12696857},
  url          = {https://doi.org/10.5281/zenodo.12696857}
}

@article{LIGOScientific:2025rid,
    author = "Abac, A. G. and others",
    collaboration = "LIGO Scientific, Virgo, KAGRA",
    title = "{GW250114: Testing Hawking{\textquoteright}s Area Law and the Kerr Nature of Black Holes}",
    eprint = "2509.08054",
    archivePrefix = "arXiv",
    primaryClass = "gr-qc",
    reportNumber = "LIGO-P2500421",
    doi = "10.1103/kw5g-d732",
    journal = "Phys. Rev. Lett.",
    volume = "135",
    number = "11",
    pages = "111403",
    year = "2025"
}

@article{Kubota:2025hjk,
    author = "Kubota, Kei-ichiro and Motohashi, Hayato",
    title = "{Resonance in black hole ringdown: Benchmarking quasinormal mode excitation and extraction}",
    eprint = "2509.06411",
    archivePrefix = "arXiv",
    primaryClass = "gr-qc",
    doi = "10.1103/pwyd-nv6v",
    journal = "Phys. Rev. D",
    volume = "113",
    number = "4",
    pages = "043053",
    year = "2026"
}

@article{Berti:2025hly,
    author = "Berti, Emanuele and others",
    editor = "Berti, Emanuele and Cardoso, Vitor and Carullo, Gregorio",
    title = "{Black hole spectroscopy: from theory to experiment}",
    eprint = "2505.23895",
    archivePrefix = "arXiv",
    primaryClass = "gr-qc",
    doi = "10.1088/1361-6382/ae59e2",
    journal = "Class. Quant. Grav.",
    volume = "43",
    number = "12",
    pages = "123001",
    year = "2026"
}

@article{Bini:2023gil,
    author = "Bini, Sophie and Vedovato, Gabriele and Drago, Marco and Salemi, Francesco and Prodi, Giovanni Andrea",
    title = "{An autoencoder neural network integrated into gravitational-wave burst searches to improve the rejection of noise transients}",
    eprint = "2303.05986",
    archivePrefix = "arXiv",
    primaryClass = "gr-qc",
    doi = "10.1088/1361-6382/acd981",
    journal = "Class. Quant. Grav.",
    volume = "40",
    number = "13",
    pages = "135008",
    year = "2023"
}

@article{Sun:2025cbo,
    author = "Sun, Hui and Wang, He and He, Jibo",
    title = "{Accelerating Bayesian sampling for massive black hole binaries with prior constraints from conditional variational autoencoder}",
    eprint = "2502.09266",
    archivePrefix = "arXiv",
    primaryClass = "astro-ph.IM",
    doi = "10.1103/8r8b-hckx",
    journal = "Phys. Rev. D",
    volume = "111",
    number = "10",
    pages = "103053",
    year = "2025"
}

@article{Bada-Nerin:2024wkn,
    author = "Bada-Nerin, Roberto and Bulashenko, Oleg and Gramaxo Freitas, Osvaldo and Font, Jos{\'e} A.",
    title = "{Parameter estimation of microlensed gravitational waves with conditional variational autoencoders}",
    eprint = "2412.00566",
    archivePrefix = "arXiv",
    primaryClass = "gr-qc",
    reportNumber = "P2400553, VIR-0983A-24",
    doi = "10.1103/PhysRevD.111.084067",
    journal = "Phys. Rev. D",
    volume = "111",
    number = "8",
    pages = "084067",
    year = "2025"
}

@article{Liao:2021vec,
    author = "Liao, Chung-Hao and Lin, Feng-Li",
    title = "{Deep generative models of gravitational waveforms via conditional autoencoder}",
    eprint = "2101.06685",
    archivePrefix = "arXiv",
    primaryClass = "astro-ph.IM",
    doi = "10.1103/PhysRevD.103.124051",
    journal = "Phys. Rev. D",
    volume = "103",
    number = "12",
    pages = "124051",
    year = "2021"
}

@article{Sun:2025afb,
    author = "Sun, Mengfei and Wu, Jie and Li, Jin and Mccane, Brendan and Yang, Nan and Ma, Xianghe and Wang, Borui and Zhang, Minghui",
    title = "{Conditional autoencoder for generating binary neutron star waveforms with tidal and precession effects}",
    eprint = "2503.19512",
    archivePrefix = "arXiv",
    primaryClass = "astro-ph.GA",
    doi = "10.1103/kmlw-y7yw",
    journal = "Phys. Rev. D",
    volume = "112",
    number = "8",
    pages = "084016",
    year = "2025"
}

@article{Moreno:2021fvp,
    author = "Moreno, Eric A. and Borzyszkowski, Bartlomiej and Pierini, Maurizio and Vlimant, Jean-Roch and Spiropulu, Maria",
    title = "{Source-agnostic gravitational-wave detection with recurrent autoencoders}",
    eprint = "2107.12698",
    archivePrefix = "arXiv",
    primaryClass = "gr-qc",
    doi = "10.1088/2632-2153/ac5435",
    journal = "Mach. Learn. Sci. Tech.",
    volume = "3",
    number = "2",
    pages = "025001",
    year = "2022"
}

@article{ref:Garg_2026,
  title = {Autoencoder model for fast generation of effective one-body gravitational waveform approximations},
  author = {Garg, Suyog and Lin, Feng-Li and Cannon, Kipp},
  journal = {Phys. Rev. D},
  volume = {114},
  issue = {2},
  pages = {024044},
  numpages = {14},
  year = {2026},
  month = {Jul},
  publisher = {American Physical Society},
  doi = {10.1103/h92m-k44j},
  url = {https://link.aps.org/doi/10.1103/h92m-k44j}
}

@article{ref:Gabbard_2022,
  title = {Bayesian parameter estimation using conditional variational autoencoders for gravitational-wave astronomy},
  author = {Gabbard, Hunter and Messenger, Chris and Heng, Ik Siong and Tonolini, Francesco and Murray-Smith, Roderick},
  journal = {Nature Physics},
  volume = {18},
  issue = {1},
  pages = {112},
  numpages = {5},
  year = {2022},
  month = {January},
  publisher = {Springer Nature},
  doi = {10.1038/s41567-021-01425-7},
  url = {https://doi.org/10.1038/s41567-021-01425-7}
}

@article{Lo:2025njp,
    author = "Lo, Rico K. L. and Sabani, Leart and Cardoso, Vitor",
    title = "{Quasinormal modes and excitation factors of Kerr black holes}",
    eprint = "2504.00084",
    archivePrefix = "arXiv",
    primaryClass = "gr-qc",
    doi = "10.1103/PhysRevD.111.124002",
    journal = "Phys. Rev. D",
    volume = "111",
    number = "12",
    pages = "124002",
    year = "2025"
}

@article{Leaver:1985ax,
    author = "Leaver, E.W.",
    title = "{An Analytic representation for the quasi normal modes of Kerr black holes}",
    doi = "10.1098/rspa.1985.0119",
    journal = "Proc. Roy. Soc. Lond. A",
    volume = "402",
    pages = "285--298",
    year = "1985"
}

@article{Leaver:1986gd,
    author = "Leaver, Edward W.",
    title = "{Spectral decomposition of the perturbation response of the Schwarzschild geometry}",
    doi = "10.1103/PhysRevD.34.384",
    journal = "Phys. Rev. D",
    volume = "34",
    pages = "384--408",
    year = "1986"
}

@article{Leaver:1986a,
    author = "Leaver, Edward W.",
    title = "{Solutions to a generalized spheroidal wave equation: Teukolsky’s equations in general relativity, and the two-center problem in molecular quantum mechanics}",
    doi = "10.1063/1.527130",
    journal = "J. Math. Phys.",
    volume = "27",
    pages = "1238",
    year = "1986"
}

@article{Giesler:2024hcr,
    author = "Giesler, Matthew and others",
    title = "{Overtones and nonlinearities in binary black hole ringdowns}",
    eprint = "2411.11269",
    archivePrefix = "arXiv",
    primaryClass = "gr-qc",
    reportNumber = "YITP-24-154, RIKEN-iTHEMS-Report-24",
    doi = "10.1103/PhysRevD.111.084041",
    journal = "Phys. Rev. D",
    volume = "111",
    number = "8",
    pages = "084041",
    year = "2025"
}

@article{Ma:2022wpv,
    author = "Ma, Sizheng and Mitman, Keefe and Sun, Ling and Deppe, Nils and H{\'e}bert, Fran{\c{c}}ois and Kidder, Lawrence E. and Moxon, Jordan and Throwe, William and Vu, Nils L. and Chen, Yanbei",
    title = "{Quasinormal-mode filters: A new approach to analyze the gravitational-wave ringdown of binary black-hole mergers}",
    eprint = "2207.10870",
    archivePrefix = "arXiv",
    primaryClass = "gr-qc",
    doi = "10.1103/PhysRevD.106.084036",
    journal = "Phys. Rev. D",
    volume = "106",
    number = "8",
    pages = "084036",
    year = "2022"
}

@article{Nollert:1999ji,
    author = "Nollert, Hans-Peter",
    title = "{Quasinormal modes: the characteristic `sound' of black holes and neutron stars}",
    doi = "10.1088/0264-9381/16/12/201",
    journal = "Class. Quant. Grav.",
    volume = "16",
    pages = "R159--R216",
    year = "1999"
}

@article{Kokkotas:1999bd,
    author = "Kokkotas, Kostas D. and Schmidt, Bernd G.",
    title = "{Quasinormal modes of stars and black holes}",
    eprint = "gr-qc/9909058",
    archivePrefix = "arXiv",
    doi = "10.12942/lrr-1999-2",
    journal = "Living Rev. Rel.",
    volume = "2",
    pages = "2",
    year = "1999"
}

@article{Berti:2009kk,
    author = "Berti, Emanuele and Cardoso, Vitor and Starinets, Andrei O.",
    title = "{Quasinormal modes of black holes and black branes}",
    eprint = "0905.2975",
    archivePrefix = "arXiv",
    primaryClass = "gr-qc",
    doi = "10.1088/0264-9381/26/16/163001",
    journal = "Class. Quant. Grav.",
    volume = "26",
    pages = "163001",
    year = "2009"
}

@article{Konoplya:2011qq,
    author = "Konoplya, R. A. and Zhidenko, A.",
    title = "{Quasinormal modes of black holes: From astrophysics to string theory}",
    eprint = "1102.4014",
    archivePrefix = "arXiv",
    primaryClass = "gr-qc",
    doi = "10.1103/RevModPhys.83.793",
    journal = "Rev. Mod. Phys.",
    volume = "83",
    pages = "793--836",
    year = "2011"
}

@misc{Cheung:2026gfd,
    author = "Cheung, Mark Ho-Yeuk",
    title = "{How much of the black hole ringdown is sourced within the light ring?}",
    eprint = "2608.29466",
    archivePrefix = "arXiv",
    primaryClass = "gr-qc",
    month = "8",
    year = "2026"
}

@ARTICLE{Jensen:2004,
  author={Jensen, J. and Heusdens, R. and Jensen, S.H.},
  journal={IEEE Transactions on Speech and Audio Processing},
  title={A perceptual subspace approach for modeling of speech and audio signals with damped sinusoids},
  year={2004},
  volume={12},
  number={2},
  pages={121-132},
  doi={10.1109/TSA.2003.819948}}

@misc{kafentzis2024,
      title={On the Parameter Estimation of Sinusoidal Models for Speech and Audio Signals},
      author={George P. Kafentzis},
      year={2024},
      eprint={2401.01255},
      archivePrefix={arXiv},
      primaryClass={eess.AS},
      url={https://arxiv.org/abs/2401.01255},
}
\end{document}